\documentclass[12pt]{article}
\pdfoutput=1
\usepackage{amssymb,amsmath}
\usepackage{epsfig}
\usepackage{epstopdf}
\usepackage{latexsym}
\usepackage{graphicx}
\usepackage{booktabs}

\usepackage[margin=20pt,small]{caption}
\usepackage{subcaption}

\usepackage[toc]{appendix}

\usepackage{color}
\usepackage{datetime}
\usepackage[
      colorlinks=false,
      linkcolor=darkblue,  
      urlcolor=blue,    
      filecolor=blue,     
      citecolor=red,
linktocpage=true,
      pdfstartview=FitV,
      bookmarksopen=true,
	  hidelinks
      ]{hyperref}

\usepackage{tikz}
\usetikzlibrary{decorations.markings,decorations.pathmorphing}
\tikzset{->-/.style={decoration={markings,mark=at position #1 with {\arrow{>}}},postaction={decorate}}}
\tikzset{>=latex}
 
\newcommand{\beq}{\begin{equation}}
\newcommand{\eeq}{\end{equation}}
\newcommand{\be}{\begin{equation}}
\newcommand{\ee}{\end{equation}}
\newcommand{\beqa}{\begin{eqnarray}}
\newcommand{\eeqa}{\end{eqnarray}}
\newcommand{\beqar}{\begin{eqnarray*}}
\newcommand{\eeqar}{\end{eqnarray*}}
\newcommand{\bea}{\begin{eqnarray}}
\newcommand{\eea}{\end{eqnarray}}
\newcommand{\e}{\epsilon}

\def\im{{\rm i}}

\ifpdf
\fi

\begin{document}  

\begin{titlepage}

\begin{center}

{\Large \bf  Squashed Holography with Scalar Condensates}

\bigskip
\bigskip
\bigskip
\bigskip
\bigskip

{\bf Gabriele Conti, Thomas Hertog, Yannick Vreys \\ }
\bigskip
\bigskip
{\it Institute for Theoretical Physics, KU Leuven \\
Celestijnenlaan 200D, B-3001 Leuven, Belgium}

\bigskip
\bigskip

\texttt{gabriele.conti,~thomas.hertog,~yannick.vreys~@kuleuven.be } \\
\end{center}

\bigskip
\bigskip
\bigskip
\bigskip

\begin{abstract}
\noindent

We evaluate the partition function of the free and interacting $O(N)$ vector model on a two-parameter family of squashed three spheres in the presence of a scalar deformation. We also find everywhere regular solutions of Einstein gravity coupled to a scalar field in AdS and in dS with the same double squashed boundary geometry. Remarkably, the thermodynamic properties of the AdS solutions qualitatively agree with the behavior predicted by the free $O(N)$ model with a real mass deformation. The dS bulk solutions specify the semiclassical  `no-boundary' measure over anisotropic deformations of inflationary, asymptotic de Sitter space. Through dS/CFT the partition function of the interacting $O(N)$ model yields a holographic toy model of the no-boundary measure. We find this yields a qualitatively similar probability distribution which is normalizable and globally peaked at the round three sphere, with a low amplitude for strong anisotropies.

\end{abstract}

\noindent 
\end{titlepage}


\setcounter{tocdepth}{2}
\tableofcontents

\section{Introduction}

Gauge/gravity duality allows one to use classical general relativity in asymptotically locally AdS or dS spaces to study CFTs on a range of curved backgrounds or, alternatively, to study aspects of quantum gravity by using dual CFTs defined on curved spaces.
In this paper we consider CFTs and their holographic duals on a two-parameter family of squashed three spheres in the presence of scalar excitations. In the context of AdS/CFT the scalar turns on a condensate. In the context of dS/CFT it drives inflation.

The metric on squashed spheres can be written as,
\begin{align}
ds^2= \frac{r_0^2}{4} \left((\sigma_1)^2 +\frac{1}{1+A} (\sigma_2)^2 + \frac{1}{1+B}(\sigma_3)^2 \right)  
\label{eqn:metric}\;,
\end{align}
where $r_0$ is an overall radius for which we choose the normalization $r_0=1$, and $\sigma_i$, with $i=1,2,3$, are the left-invariant one-forms of $SU(2)$ given by
\begin{equation}\label{eqn:left1forms}
 \sigma_1=-\sin\psi d\theta+\cos\psi \sin \theta d\phi \ , \qquad
 \sigma_2=\cos\psi d\theta+\sin\psi \sin \theta d\phi \ ,\qquad
 \sigma_3=d\psi+\cos \theta d\phi \ ,
\end{equation} 
with $0\leq \theta\leq \pi$, $0\leq \phi \leq 2\pi$ and $0\leq \psi \leq 4\pi$. We are interested in CFT partition functions as a function of the two squashing parameters $A$ and $B$ in \eqref{eqn:metric} and of the coupling $\tilde \alpha$ of the deformation dual to the scalar excitation in the bulk. 

An interesting CFT that is feasible to study is the three-dimensional $O(N)$ vector model, which is dual to Vasiliev higher-spin gravity in $AdS_4$ \cite{Sezgin2002,Klebanov2002,Giombi2009}. However we will not consider higher-spin gravitational theories directly. Instead we aim for a qualitative comparison between the physics of the deformed $O(N)$ model on the squashed sphere in \eqref{eqn:metric} and Einstein gravity with AdS (or dS) boundary conditions. To do so we first numerically construct new solutions in a consistent truncation of M-theory compactified on $AdS_4 \times S^7$ with a single $m^2=-2l_{AdS}^2$ scalar moving in a negative exponential potential $V$ and with a double squashed sphere of the form \eqref{eqn:metric} as their boundary. Our solutions are generalizations of the AdS Taub-NUT and Taub-Bolt solutions \cite{Taub1951, Newman1963} to two squashings and with an additional scalar condensate turned on. Comparing the thermodynamic properties of these with the partition function of the free $O(N)$ model we find that both systems exhibit a qualitatively similar behavior over much of the boundary configuration space. On the other hand they differ in specific features such as the NUT to Bolt transition at large positive values of the squashing parameters, which is evidently absent in the free dual theory.   

In the context of dS/CFT \cite{Strominger2001} the squashed spheres \eqref{eqn:metric} enter as the future boundary of homogeneous but anisotropic deformations of de Sitter space. In the second part of this paper we first find complex generalizations of the solutions above that are regular everywhere and in the large volume limit describe anisotropic deformations of real Lorentzian de Sitter space with a scalar field driving (eternal) inflation and an effective potential $-V$. These solutions are saddle points of the no-boundary wave function.
At the semiclassical level, dS/CFT conjectures that the no-boundary wave function with future de Sitter boundary conditions is dual to the partition function of complex deformations of Euclidean AdS/CFT duals defined on the future boundary \cite{Maldacena2002,Harlow2011,Hertog2011}\footnote{The applicability of Euclidean AdS/CFT stems from the observation \cite{Hertog2011} that all complex no-boundary saddle points in models with a positive scalar potential $V$ admit a geometric representation in which their amplitude is fully specified by an interior, locally AdS, domain wall region governed by an effective negative potential $-V$. This resonates with an alternative formulation of dS/CFT developed and explored in \cite{Skenderis:2007sm,McFadden2009,Bzowski:2012ih} and based on the analytic continuation of real AdS domain wall solutions to real inflationary histories. This formulation evidently does not  yield a measure over backgrounds but its predictions for the fluctuation spectra agree to leading order in the slow roll parameters with those of the framework considered here \cite{Hertog:2015nia}.}, yielding the following holographic form of the no-boundary wave function
\be
\Psi_{\rm NB} [h_{ij}, \phi]= Z^{-1}_{\rm QFT}[\tilde h_{ij}, \tilde \alpha] \exp(iS_{\rm ct}[h_{ij}, \phi]/\hbar)   \ ,
\label{dSCFT}
\ee
Here the sources $(\tilde h_{ij}, \tilde \alpha)$ are conformally related to the argument $(h_{ij}, \phi)$, and $S_{\rm ct}$ are the usual surface terms. The partition functions $Z_{\rm QFT}$ in \eqref{dSCFT} are complex deformations of Euclidean AdS/CFT duals. This form of dS/CFT reduces in minisuperspace to formulations based on analytic continuation  \cite{Maldacena2002,McFadden2009,Maldacena2011} and agrees to leading order with the higher-spin realization \cite{Anninos2011} where the $Sp(N)$ and $O(N)$ partition functions are inversely related. It is tempting indeed to view Euclidean AdS/CFT and dS/CFT as two real domains of a single complexified theory \cite{Skenderis:2007sm,Hartle2012a,Hartle2012b,Hull:1998vg,Dijkgraaf:2016lym}.

The dependence of the partition function in \eqref{dSCFT} on the values of the sources yields a holographic measure on the space of asymptotically locally de Sitter universes. General field theory results imply that the holographic amplitude of the undeformed CFT on the round $S^3$ is a local maximum with respect to scalar deformations \cite{Jafferis:2011zi, Klebanov2011} and deformations of the geometry \cite{Bobev2017,Fischetti:2017sut}. 

We also study dS/CFT for large deformations by comparing the saddle point no-boundary wave function evaluated using the complex bulk solutions above, with the partition function of the interacting $O(N)$ vector toy model on a two-parameter family of squashed three spheres in the presence of a mass deformation modeling the bulk scalar driving inflation. We consider the interacting model because according to \eqref{dSCFT} the bulk scalar now sources a scalar deformation by an operator ${\cal O}$ of dimension one, with coupling $\tilde \alpha$. This is a relevant operator which in the $O(N)$ model induces a flow from the free to the critical $O(N)$ model. Moreover, the coupling is imaginary in the dS domain of the theory as discussed above. Hence we evaluate the partition function of the critical $O(N)$ model as a function of the squashing parameters $A$ and $B$ and an imaginary mass deformation.
We find the holographic measure is normalizable and globally peaked at the round three sphere. The implications of this for eternal inflation are explored in an accompanying paper \cite{Hawking:2017wrd}.

\begin{figure}[ht!]
\centering
\includegraphics[width=0.45\textwidth]{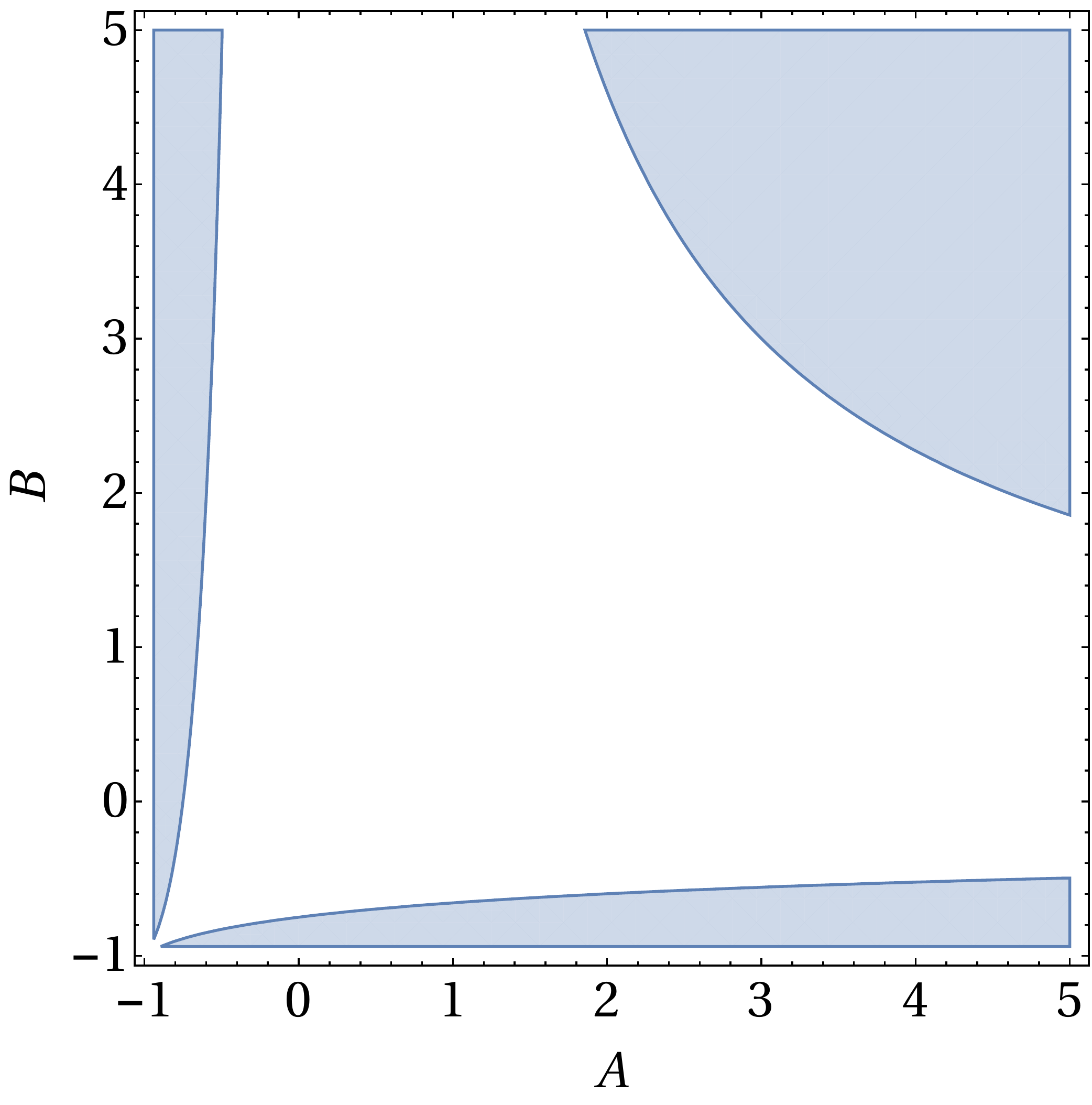}
\caption{The shaded blue region corresponds to values of the squashing parameters $(A,B)$ of $S^3$ for which the scalar curvature $R$ is negative. This includes the region of parameter space where its absolute value is large. }\label{fig:ricci}
\end{figure}

The region of the configuration space of boundary geometries with negative Ricci scalar is particularly intriguing. The Ricci scalar of a double squashed three sphere of the form \eqref{eqn:metric} is given by
\begin{align}
 R=\frac{6+8A+8B+2AB (6- A B)}{(1+A)(1+B)},
 \label{Ricci}
\end{align}
which is symmetric in $A$ and $B$. For $B=0$ there is a single region $A < -3/4$ where $R$ is negative. Adding a second squashing leads to an additional $R<0$ region associated with large positive values of both $A$ and $B$ as illustrated in Fig. \ref{fig:ricci}. In the context of dS/CFT solutions with $R<0$ boundaries can be viewed as toy models for bubble like geometries in eternal inflation. The holographic measure we compute tells us something about the likelihood to develop such surfaces of constant density in eternal inflation.

\section{Scalar Excitations of Squashed AdS Taub-NUT/Bolt }
\label{sec:AdSsols}

We are interested in four-dimensional solutions governed by the following action
\begin{equation}\label{eqn:GRaction}
I_{\rm E} = -\frac{1}{16\pi G} \int_\mathcal{M} d^4x\sqrt{g} ( R - (\nabla \Phi)^2 - 2 V(\Phi) ) -\frac{1}{8\pi G}\int_{\partial \mathcal{M}} d^3x\sqrt{h}K \,, 
\end{equation}
where $h$ and $K$ are respectively the induced metric on the boundary \eqref{eqn:metric} and its extrinsic curvature. 
For reasons that will become clear below we consider the consistent truncation of M-theory on $AdS_4 \times S^7$ consisting of gravity coupled to a single scalar $\Phi$ with potential
\begin{equation}\label{pot}
V(\Phi)= -2-\cosh(\sqrt{2}\Phi)\,,
\end{equation}
in units where $\Lambda=-3$ and hence $l^2_{AdS}=1$. For $\Phi=0$ and a single squashing, i.e. $B= 0$ in \eqref{eqn:metric}, the solutions that asymptotically tend to \eqref{eqn:metric} are well-known and can be thought of as generalizations of the asymptotically flat Taub-NUT and Taub-Bolt solutions \cite{Taub1951,Newman1963,Emparan1999}. These are two sets of topological distinct solutions that are asymptotically AdS. The NUT solutions have a zero-dimensional fixed point set, the NUT, around which the solutions are topologically $\mathbb{R}^4$. The second set, the Bolt solutions, have a two-dimensional fixed point set, the Bolt. These solutions are locally $\mathbb{R}^2 \times S^2$ in the neighbourhood of the Bolt. 

The metric of solutions that have the same NUT/Bolt topology in the interior and that asymptote to the squashed sphere \eqref{eqn:metric} with two non-vanishing squashing parameters $A$ and $B$ can be written in the following form,
\begin{align}
ds^2 = l_0(r)^2dr^2 + l_1(r)^2 \sigma _1 ^2+l_2(r)^2 \sigma _2 ^2+l_3(r)^2 \sigma _3 ^2\, , \label{eqn:doublesqansatz}
\end{align}
together with a radial scalar profile $\Phi(r)$.

Plugging this Ansatz into the equations of motion derived from the action \eqref{eqn:GRaction} one finds a system of non-linear second order differential equations for the metric functions $l_a(r)$ and the scalar $\Phi(r)$ which are given in Appendix \ref{App:AdSexpansions}. Numerical solutions to this system with the scalar set to zero were found in \cite{Bobev2016}. Here we generalize these by including a scalar excitation and its backreaction on the geometry.

We start by considering an expansion at large values of $r$ which, employing holographic terminology, we call UV expansion. The UV expansion is of the Fefferman-Graham type and the same for both the NUT and Bolt solutions since in both cases the non-trivial information is encoded in the interior of the solutions, i.e. in the IR. The leading order terms in the metric for large $r$ are given by 
\begin{equation}
ds^2= dr^2 + e^{2r} \left(A_0 \sigma_1^2+ B_0 \sigma_2^2+ C_0 \sigma_3^2 \right) \label{eqn:UVmetric1} \ .
\end{equation}
Notice that we have implemented the gauge $l_0(r)=1$. The next terms in the UV expansion of the solutions read 
\begin{equation}\label{eqn:genUV}
\begin{split} 
l_1(r)=A_0 e^{r} + A_k e^{(1-k) r} \;, \qquad 
l_2(r)=B_0 e^{r} + B_k e^{(1-k) r} \;, \qquad
l_3(r)=C_0 e^{r} + C_k e^{(1-k) r} \;,
\end{split}
\end{equation}
\begin{equation}\label{UVscalar2}
\Phi(r) = \frac{\alpha}{(A_0 B_0 C_0 )^{1/3}} e^{- r}+ \frac{\beta}{(A_0 B_0C_0)^{2/3}} e^{-2r} + D_k e^{-(2+k)r}\;,
\end{equation}
where the sum over $k$ goes over all positive integers.

We plug the series expansions \eqref{eqn:genUV}-\eqref{UVscalar2} into the Einstein equations and solve them order by order in powers of $e^r$. The results of this procedure are summarized in Appendix \ref{App:AdSexpansions}. The important upshot is that the UV expansion is controlled by seven independent parameters $\{A_0,B_0,C_0,A_3,B_3, \alpha, \beta\}$. It turns out that the Einstein equations are invariant under constant shifts of $r$ which we use to eliminate one of the parameters, setting $A_0=\frac{1}{4}$. Comparing the asymptotic form of the metric with the metric \eqref{eqn:metric} on the double squashed sphere one can find the following relation between the squashing parameters $A$ and $B$ and the leading order coefficients $B_0$ and $C_0$
\begin{equation}\label{eqn:abBC}
A = \frac{1}{4C_0} - 1\;, \qquad\qquad B = \frac{1}{4B_0} - 1\;.
\end{equation}
The leading coefficients $B_0$, $C_0$ and $\alpha$ specify the asymptotic values of metric and field. As we discuss in Appendix \ref{App:AdSexpansions} the values of the subleading coefficients ($A_3$, $B_3$ and $\beta$) are fixed by imposing regularity conditions (either on a NUT or a Bolt) in the bulk of the full solution of the nonlinear equations of motion. 

\begin{figure}[ht!]
\centering
\includegraphics[width=0.85\textwidth]{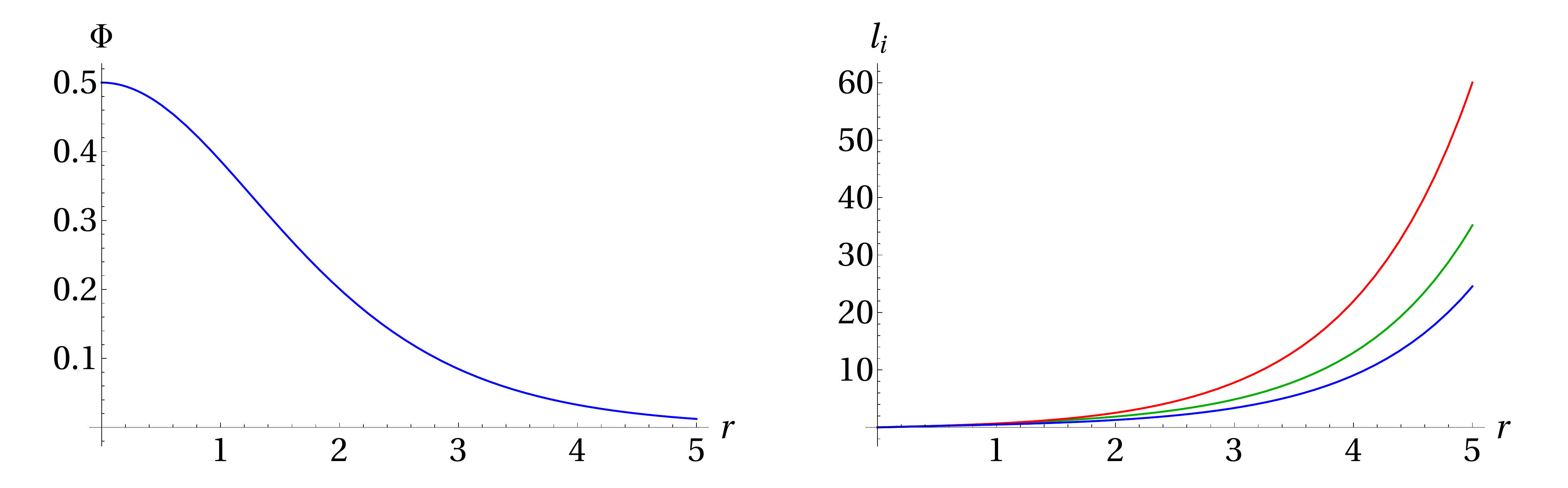}
\caption{A typical solution with a NUT in the IR and with a non-trivial scalar profile in the radial direction. The geometry asymptotes to a double squashed sphere in the UV.}\label{fig:AdSsol} 
\end{figure}

In practice we use the IR expansions (cf. \eqref{icnut} and \eqref{icbolt}) as initial conditions to integrate the equations of motion numerically to the UV. This yields a three-parameter family of solutions that are controlled by two coefficients specifying the IR behavior of the scale factors $l_a(r)$ and by the initial value $\Phi_0$ of the scalar field. There are two distinct classes of solutions. The first class consists of regular solutions for which the metric functions $l_{a}(r)$ grow exponentially, the scalar field gradually decays and the boundary metric is a sphere with two non-trivial squashing parameters as in \eqref{eqn:metric}. A representative example of a NUT solution of this kind is shown in Fig. \ref{fig:AdSsol}. We also find a class of singular solutions for which one or more of the metric functions $l_{a}(r)$ vanish at some finite value of $r$, leading to a curvature singularity. We will ignore the second class of solutions since they do not contribute to the wave function in the large three-volume regime.

The Bolt solutions only exist for sufficiently large, positive squashings. In this regime there is often more than one combination of IR parameters that yields the same values of the leading asymptotic parameters $A$, $B$ and $\alpha$.

The regularity condition on the scalar field in the interior yields a relation $\beta(\alpha)$ between the coefficients of its UV profile which depends on the squashings and encodes information about the scalar potential. In Section \ref{holo} we will compare our results with the free $O(N)$-model using the AdS/CFT duality. Under the holographic dictionary this relation can be translated to a relation between the source and vev of the dual theory. To do so we match the conformal dimensions of the deformations on both sides. Because on the CFT side the conformal dimension of the source is two, we have to use the alternate quantization of AdS, which means we fix $\beta$ on the boundary instead of $\alpha$ by applying a Legendre transform \cite{Klebanov1999}, i.e. we have Neumann boundary conditions. The exact procedure requires an analysis of the action using holographic renormalization\cite{Henningson1998,deHaro2000,Bianchi2001,Papadimitriou2007} of which the precise details can be found in Appendix \ref{App:AdSexpansions}. This immediately gives the relation between $\alpha$ in the bulk and the vev on the boundary for a source $J=\beta$,
\begin{align}
	\langle \mathcal{O}\rangle = \alpha \ .
\end{align}

The potential \eqref{pot} is special in the sense that, at least for sufficiently small squashings, $\beta$ tends to a constant when $\alpha$ (or equivalently $\Phi_0$) is taken larger. This property depends delicately on the large field regime of the potential. From a dual viewpoint this means there is a critical deformation $\beta$ at which the expectation value $\alpha$ of the operator dual to $\Phi$ diverges. This is also a feature of the vector model we consider in Section \ref{holo} below \cite{Bzowski2015}, which serves to justify the bulk boundary comparison we explore there.

\begin{figure}[ht!]
\centering
    \begin{subfigure}[t]{0.32\textwidth}
    \includegraphics[width=\textwidth]{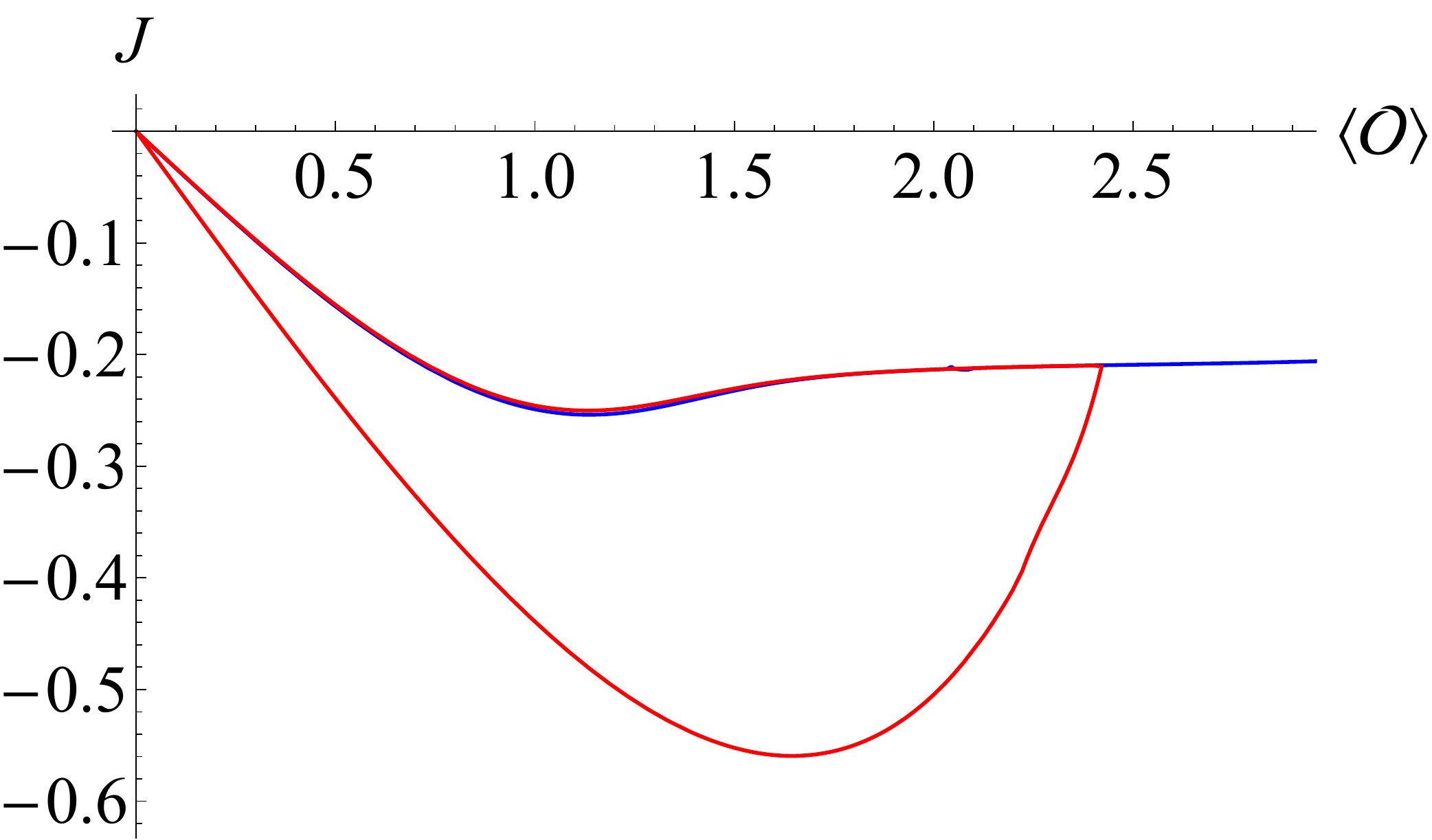}
    \caption{$A=40.0$}
    \label{fig:alphavsbetabp50}
    \end{subfigure}
	\begin{subfigure}[t]{0.32\textwidth}
	\includegraphics[width=\textwidth]{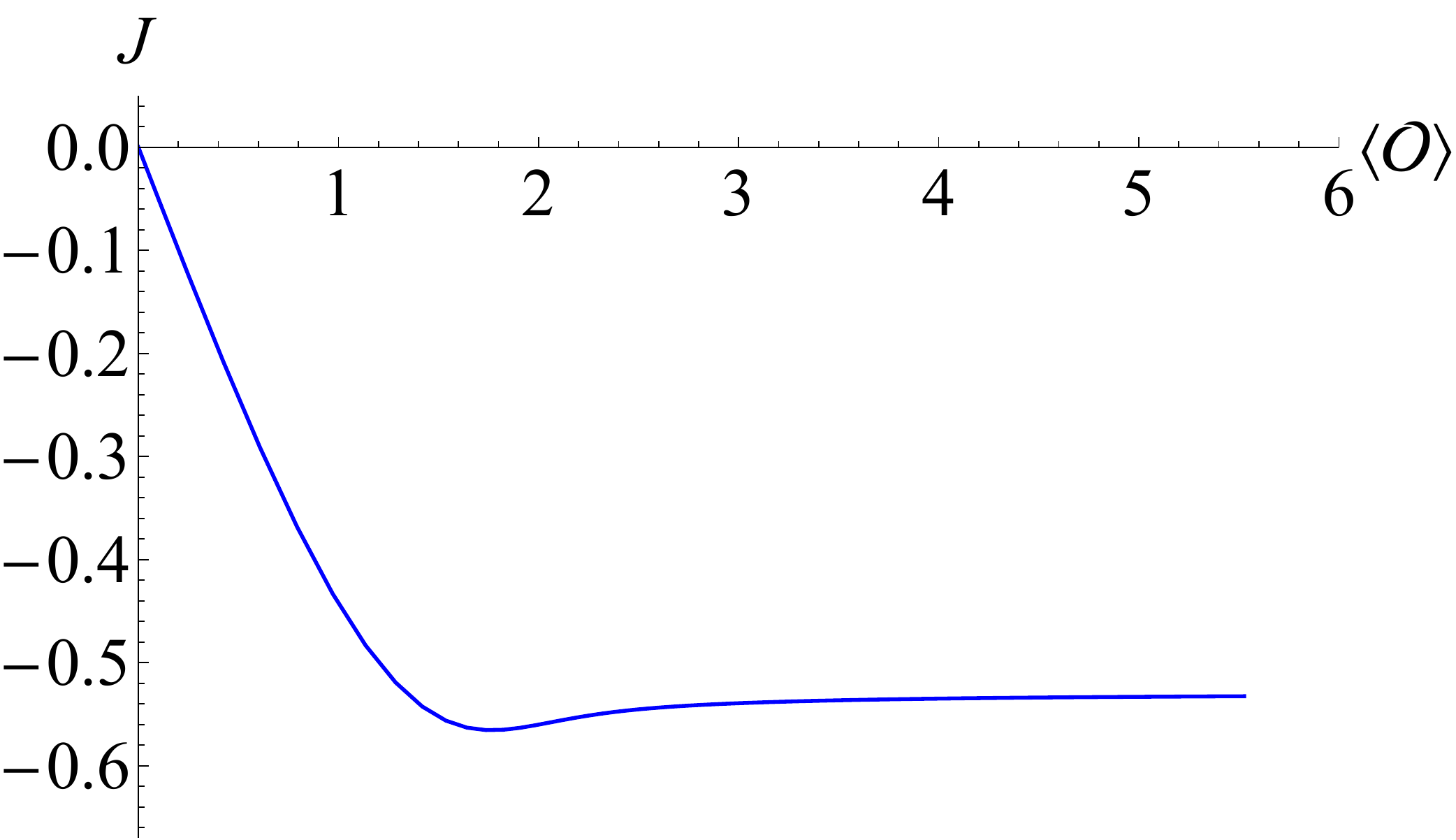} 
	\caption{$A=0$}
	\end{subfigure}
	\begin{subfigure}[t]{0.32\textwidth}
    \includegraphics[width=\textwidth]{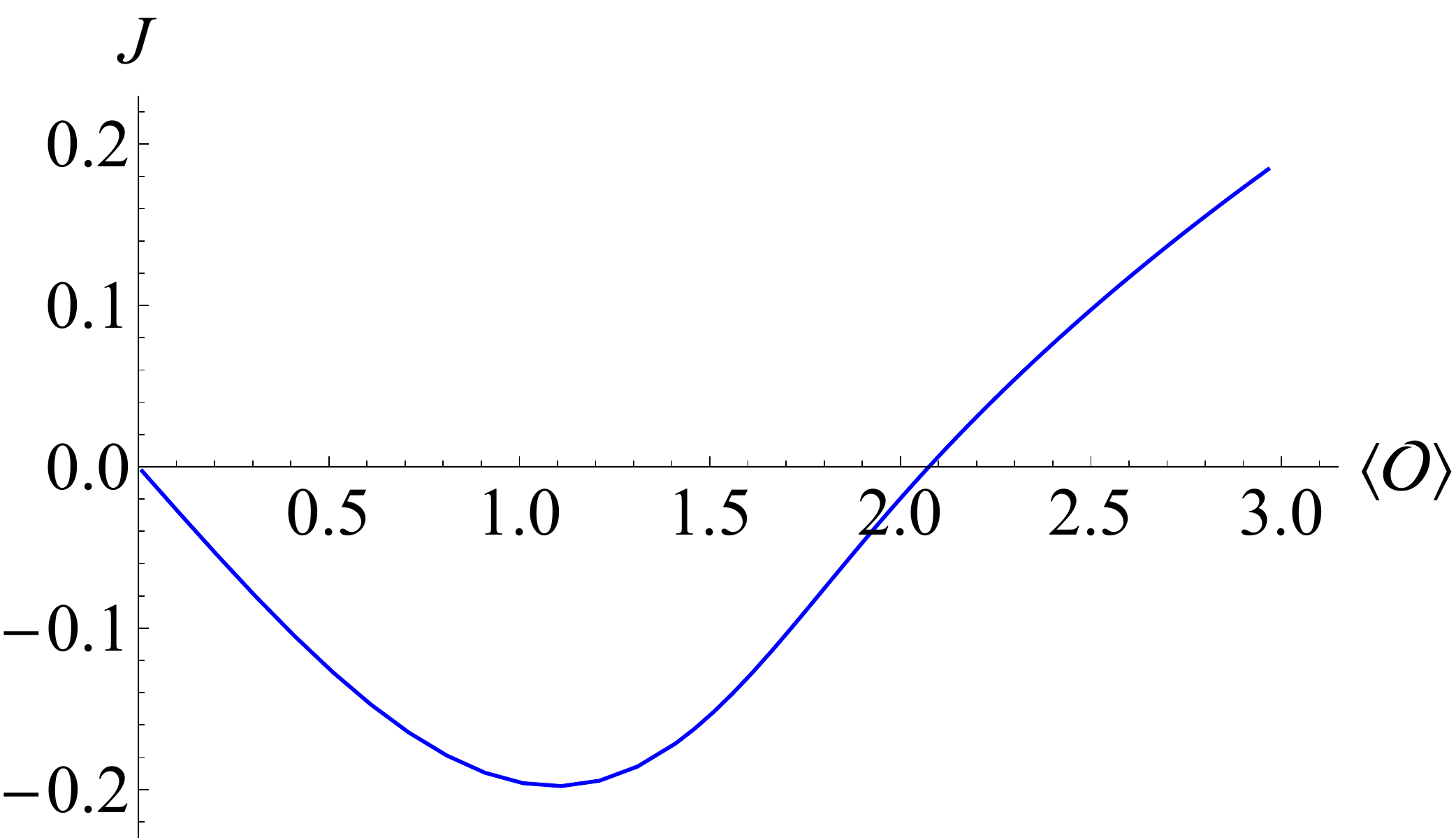}
    \caption{$A=-0.85$}
    \end{subfigure}
  \caption{The relation $J(\langle \mathcal{O}\rangle)=\beta(\alpha)$ that characterizes the asymptotic scalar profile is shown for $B=0$ and three different values of the squashing parameter $A$. The blue curves correspond to NUT solutions, which exist for all three values of the squashing, while the red curves in panel (a) represent the two branches of Bolt solutions. }\label{fig:alphavsBbetaall}
\end{figure}

Fig. \ref{fig:alphavsBbetaall} shows the relation $\beta(\alpha)$, or in holographic notation $J(\langle \mathcal{O}\rangle)$, for $B=0$ and for three different values of the squashing parameter $A$. The third panel indicates that the behavior of $\beta(\alpha)$ is qualitatively different for sufficiently negative squashings. Specifically, we find there is a phase transition at $A=-3/4$, precisely where the Ricci scalar on the boundary changes sign, such that for $A \leq -3/4$ the parameter $\beta$ no longer converges. A second solution with $\alpha \neq 0$ comes into play in this regime even at $\beta=0$. The new solution is thermodynamically subdominant, as we will see below, but may nevertheless contribute to certain observables \cite{Maldacena:2001kr,Bzowski2015}. A similar behavior of $\beta(\alpha)$ is found in the entire region of configuration space $(A,B)$ where the Ricci scalar of the boundary geometry is negative (cf. Fig. \ref{fig:ricci}). We note also that the relation $\beta(\alpha)$ associated with the generalized Bolt solutions, shown in the first panel in Fig. \ref{fig:alphavsBbetaall}, is reminiscent of the constant temperature relation found for black holes with scalar hair in this theory, as expected \cite{Hertog:2005hu,Hertog2004}.
 
The thermodynamic behavior of our set of solutions can be studied by evaluating their regularized, Euclidean on-shell action.
Since we do not have analytic solutions we evaluate the regularized on-shell action numerically following the accurate procedure developed in \cite{Bobev2016} and summarized in Appendix \ref{App:AdSexpansions}. 
Fig. \ref{fig:actionsifoA} shows the resulting free energy for a number of representative slices of constant $\beta$ through the three-dimensional phase space of solutions. These indicate that the on-shell action exhibits a maximum at zero squashing and scalar field when the scalar curvature is positive. 

Without a scalar field it was found in \cite{Bobev2016} that the well known Hawking-Page type phase transition from the NUT to the Bolt solutions that occurs as one increases the value of the squashing, qualitatively generalizes to the case of two squashings. 
We find this remains true in the presence of a scalar field, except for the fact that the range of squashings for which the NUT solutions exist gradually shrinks and becomes centered around zero squashing for large values of $\Phi_0$. At the same time the minimum squashing required for Bolt solutions to exist increases for increasing $\Phi_0$, leading to a critical value above which there is a regime of squashings in which no regular solutions exist.

\begin{figure}[ht!]
\centering
\begin{subfigure}[t]{0.32\textwidth}
\includegraphics[width=\textwidth]{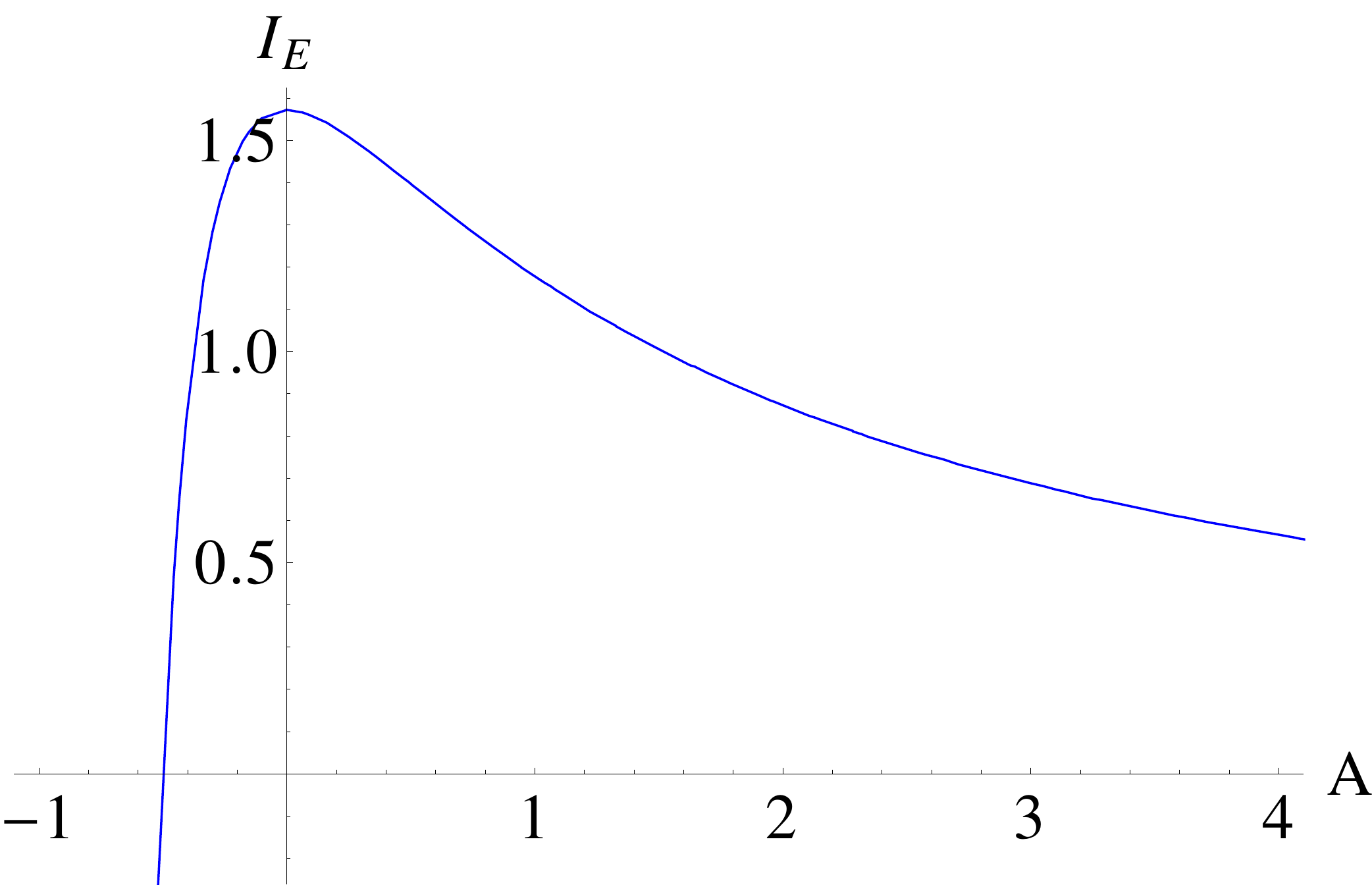} 
\caption{$\beta=0.0$}
\end{subfigure}
\begin{subfigure}[t]{0.32\textwidth}
\includegraphics[width=\textwidth]{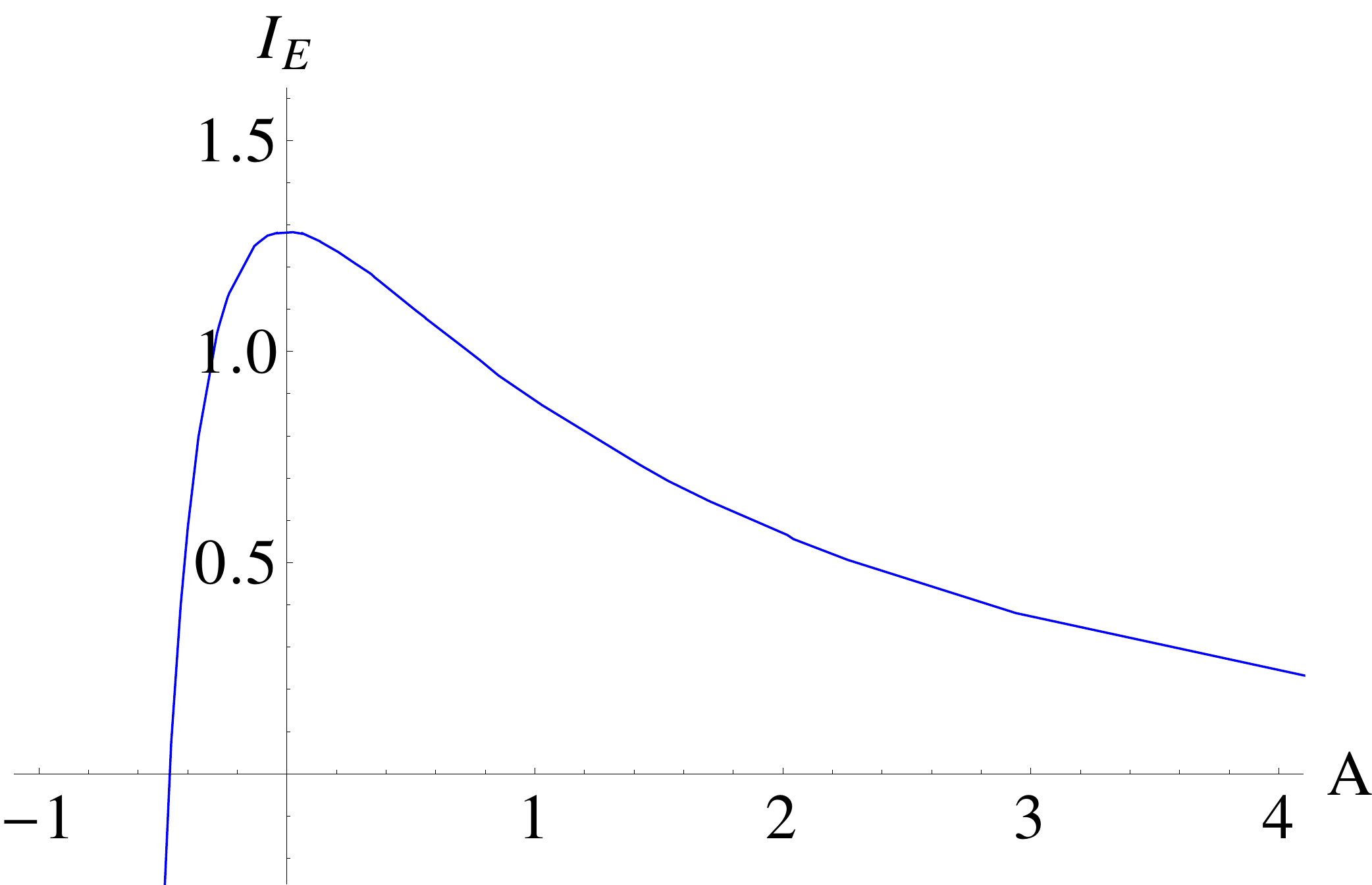} 
\caption{$\beta=-0.21$} \label{fig:actionsbetasmallU}
\end{subfigure}
\begin{subfigure}[t]{0.32\textwidth}
\includegraphics[width=\textwidth]{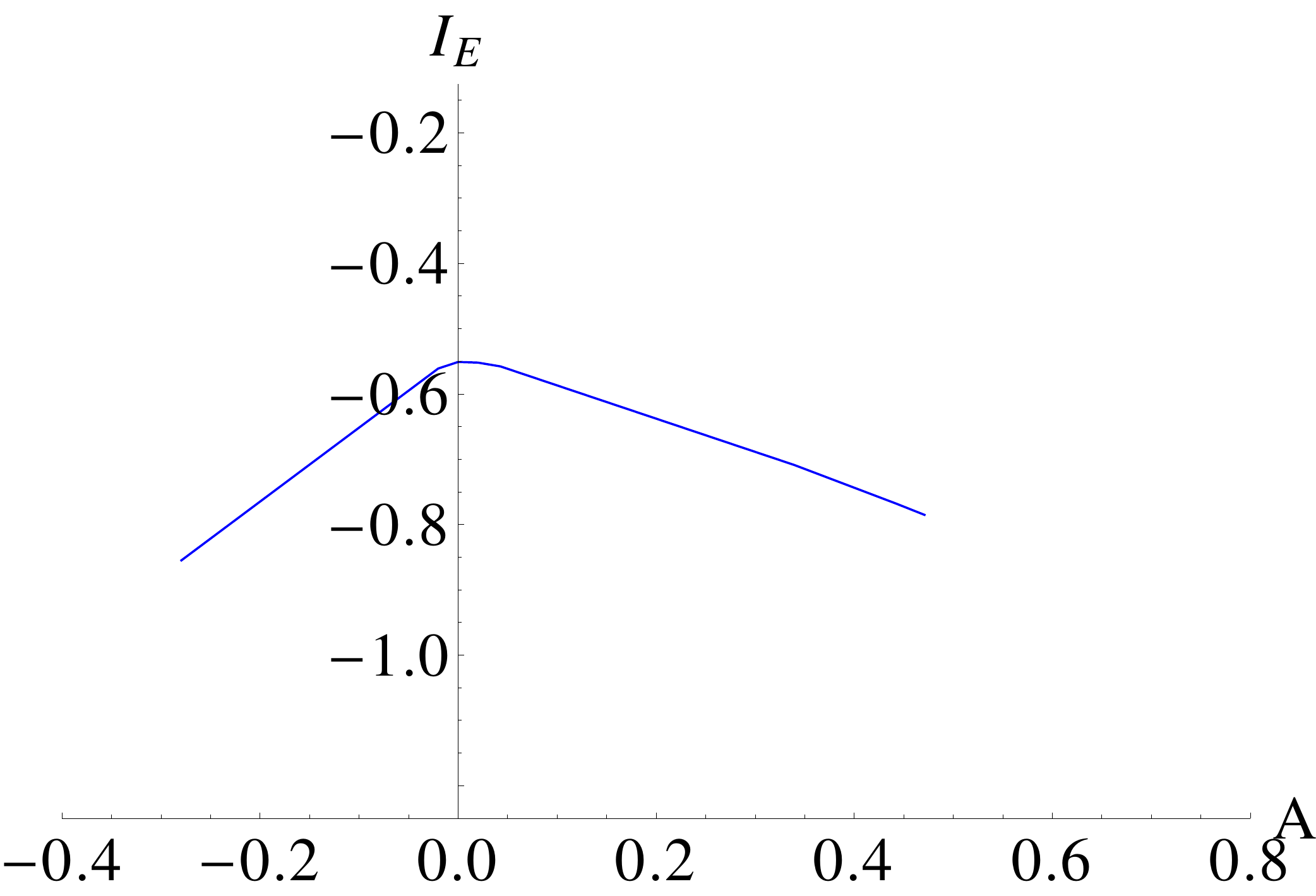} 
\caption{$\beta=-0.56$}
\end{subfigure} \\
\begin{subfigure}[t]{0.32\textwidth}
\includegraphics[width=\textwidth]{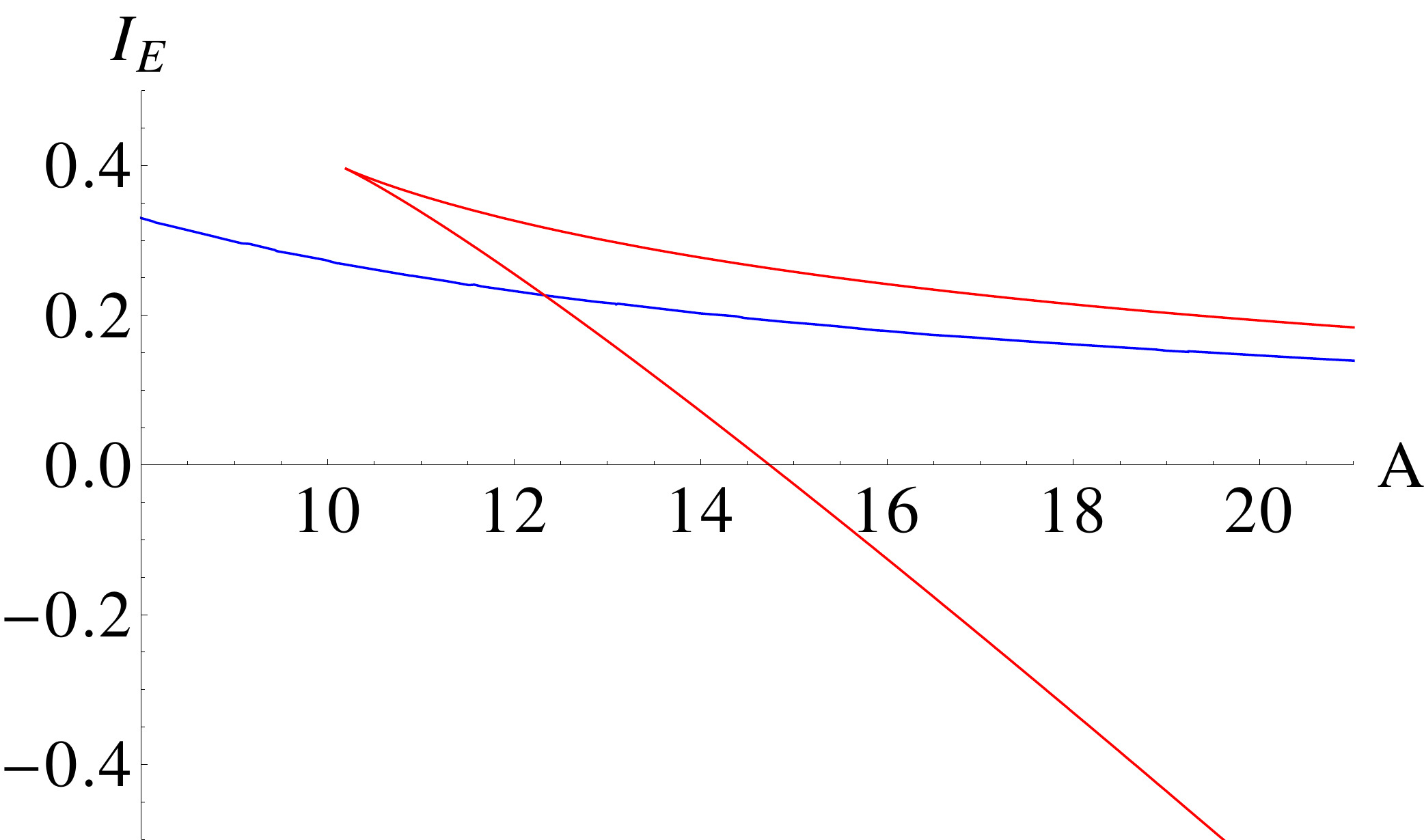} 
\caption{$\beta=0.0$}
\end{subfigure}
\begin{subfigure}[t]{0.32\textwidth}
\includegraphics[width=\textwidth]{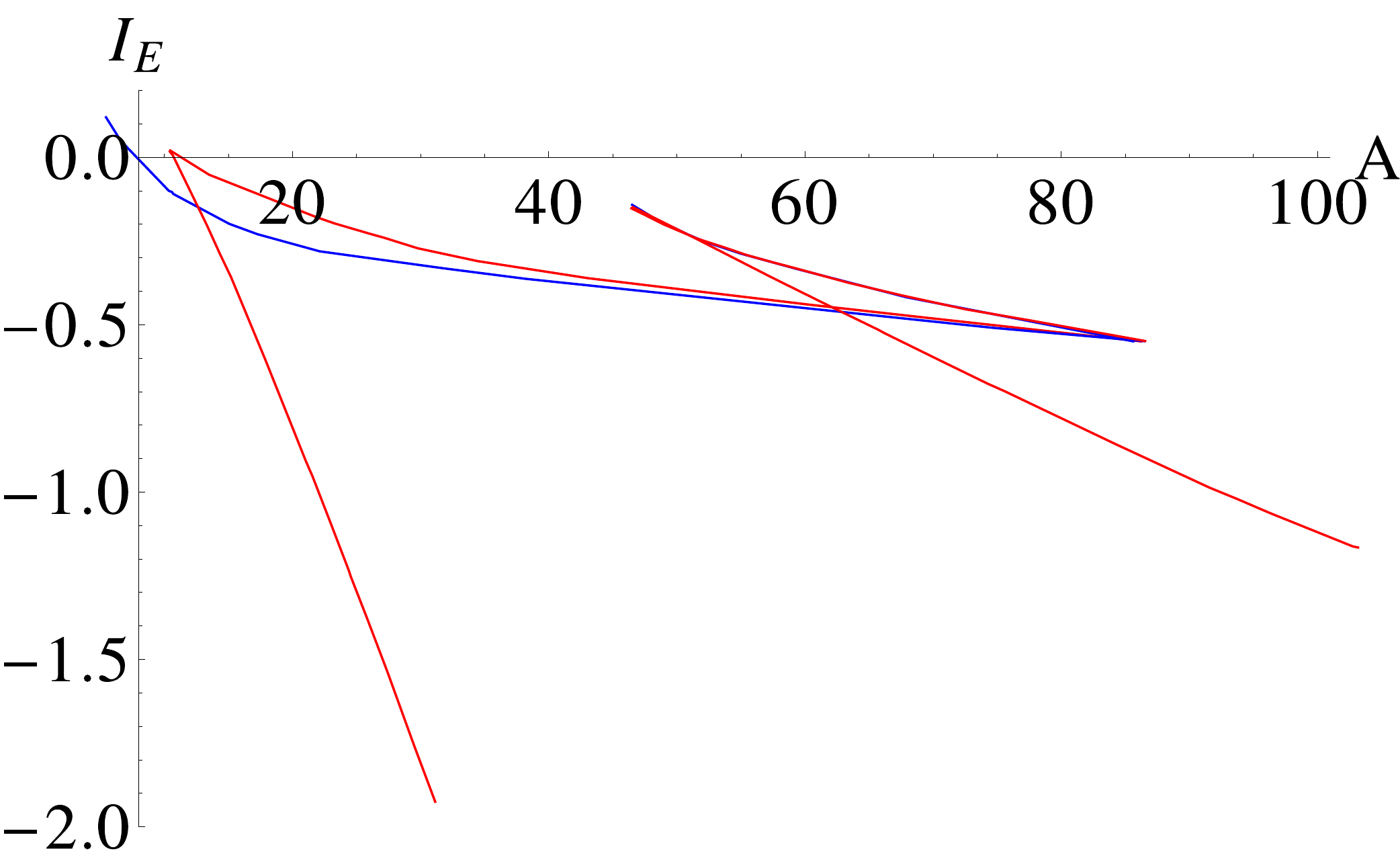} 
\caption{$\beta=-0.21$}\label{fig:actionsbetasmallL}
\end{subfigure}
\begin{subfigure}[t]{0.32\textwidth}
\includegraphics[width=\textwidth]{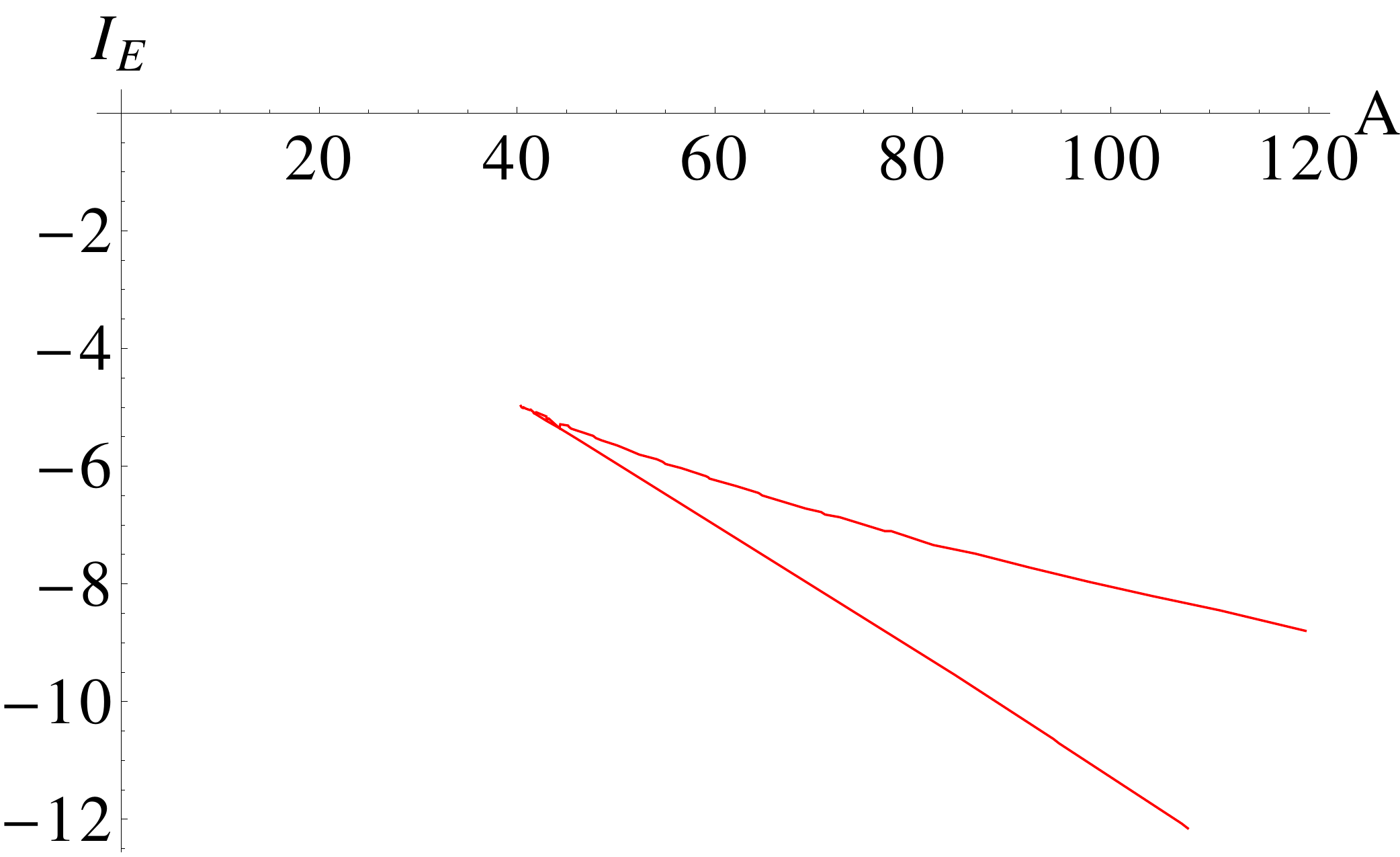} 
\caption{$\beta=-0.56$}
\end{subfigure}
\caption{The free energy as a function of the squashing for different values of the UV deformation parameter $\beta$. The blue (red) curves represent the NUT (Bolt) solutions. In the top row the behavior around zero squashing is shown, while in the bottom row the transition from NUT to Bolt is highlighted.}
\label{fig:actionsifoA}
\end{figure}

\section{Mass deformed $O(N)$ model on double squashed sphere} \label{holo}

Our bulk model is a consistent truncation of M-theory compactified on $AdS_4\times S^7$. Therefore the dual field theory is the ABJM SCFT and we are faced with the problem of evaluating the partition function of supersymmetry breaking deformations of this theory. We will not attempt to do this here. Instead we will focus on a simplified model of this setup where we consider an $O(N)$ vector theory conjectured to be dual to higher-spin Vassiliev gravity in four dimensions \cite{Sezgin2002,Klebanov2002,Giombi2009,Anninos2011}. Higher-spin theories are very different from pure Einstein gravity. However, it is plausible that the behavior of the free energy of vector models qualitatively captures that of duals to Einstein gravity when one restricts to spin 0 and spin 2 deformations \cite{Hartnoll2005, Conti2015,Bobev2016,Anninos2012}. We therefore view these vector theories as dual toy models in this section and proceed to evaluate their partition functions and the one-point functions associated with the scalar condensates in the bulk.

The mass deformed free model partition function is given by
\begin{align} 
	Z_{\rm free}[m^2]=\int \mathcal{D}\phi e^{-I_{\rm free}+\int d^3x\sqrt{g} m^2\mathcal{O}(x)} \ , \label{eqn:Zfree}
\end{align}
where $I_{\rm free}$ is the action of the conformal, free $O(N)$ model, 
\begin{align}
I_{\rm free}=\frac{1}{2}\int d^3x \sqrt{g} \left( \partial_{\mu} \phi_a \partial^{\mu}\phi^a +\frac{1}{8} R\phi_a \phi^a\right).
\end{align}
Here $\phi_a$ is an $N$-component field transforming as a vector under $O(N)$ rotations and $R$ is the Ricci scalar of the boundary geometry.

We calculate the partition function \eqref{eqn:Zfree} on a double squashed sphere. Evaluating the Gaussian integral in \eqref{eqn:Zfree} amounts to computing the following determinant
\begin{align} 
 -\log Z_{\rm free}=F=\frac{N}{2}\log \left(\textrm{det}\left[ \frac{-\nabla^2 + m^2+\frac{R}{8}}{ \Lambda^2}\right]\right) \ , \label{eqn:LogZGeneral}
\end{align}
where $\Lambda$ is a cutoff that we will use to regularize the UV divergences in this theory.
For a single squashing $A$ the eigenvalues of the operator in \eqref{eqn:LogZGeneral} can be found in closed analytic form  \cite{Hu1973},
 \begin{align}
	\lambda_{n,q}= n^2+A (n-1-2q)^2-\frac1{4(1+A)} +m^2 \ , \qquad q=0,1,\ldots , n-1,\ n=1,2,\ldots \ .
\end{align}
To find the eigenvalues on double squashed spheres we apply the numerical techniques developed in \cite{Hu1973, Bobev2016}. These enable us to determine the spectrum numerically to (in principle) any desired accuracy.

To regularize the infinite sum in \eqref{eqn:LogZGeneral} one may be tempted to use an analytic approach like $\zeta$-function regularization. However, this method is not well-adapted to situations where the spectrum of the Laplacian is only known numerically. Therefore we use a heat-kernel type regularization which can be implemented numerically and was discussed in detail in \cite{Anninos2012, Bobev2016}. 

\begin{figure}[ht!]
\begin{subfigure}[t]{0.32\textwidth}
\includegraphics[width=\textwidth]{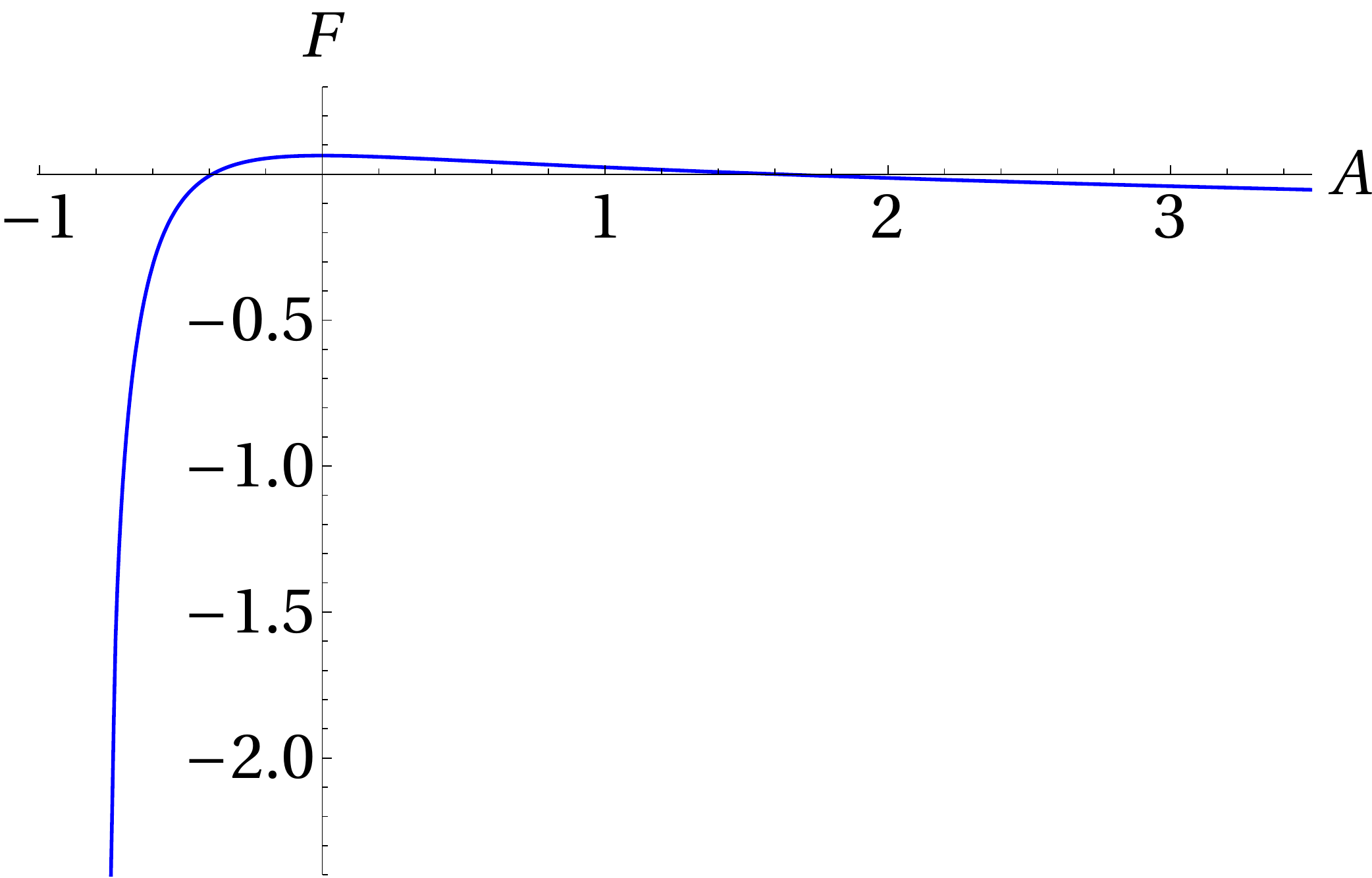}
\caption{$B=0$, $m^2=0.0$}
\end{subfigure}
\begin{subfigure}[t]{0.32\textwidth}
\includegraphics[width=\textwidth]{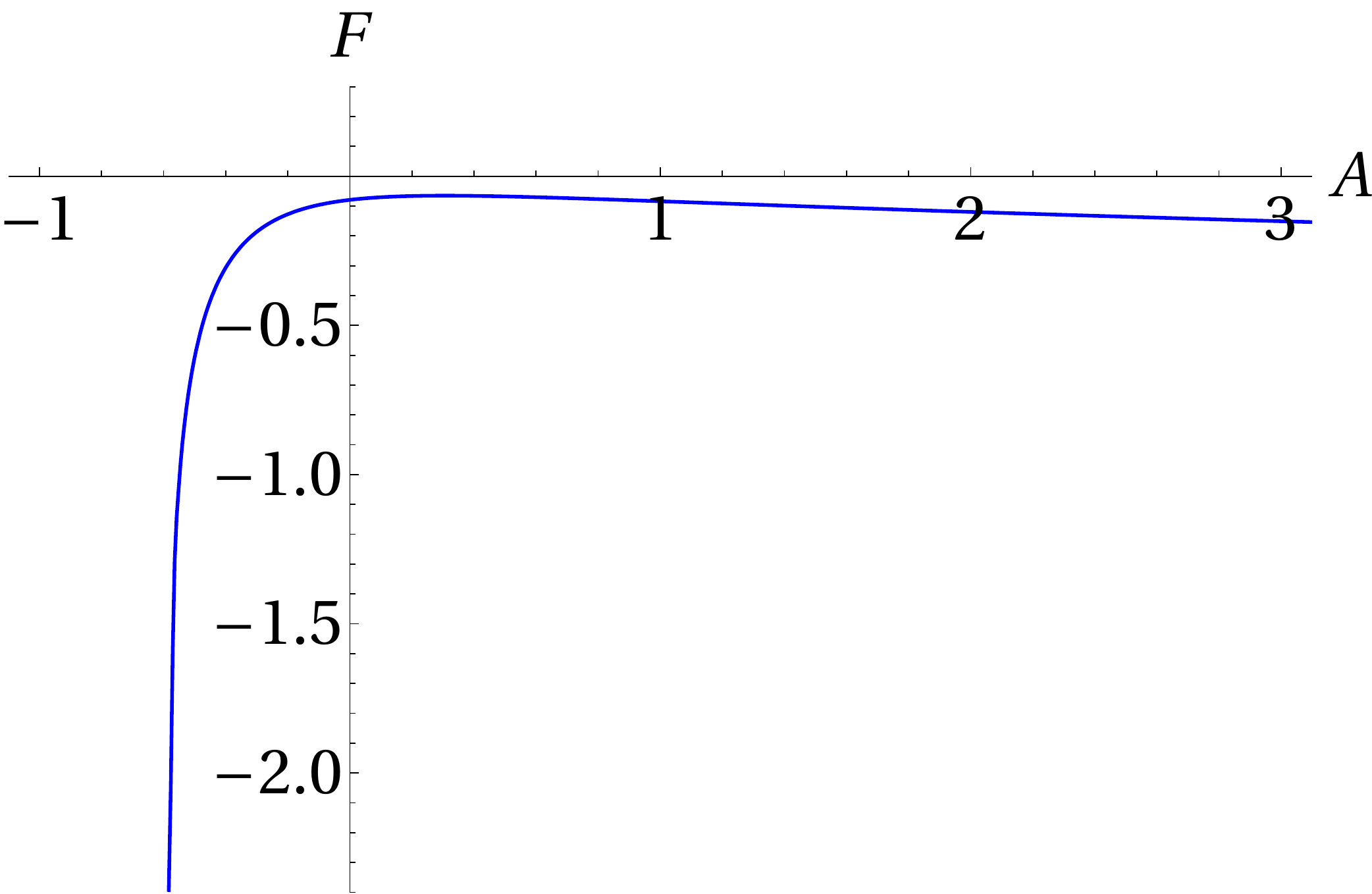}
\caption{$B=0$, $m^2=-0.4$}
\label{fig:Falphavsbetabp50ON}
\end{subfigure}
\begin{subfigure}[t]{0.32\textwidth}
\includegraphics[width=\textwidth]{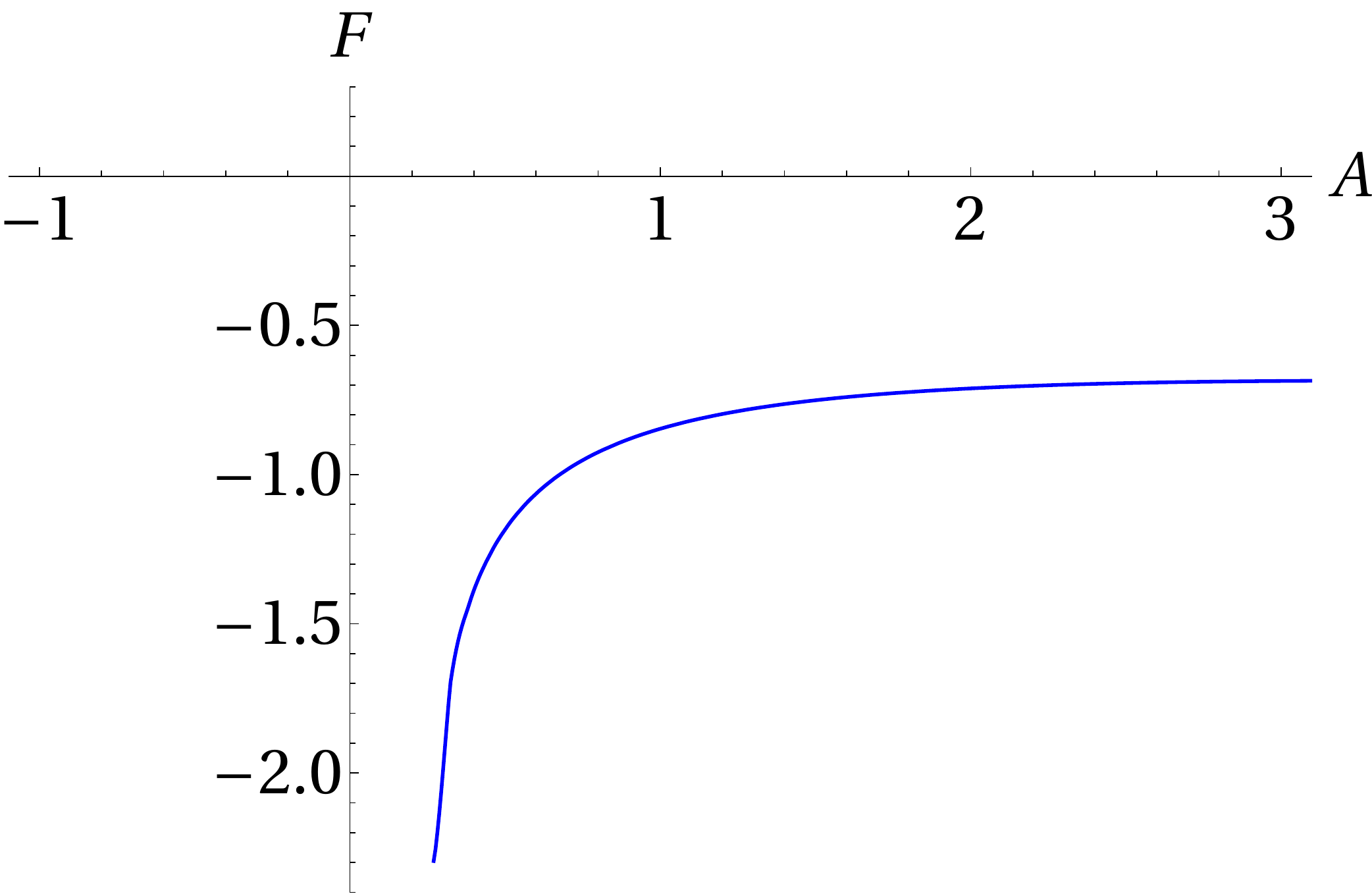}
\caption{$B=0$, $m^2=-0.8$}
\label{fig:Falphavsbetabm08ON}
\end{subfigure}\\
\begin{subfigure}[t]{0.32\textwidth}
\includegraphics[width=\textwidth]{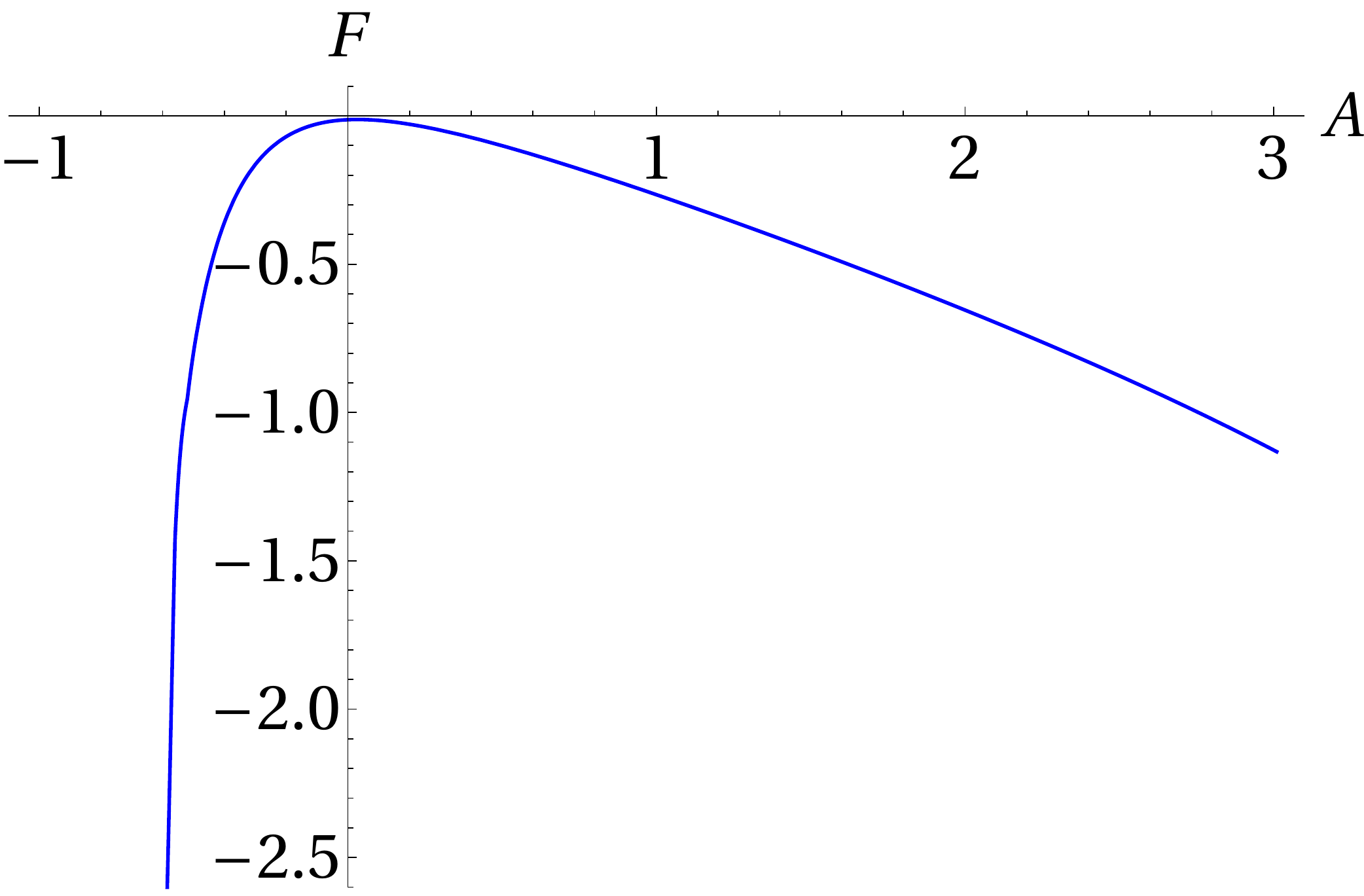}
\caption{$B=2$, $m^2=0.0$}
\end{subfigure}
\begin{subfigure}[t]{0.32\textwidth}
\includegraphics[width=\textwidth]{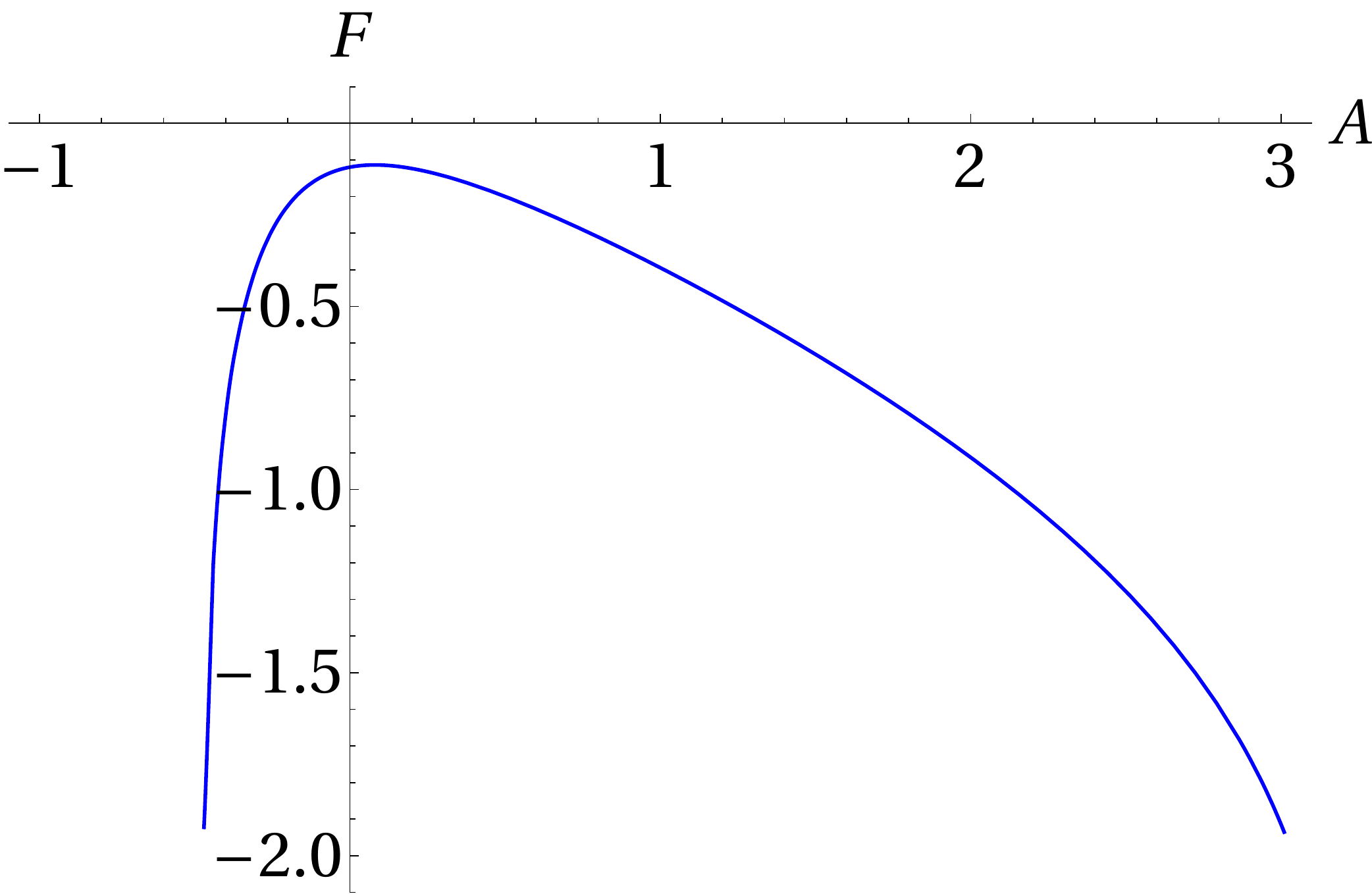}
\caption{$B=2$, $m^2=-0.4$}
\label{fig:Falphavsbetabp50ONB2}
\end{subfigure}
\begin{subfigure}[t]{0.32\textwidth}
\includegraphics[width=\textwidth]{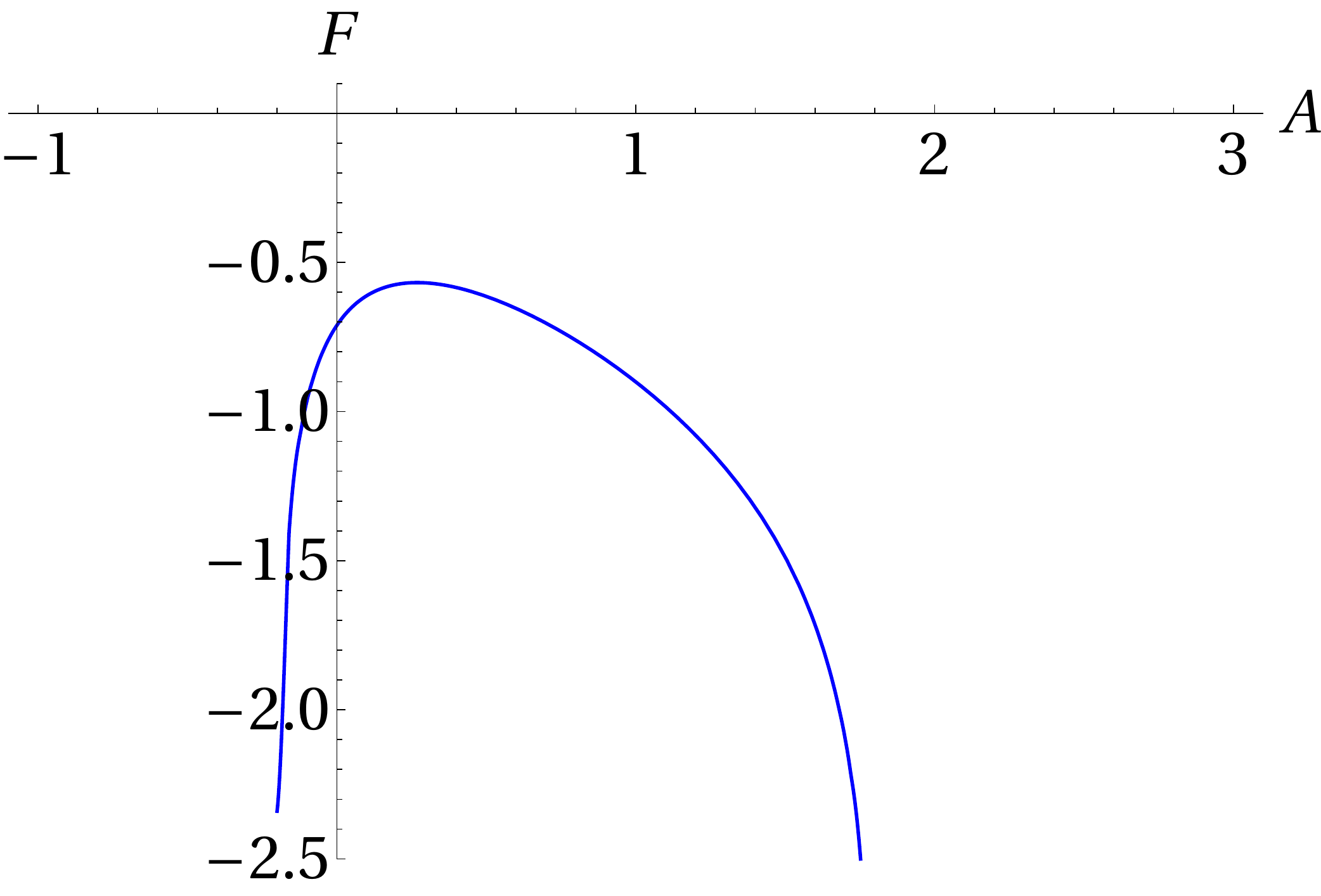}
\caption{$B=2$, $m^2=-0.8$}
\label{fig:Falphavsbetabm08ONB2}
\end{subfigure}
\caption{The free energy of the free $O(N)$ model as a function of the squashing $A$ for three different, real mass deformations. In the top row $B=0$ while for the bottom row $B=2$.}\label{fig:FvsBbetaallON}
\end{figure}

Using a heat-kernel the sum over eigenvalues divides in a UV and an IR part. The latter converges and can readily be done numerically whereas the former contains all the divergences and should be treated with care. When the eigenvalues are known analytically, it is possible to find the divergent behavior using the Euler-Maclaurin formula \cite{Anninos2012}, however, for two squashings a numerical procedure is necessary. For the details of the latter method we refer to Appendix \ref{app:CFTreg}. 

In Fig. \ref{fig:FvsBbetaallON} we show a few slices of the resulting free energy as a function of a squashing $A$ and a real mass deformation $m^2$ for two values of $B$. This can be compared with the action of the real, asymptotically locally AdS solutions discussed in Section \ref{sec:AdSsols} and shown in Fig. \ref{fig:actionsifoA}. An important feature of this model is that the free energy diverges when $R/8 + m^2 \rightarrow 0$. This is a generalization to mass deformed theories of the divergences found in \cite{Bobev2016} and can be understood by inspecting \eqref{eqn:LogZGeneral} in more detail. The determinant, which is a product over all eigenvalues of the operator $-\nabla^2 +m^2+R/8$, vanishes when the operator has a zero eigenvalue, leading to a divergent free energy. Since the lowest eigenvalue of the Laplacian $\nabla^2$ is always zero, the first eigenvalue $\lambda_1$ of the operator in \eqref{eqn:LogZGeneral} is zero when $R/8 + m^2  =0$. In the region of configuration space where the operator has one or more negative eigenvalues the Gaussian integral \eqref{eqn:Zfree} does not converge, and \eqref{eqn:LogZGeneral} does not apply. This is more obvious in the bottom row of Fig. \ref{fig:FvsBbetaallON} where we took a non-zero value for $B$, because in this case $R/8 + m^2$ becomes negative for large negative values and large positive values of $A$.

\begin{figure}[ht!]
\centering
\begin{subfigure}[t]{0.32\textwidth}
\includegraphics[width=\textwidth]{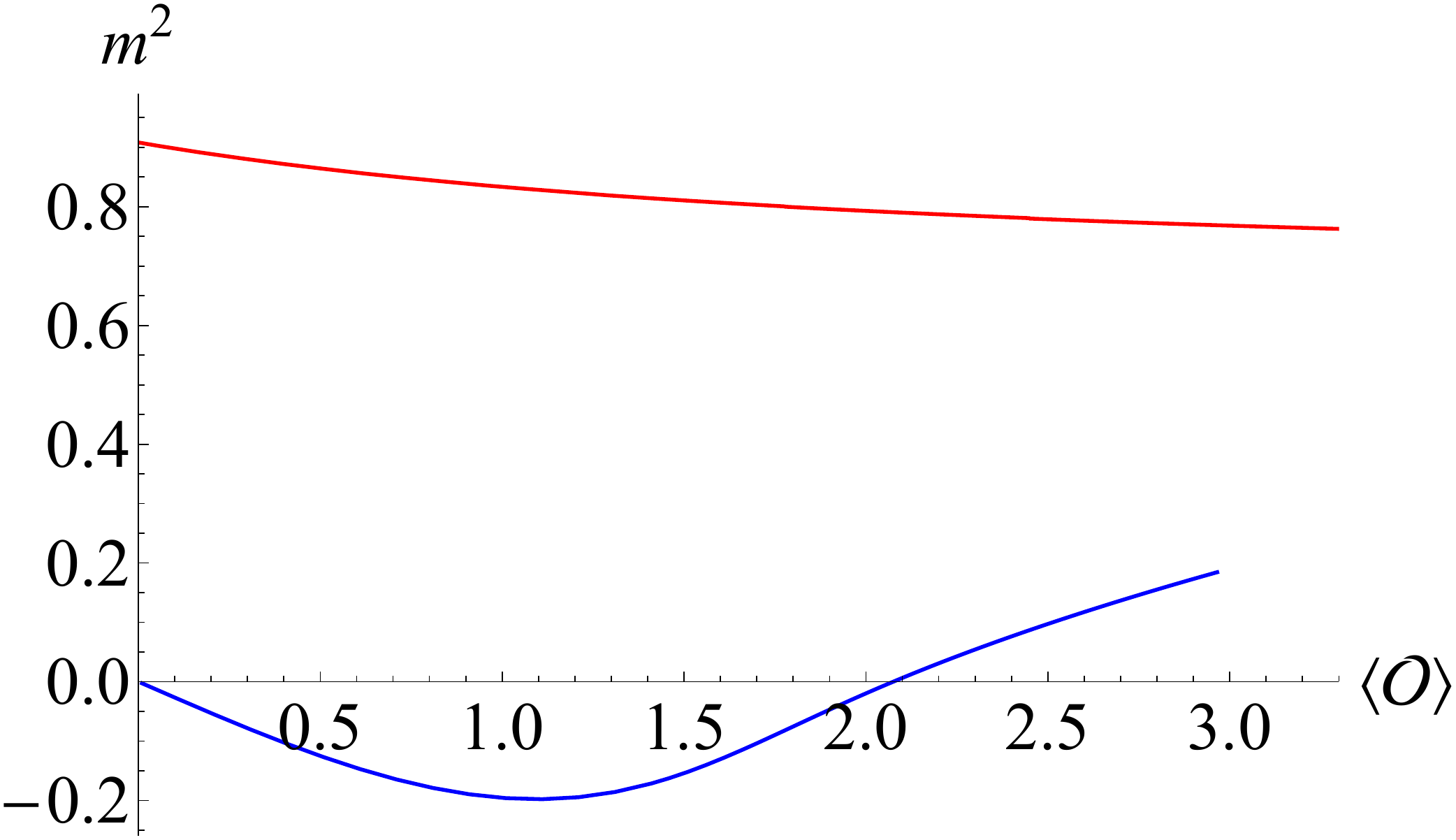}
\caption{$A=-0.85$}
\end{subfigure}
\begin{subfigure}[t]{0.32\textwidth}
\includegraphics[width=\textwidth]{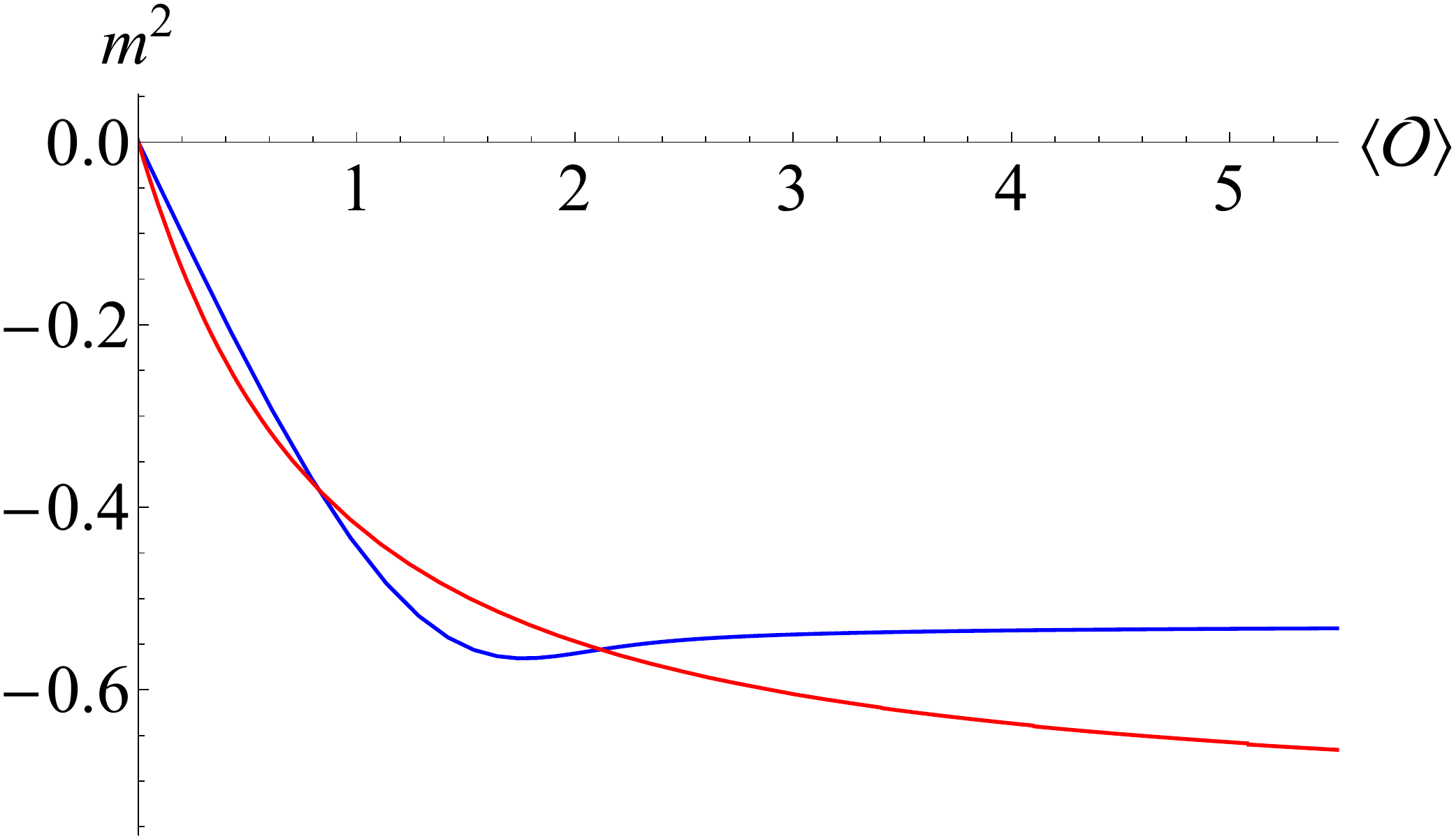}
\caption{$A=0$}
\label{fig:alphavsbetabp50ON}
\end{subfigure}
\begin{subfigure}[t]{0.32\textwidth}
\includegraphics[width=\textwidth]{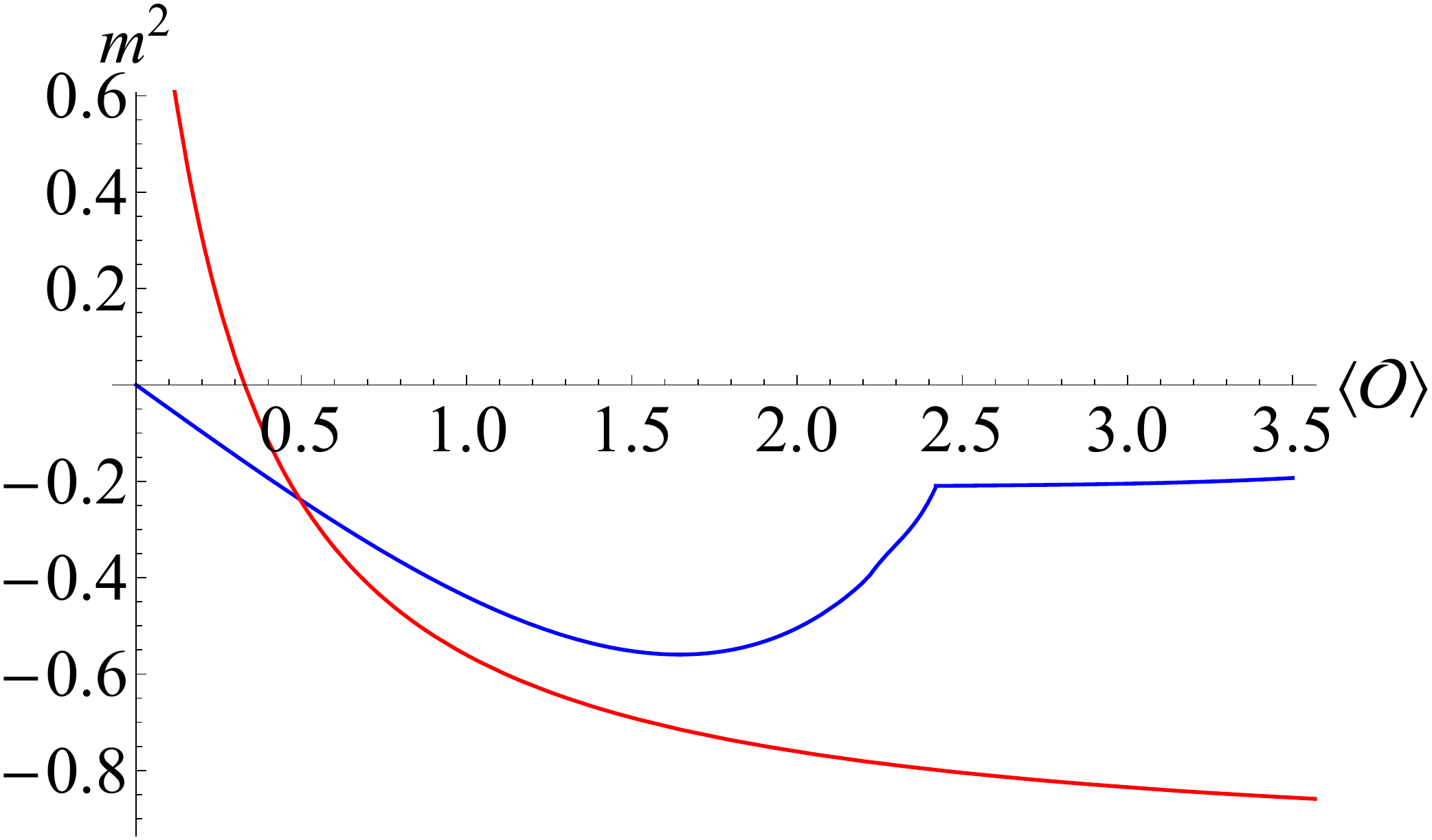}
\caption{$A=40$}
\label{fig:alphavsbetabm08ON}
\end{subfigure}
\caption{Both the bulk and the boundary relation between the vev on the x-axis and the source on the y-axis. In blue the bulk result, calculated via holographic renormalization, is shown and in red the boundary result for three different values of the squashing $A$. 
}\label{fig:vevvssource}
\end{figure}

From the free energy we can find the relation between the vev and its source, the mass of the deformation. In Fig. \ref{fig:vevvssource} we show the resulting vev-source relation and we compare this with the corresponding gravity result obtained from holographic renormalization. It is remarkable that on both sides of the duality, the vev diverges at a finite value of the source. This feature strongly depends on the particular form of the scalar potential \eqref{pot} in the bulk and motivates the particular potential we used. 
Another similarity between the vev-source relations, is the fact that for squashings with $A<-3/4$ the sources attain positive values for large vevs. A clear difference, however, is that for the $O(N)$ model a zero vev can correspond to a non-zero source, this feature is absent in Einstein gravity and can only be explained by the peculiarities of Vasiliev gravity. 

\section{Anisotropic inflationary minisuperspace} \label{dscft1}

In section \ref{sec:AdSsols} we considered real, asymptotically locally AdS solutions, with real radial scalar field profiles. 
We now turn to the de Sitter domain of the theory. At the semiclassical level this is specified by complex solutions of the same theory, given by the action \eqref{eqn:GRaction}, that asymptotically tend to real, Lorentzian, locally de Sitter space. The asymptotically Lorentzian behavior of this set of solutions provides a large imaginary contribution to their Euclidean action resulting in a rapidly oscillating wave function exhibiting classical WKB behavior.

\medskip
{\it Complex saddle points}
\medskip

We consider the same anisotropic minisuperspace model as before, consisting of squashed sphere boundary surfaces \eqref{eqn:metric} with spatially homogeneous scalar field configurations. The metric of the interior saddle point solutions can thus again be written in the form \eqref{eqn:doublesqansatz}, but now with complex scale factors $l_a$ and a complex scalar field profile $\Phi$.

\begin{figure}[htb]
\begin{center}
\begin{tikzpicture}
    \draw[->, thick] (-10pt,0pt) -- (180pt,0pt) node[below] {r};
    \draw (-12pt, 0pt) node[left] {SP};
    \draw[->, thick] (0pt, -10pt) -- (0pt,100pt) node[left] {y};
    \draw[->-=.5, ultra thick] (-.8pt,0pt) -- (150.8pt,0pt);
    \draw[->-=.5, ultra thick] (150pt,-.8pt) to (150pt,60.8pt);
   \draw[->-=.5, ultra thick, dashed] (0pt,0pt) -- (0pt,60.8pt);
   \draw[->-=.5, ultra thick, dashed] (0pt,60pt)-- (150pt,60pt);
    \draw[solid,fill] (150pt,60pt) circle (2pt) node[right] {$\upsilon$};
    \draw (80pt, 65pt) node[above] {${\cal C}'$}; 
    \draw (80pt, -10pt) node[below] {${\cal C}$}; 
    \draw (-5pt, 60pt) node[left] {$\pi/2$};
    \draw (180pt, 100pt) -- (170pt, 100pt) -- (170pt, 115pt) node[below right] {$\tau$};
\end{tikzpicture}
\end{center}
\caption{Two representations in the complex $\tau$-plane of the same no-boundary saddle point associated with an inflationary universe. Along the horizontal part of the AdS contour ${\cal C}$ the geometry is an asymptotically AdS, spherically symmetric domain wall with a complex scalar field profile. Along the horizontal branch of the dS contour ${\cal C}'$ the saddle point behaves as a Lorentzian, inflationary universe.}
\label{contour}
\end{figure}
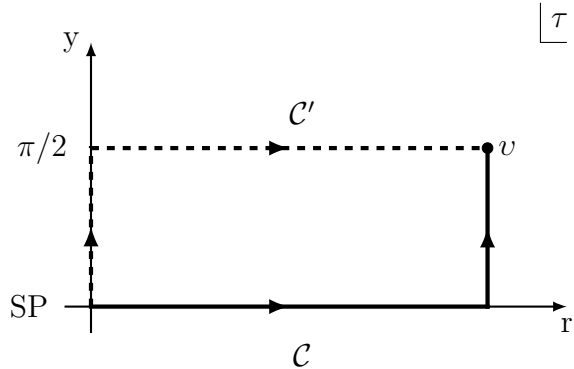

To represent the solutions it is useful to introduce a complex time coordinate $\tau(r)$ defined by
\begin{equation}
\tau (r) \equiv \int_0^r dr' l_0(r')\ . \label{eqn:tau}
\end{equation}
In terms of the variable $\tau$ the asymptotically dS domain of the wave function is to be found along the asymptotically horizontal line $\tau = t+i\pi/2$ in the complex $\tau$-plane. Along this line the leading order Fefferman - Graham - Starobinsky expansion of the metric \eqref{eqn:UVmetric1} becomes
\begin{equation} 
ds^2= dt^2 - e^{2t} \left(A_0 \sigma_1^2+ B_0 \sigma_2^2+ C_0 \sigma_3^2 \right) \label{eqn:UVmetric2} \ .
\end{equation}
Here ${A_0,B_0,C_0}$ are real constants specifying the degree of asymptotic anisotropy. The Lorentzian signature of the asymptotic metric means the original scale factors $l_i$ defined in \eqref{eqn:doublesqansatz} are to leading order purely imaginary in the dS domain. Their subleading behavior can be deduced from an asymptotic analysis of the equations of motion and is given in Appendix A. We illustrate the representation of the saddle point solutions in the complex $\tau$-plane in Fig. \ref{contour}. The semiclassical AdS domain of the theory is specified by solutions that are regular in the IR and real along the real $\tau=r$ axis. The semiclassical dS domain, by contrast, involves everywhere regular complex geometries that tend to asymptotically real, Lorentzian solutions along the $\tau = t+i\pi/2$ line. The AdS contour labeled as $\mathcal{C}$ in Fig. \ref{contour} provides a geometric representation of these complex solutions in which their interior geometry consists of a Euclidean AdS domain wall that makes a smooth (but complex) transition to a Lorentzian asymptotically dS universe. The signature of the asymptotic metric \eqref{eqn:UVmetric2} means that the potential \eqref{pot} in the original Euclidean action \eqref{eqn:GRaction} acts in the dS regime as a positive effective potential
\begin{equation}\label{potdS}
V^{dS}_{\rm eff}(\Phi)= -V = 2+\cosh(\sqrt{2}\Phi) \ .
\end{equation}

\begin{figure}[ht!]
\centering  
    \includegraphics[width=0.45\textwidth]{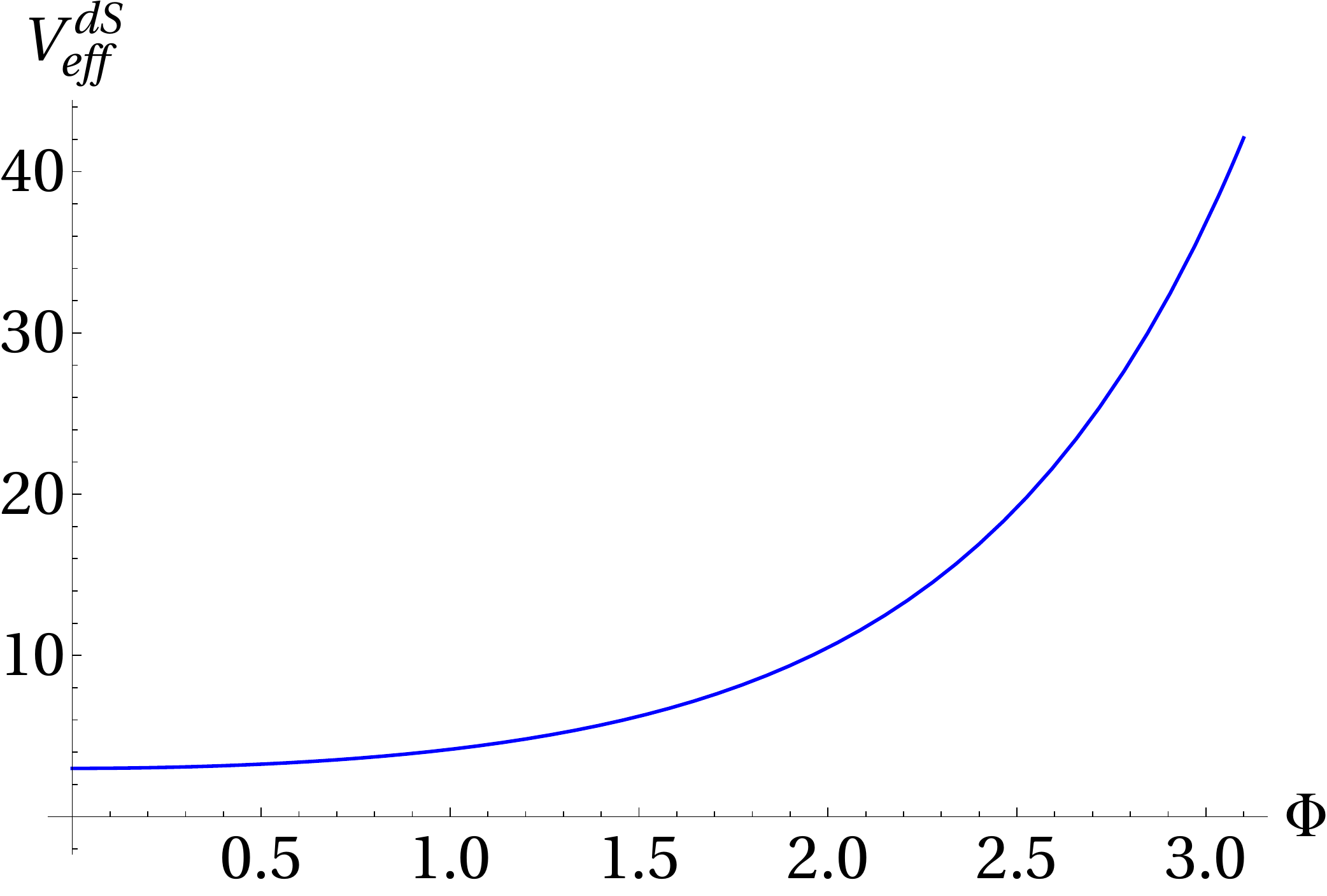}
    \includegraphics[width=0.45\textwidth]{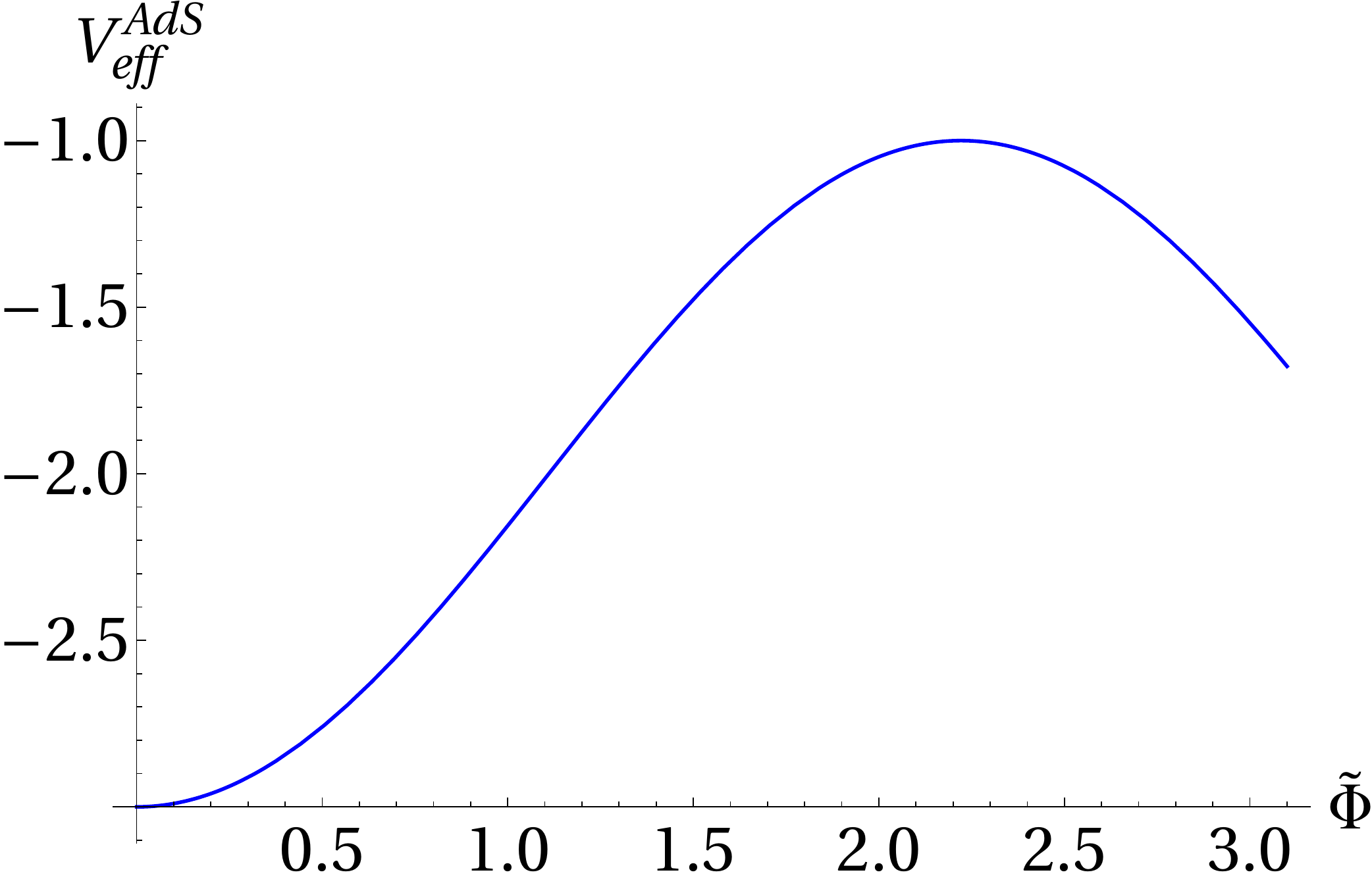}
\caption{\textit{Left panel}: The effective potential in the asymptotic Lorentzian dS region. \textit{Right panel}: The effective potential in the interior AdS domain wall region of the complex saddle points specifying the dS domain of the wave function.}\label{fig:potential}
\end{figure}

The argument of the wave function is real. This means that in order for the above complex solutions to be valid saddle points specifying the semiclassical wave function, the scalar field must also become real along the same line in the $\tau$-plane. The UV expansion 
\eqref{UVscalar2} shows this requires its leading coefficient $\alpha$ to be imaginary, which in turn means that the scalar profile is imaginary all along the AdS part of the contour ${\cal C}$ shown in Fig. \ref{contour}. The interior region of the saddle points specifying the Lorentzian dS domain of the wave function thus involves complex generalizations of Euclidean AdS domain walls.  

The effective potential in this AdS domain wall regime is therefore 
\begin{equation}
V^{AdS}_{\rm eff}(\tilde \Phi)= -2-\cos(\sqrt{2}\tilde \Phi)\ ,
\end{equation}
where $\tilde \Phi \equiv \im \Phi$, and is illustrated in the right panel of Fig. \ref{fig:potential}.  This shows that the asymptotic dS domain of the wave function corresponds to a finite domain of IR values $\Phi_0$ of the scalar field bounded by $\vert \Phi_0 \vert < \sqrt{2}\pi/2 \equiv \Phi_c$. From an AdS perspective this is simply a consequence of the shape of the effective potential governing the inner AdS region of the saddle point solutions. From a dS perspective this bound signals the boundary of the inflationary regime of the effective potential \eqref{potdS} in the dS domain. The potential \eqref{potdS} clearly admits inflationary solutions near its minimum. For large values of the scalar field however it is too steep. The IR regularity condition of the Hartle-Hawking saddle points selects those patches of scalar potentials where the conditions for inflation hold \cite{Hartle2008}. The semiclassical wave function has no support outside these inflationary patches.

\begin{figure}[ht!]
\centering  
    \includegraphics[width=0.82\textwidth]{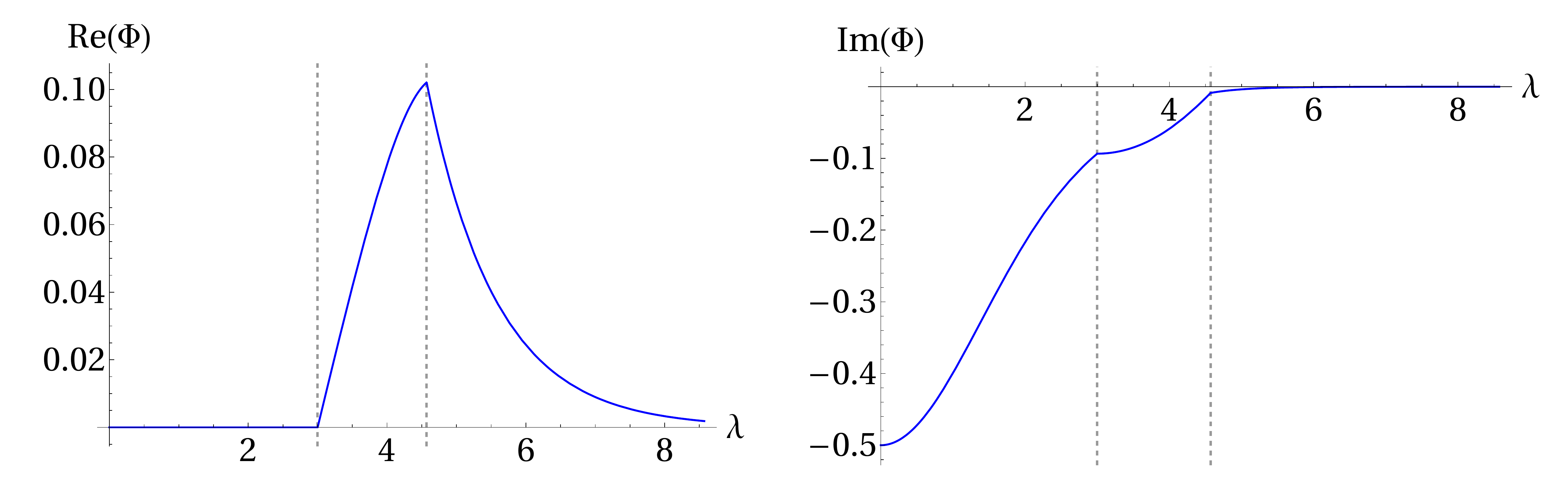}\\
    \includegraphics[width=0.82\textwidth]{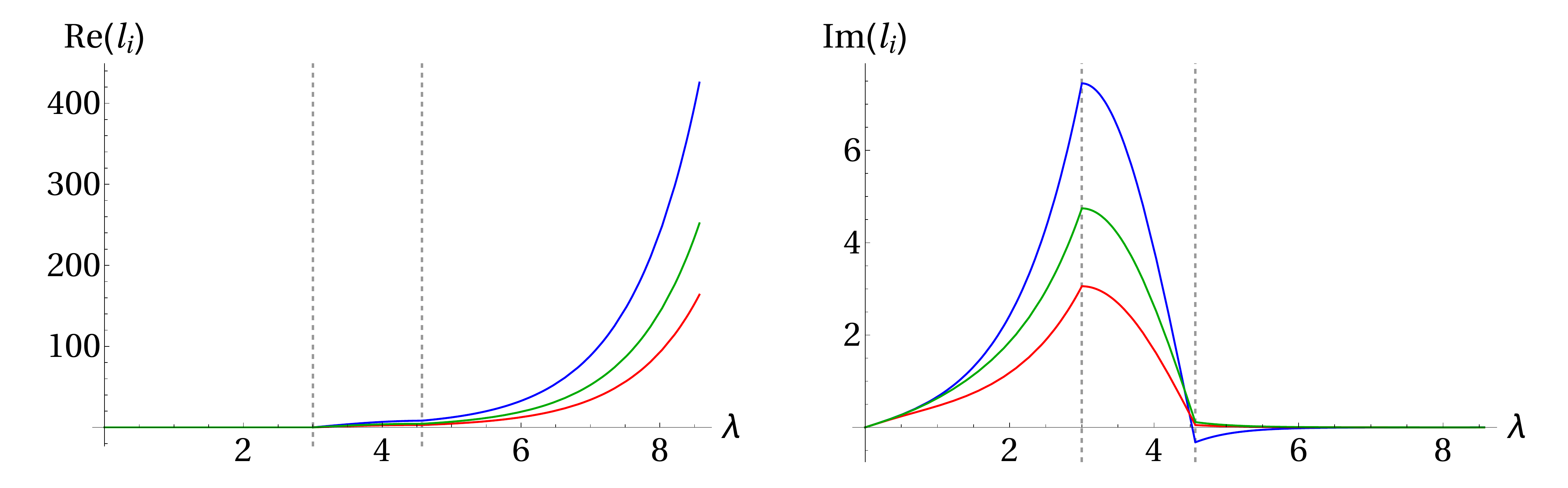}
\caption{A representative example of a complex solution with a NUT in the interior and a scalar field profile, that is asymptotically locally dS with a double squashed sphere future spatial boundary. The solution is shown here along a contour ${\cal C}$ like the one depicted in Fig. \ref{contour}, consisting of three segments parameterized by $\lambda$. The third segment is along $\tau=i \pi/2+t$  and  clearly shows the solution tends to asymptotic Lorentzian dS. The top panels show the evolution of the real and imaginary components of the scalar field along this contour. The bottom panels show the real and imaginary components of the three scale factors. The values of the IR parameters are $\phi_0=1/2$, $\beta_4= -1/6$ and $\gamma_4= -11/60$.}\label{fig:classols1}
\end{figure}

To determine for which real boundary conditions in the asymptotic dS domain regular complex solutions exist we must numerically solve the complex equations of motion derived from the action \eqref{eqn:GRaction}. The regularity conditions on geometry and field in the IR, either at a NUT or a Bolt, leave three free parameters; two associated with the IR behavior of the scale factors and one for the complex value $\Phi_0$ of the scalar field.  Varying these and numerically integrating the Einstein equation in the complex $\tau$-plane to the asymptotic dS regime in the UV yields a three-parameter family of complex solutions whose action specifies the semiclassical no-boundary wave function in the dS domain. Details of this procedure are given in Appendix \ref{App:AdSexpansions}.\footnote{The same method was already employed in \cite{Bramberger2017} to find anisotropic saddle points of the no-boundary wave function for a different potential.} Fig. \ref{fig:classols1} shows a representative example of a solution with two squashings and the scalar field turned on, along a contour ${\cal C}$ along which the solution exhibits an inner Euclidean AdS domain wall region.

\begin{figure}[ht!]
\centering  
    \includegraphics[width=0.45\textwidth]{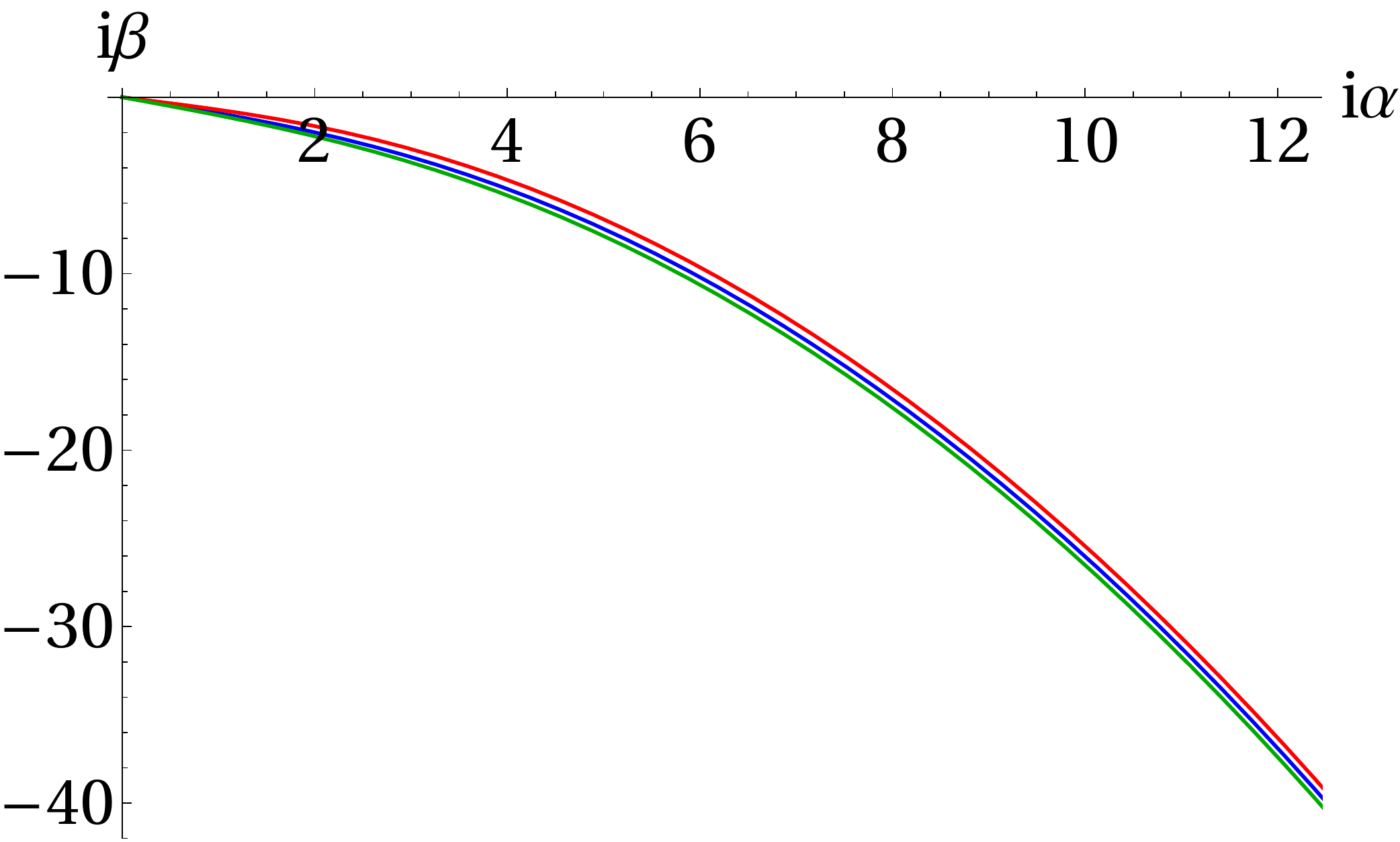}
    \includegraphics[width=0.40\textwidth]{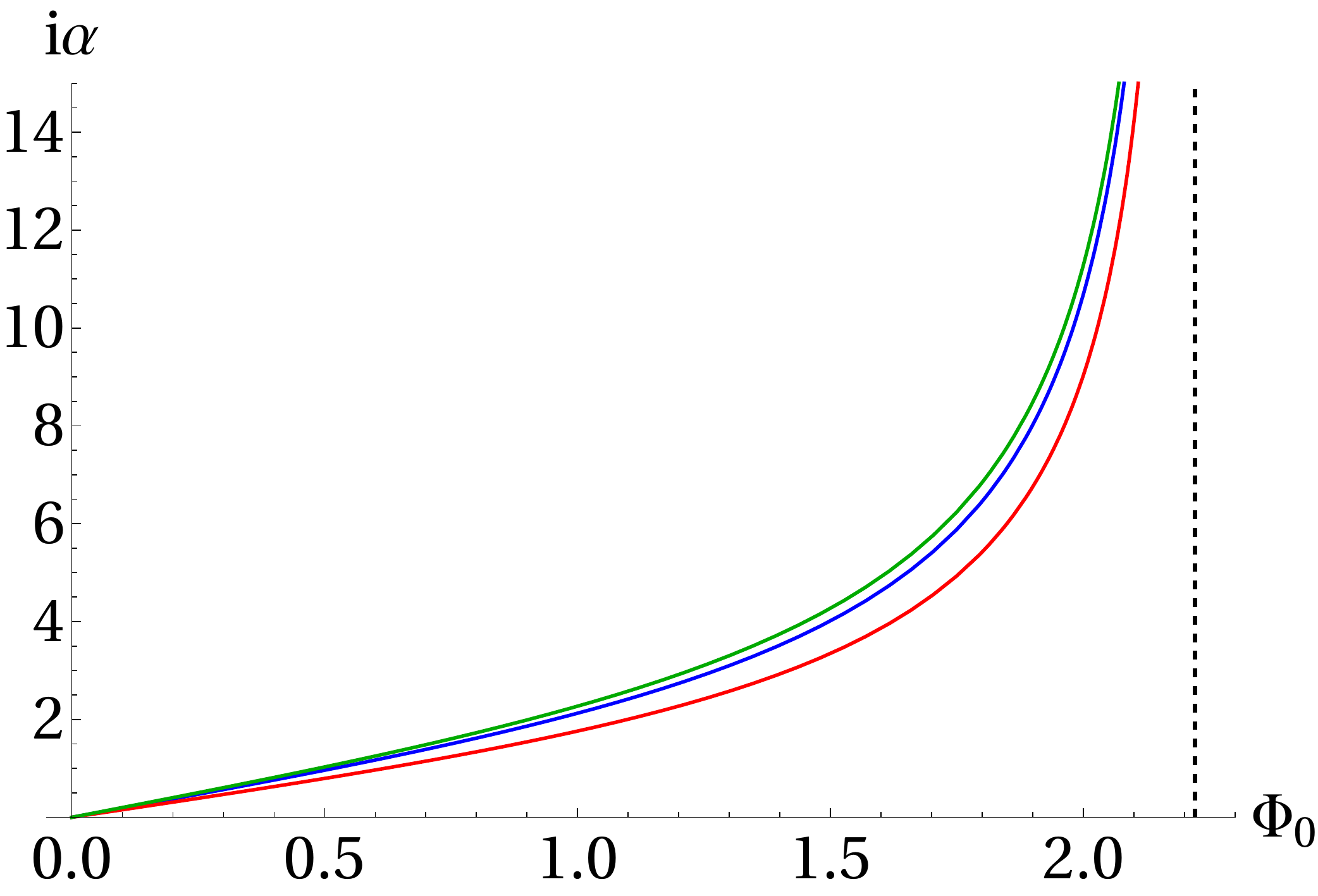}
\caption{\textit{Left panel:} The behavior of $\beta$ as a function of $\alpha$ for different degrees of anisotropy. \textit{Right panel:} The behavior of $\alpha$ as a function of $\Phi_0$. All curves have $B=0$. The blue curve has $A=5$, the red one $A=40$ and the green one $A=-0.85$. When $\Phi_0 \rightarrow \sqrt{2}\pi/2$, $\alpha$ diverges. }\label{fig:alphavsphi0andbeta}
\end{figure}

The regularity condition on the scalar field in the interior yields a relation $\beta (\alpha)$ between the coefficients of its asymptotic profile. We show this for three different combinations of squashings in the left panel of Fig. \ref{fig:alphavsphi0andbeta}. A characteristic feature of this model is that $\beta$ is imaginary over the entire range of parameter space. The dual interpretation of this suggests that the corresponding cosmological histories behave only approximately classically in the large volume regime \cite{Hertog2016}. This may seem surprising but is perhaps related to the fact that the potential $V^{dS}_{\rm eff}(\Phi)$ describes a regime of eternal inflation and not slow roll inflation. One might have thought that the boundedness of the range of values $\Phi_0$ in the IR would mean the semiclassical wave function has support over a limited range of values $\alpha$ in the UV. This is not the case. The right panel of Fig. \ref{fig:alphavsphi0andbeta} plots $\alpha$ as a function of $\Phi_0$ for three different combinations of squashings. One sees $\alpha$ diverges as $\Phi_0\rightarrow \Phi_c$.

\medskip
\noindent
{\it Anisotropic inflationary histories}
\medskip

The Euclidean action \eqref{eqn:GRaction} of the above solutions specifies the semiclassical no-boundary wave function in the asymptotic dS domain. The complex nature of the solutions means that in the large three-volume region of superspace the wave function takes the form 
\begin{equation}
\Psi[a,A,B,\Phi_f] \approx  \exp\{(-I_{\rm R}[a,A,B,\Phi_f] +i S[a,A,B,\Phi_f])/\hbar\} .
\label{semiclass}
\end{equation}
where $a\equiv e^{t}$ is the overall volume scale factor. Here $I_{\rm R}[a,A,B,\Phi_f]$ and $-S[a,A,B,\Phi_f]$ are the real and imaginary parts of the Euclidean action $I_{\rm E}$ of the regular complex saddle point solution that matches the real boundary data $(a,A,B,\Phi_f)$, with $(A,B)$ the squashing parameters and $\Phi_f \approx -i\alpha/2a(A_0 B_0 C_0 )^{1/3}$. In the large volume regime the phase factor $S$ varies rapidly compared to $I_{\rm R}$,
\be
| \vec{\nabla} I_{\rm R}|\ll |\vec{\nabla} S|    \ .
\label{eq:classicalityIntro}
\ee
Hence the wave function predicts that the boundary configuration evolves classically \cite{Hartle2008}. This is analogous to the prediction of the classical behavior of a particle in a WKB state in non-relativistic quantum mechanics. 
Thus the NBWF in the dS domain predicts an ensemble of classical, asymptotically locally de Sitter histories that are the integral curves of $S$ in superspace, with relative probabilities that are proportional to $\exp[-2 I_{\rm R}(A,B,\Phi_f)]$. The latter are conserved  under scale factor evolution as a consequence of the Wheeler-DeWitt equation \cite{Hartle2008}. 

\begin{figure}[ht!]
\centering  
    \includegraphics[width=0.45\textwidth]{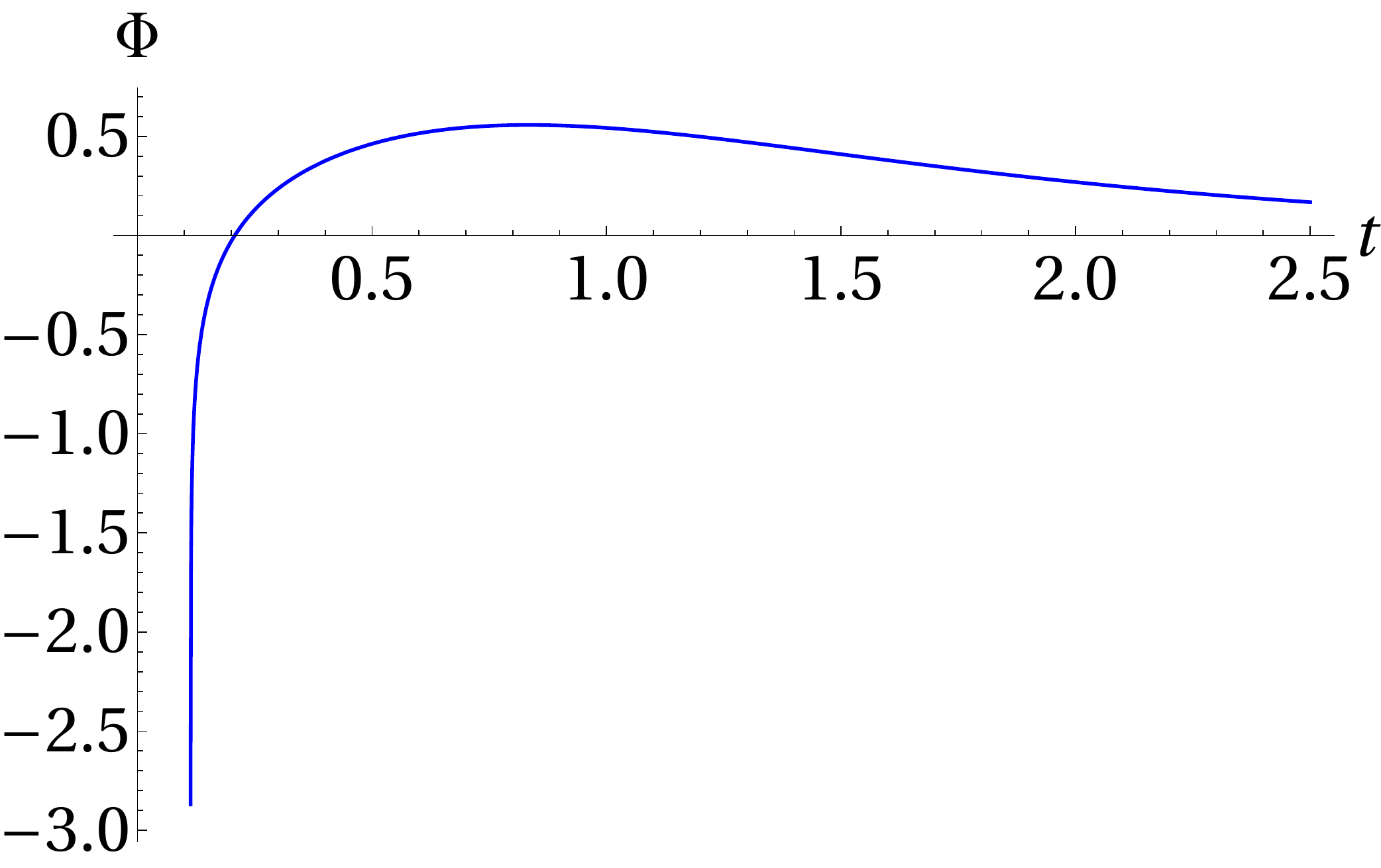}
    \includegraphics[width=0.45\textwidth]{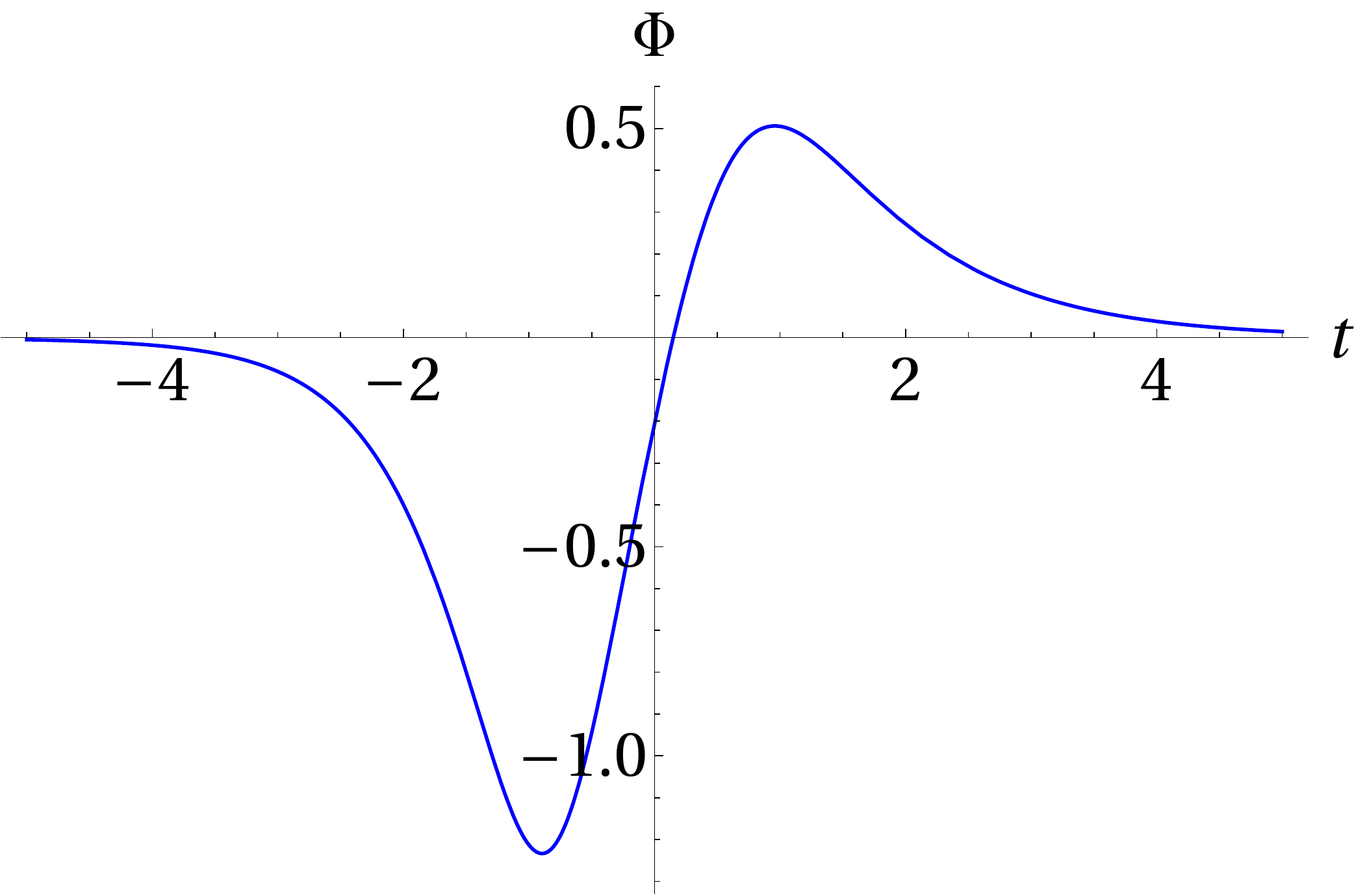}
\caption{\textit{Left panel:} An asymptotically classical history of $\phi$ that is initially classically singular, for $\Phi_0=0.5$, $A=2.3$ and $B=0$. \textit{Right panel:} An asymptotically classical history of $\phi$ with a bounce in the semiclassical domain, for $\Phi_0=0.5$, $A=0.16$ and $B=0$.}\label{fig:clashistories}
\end{figure}

The classical ensemble consists of a three-parameter family of eternally inflating histories that are asymptotically dS and have a certain degree of anisotropy, parameterized by $(A,B)$ on the future boundary. The histories in this model do not exhibit a phase of reheating and slowing expansion. Instead they transition from a phase of scalar field driven inflation to a phase of accelerated expansion driven by the cosmological constant. The potential is such that the scalar field inflation is of the type of slow roll eternal inflation. Hence if one were to include inhomogeneous fluctuations, one would find that the wave function became broadly distributed, predicting an ensemble of histories with exceedingly large or even infinite constant scalar density surfaces \cite{Linde:1996hg,Hartle:2010vi}. 

\begin{figure}[ht!]
\centering
\includegraphics[width=0.48\textwidth]{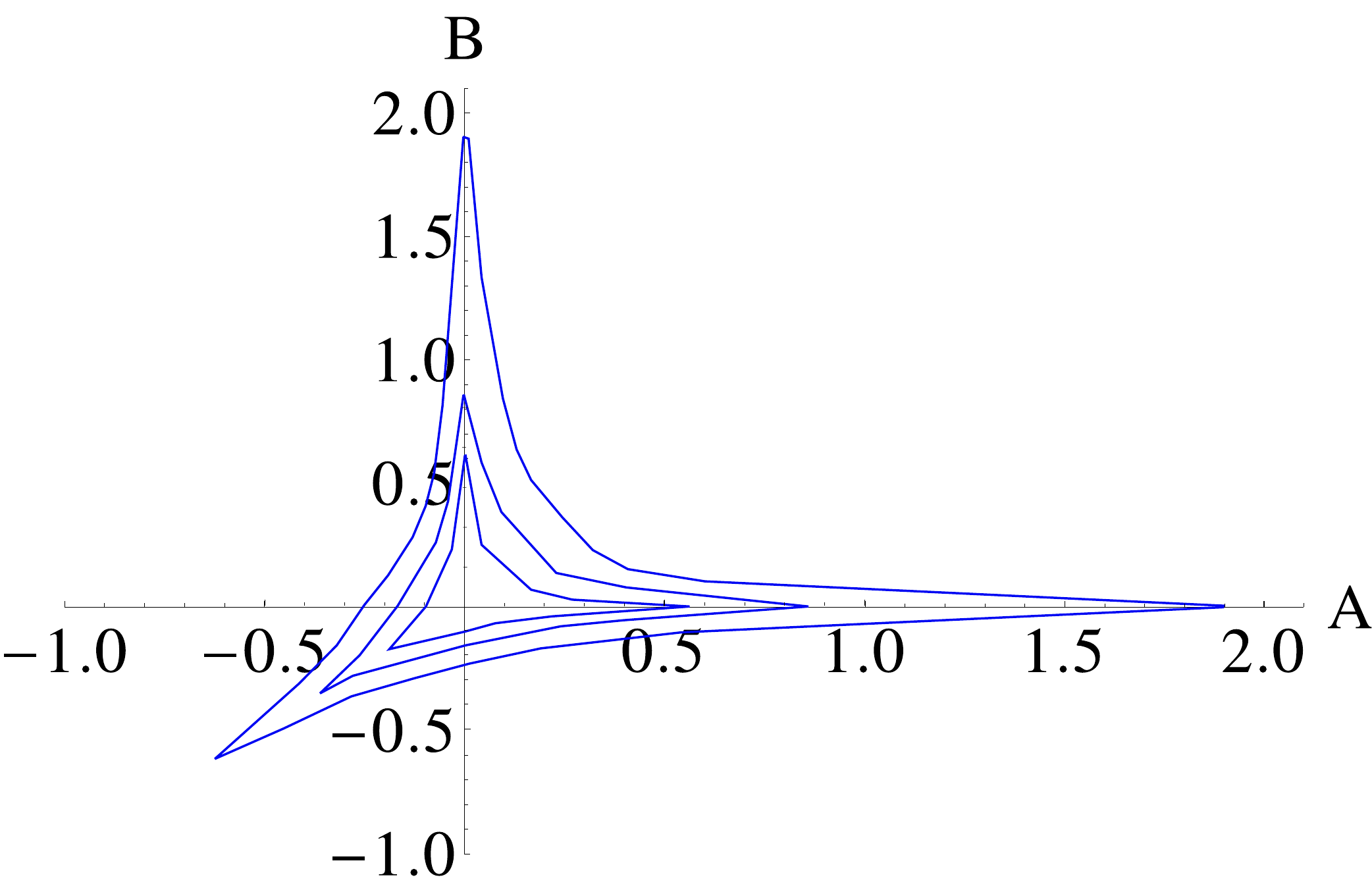}
\includegraphics[width=0.48\textwidth]{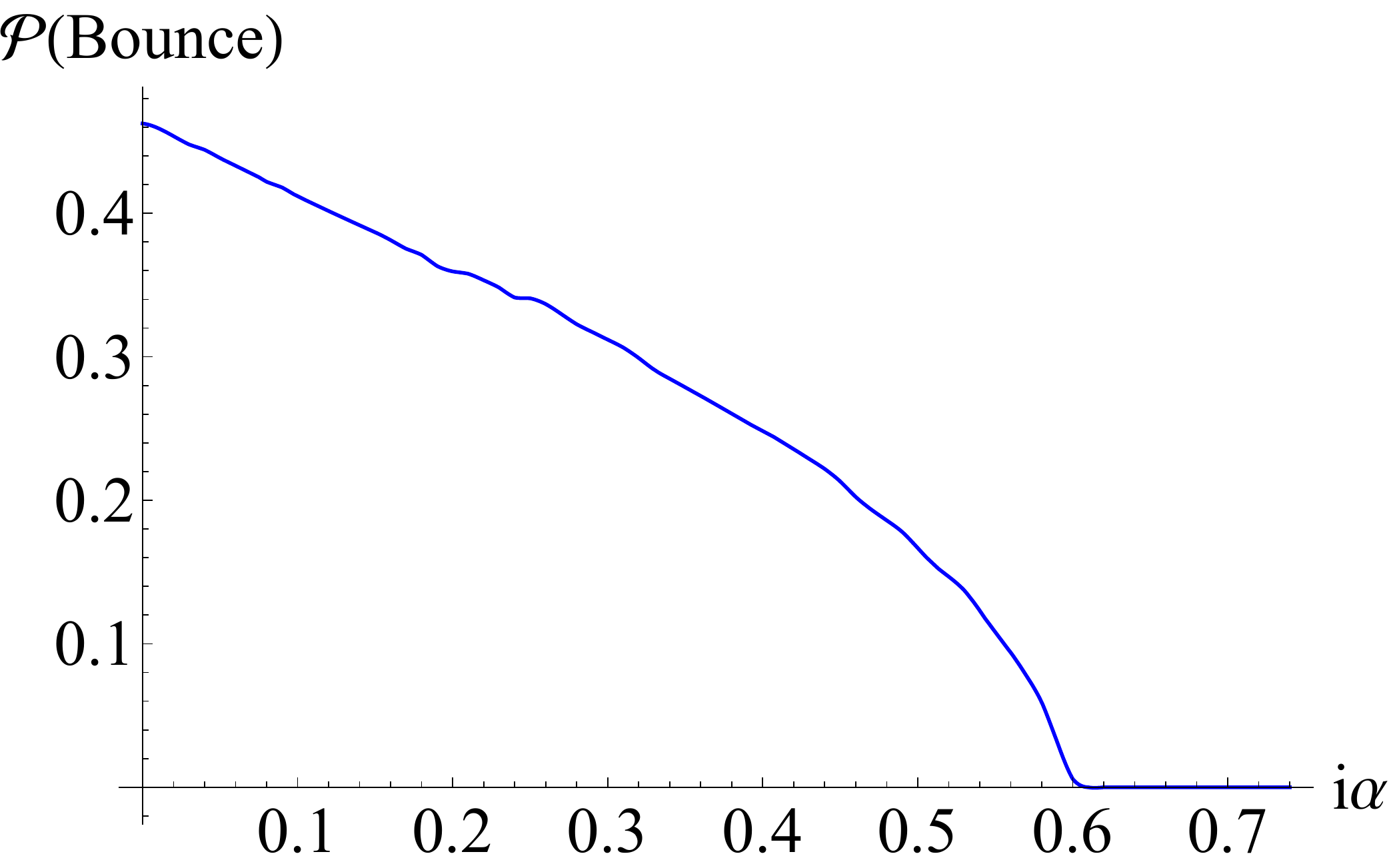}
\caption{\textit{Left panel:} The three contours bound the region in the $(A,B)$ plane for which an approximately classical bounce occurs for (from small to large) resp. $\alpha=3/8i$, $\alpha=i/4$ and $\alpha=0$.
\textit{Right panel:} The total probability for a semiclassical bounce as a function of $\alpha$.}\label{fig:parspace}
\end{figure}

Within the minisuperspace model the classical extrapolation of the histories backwards in time is justified as long as the classicality conditions \eqref{eq:classicalityIntro} hold. We find two distinct classes of past evolutions. For reasonably small values $(A,B,\Phi_f)$ the classical extrapolation backwards exhibits a de Sitter like bounce to approximately the same (time reversed) history on the other side. By contrast, for large values $(A,B,\Phi_f)$ the histories are classically past singular. Fig. \ref{fig:clashistories} shows a representative example in each class. The classical extrapolations of all Bolt saddle points, which only exist for large squashings, are past singular. The range of squashings $(A,B)$ for which the classical histories bounce in the past decreases for increasing $\Phi_f$. This is in line with our expectations for this particular scalar potential, which becomes too steep at large $\Phi$ to sustain inflation. We illustrate this in the left panel of Fig. \ref{fig:parspace}  where we show the region in the $(A,B)$ phase space for three different values of $\Phi_f$ within which the classically extrapolated histories bounce.

\medskip
\noindent {\it Semiclassical Wave Function}
\medskip

The relative probabilities of the individual histories in the classical ensemble are fully specified by the regularized action of the interior AdS domain wall regime of the saddle points. Specifically in the large three-volume region we have \cite{Hertog2011}
\begin{equation}
I_{\rm R}[a,A,B,\Phi_f] = -I_{AdS}^{\rm reg}[A,B,\alpha_f] \ ,
\end{equation}
where $\alpha_f \equiv \alpha (\Phi_f)$ is defined in the AdS regime of the saddle points (cf.  \eqref{UVscalar2}) and purely imaginary for real boundary values $\Phi_f$ in the dS domain \cite{Hertog2011}. To compute the regularized action one can perform the regularization procedure numerically as detailed in the Appendix for the real AdS solutions. However it is more convenient to consider the complex saddle points along a different contour, indicated with ${\cal C'}$ in Fig. \ref{contour}. This yields a geometric representation of the solutions in which a Euclidean deformed four sphere gradually transitions to a Lorentzian asymptotically locally de Sitter space. The Lorentzian behavior of the solution along the second leg of ${\cal C'}$ means the real part of the Euclidean action stabilizes automatically along ${\cal C'}$ \cite{Hartle2008}.

 \begin{figure}[ht!] 
 \centering
 \includegraphics[width=0.7\textwidth]{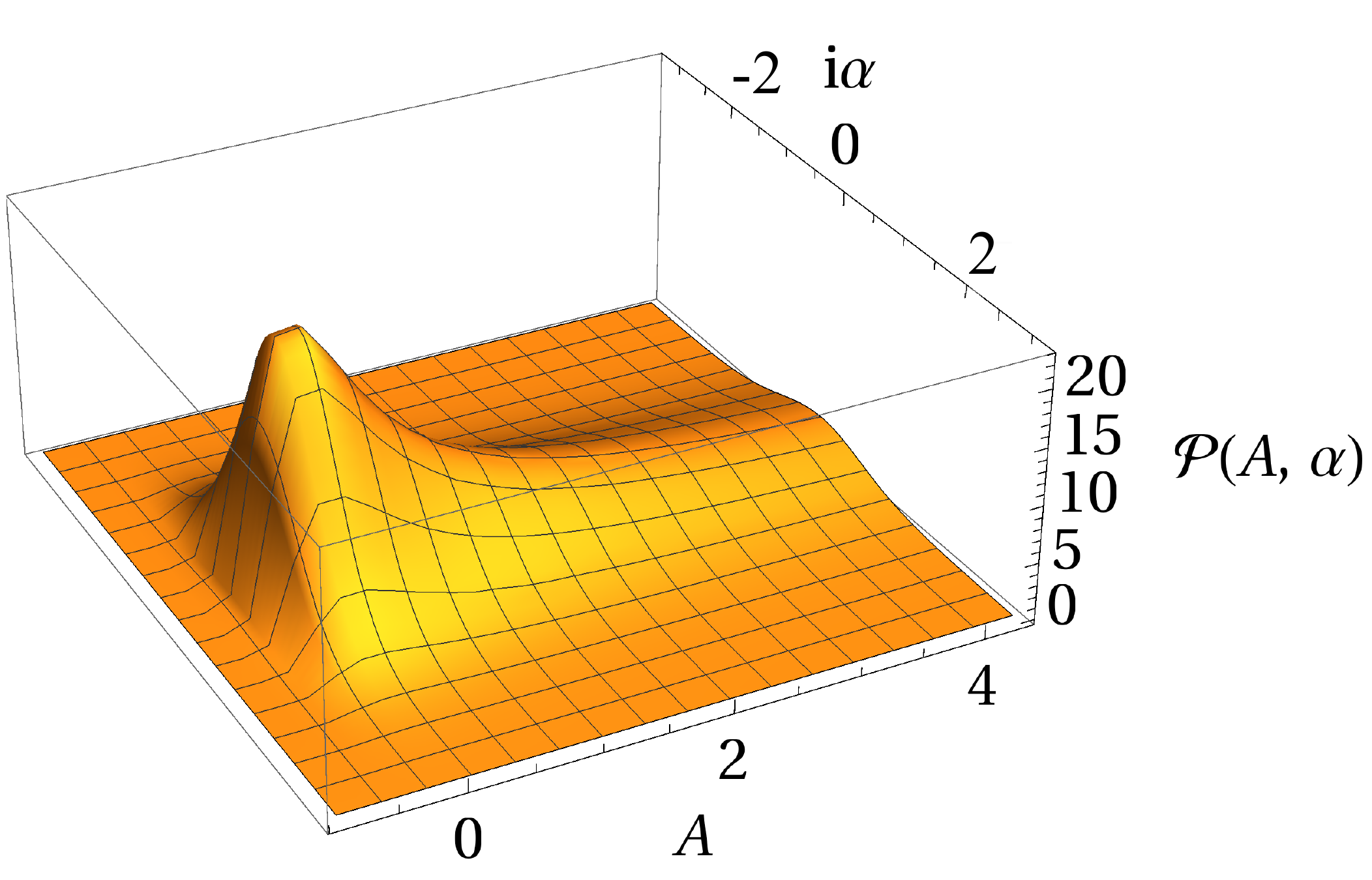}
\caption{The probability distribution over anisotropic minisuperspace as a function of the squashing $A$ and the amount of scalar field inflation parameterized by $i\alpha$.}
\label{fig:PBdS}
\end{figure}

\begin{figure}[ht!] 
 \centering
\begin{subfigure}[t]{0.32\textwidth}
 \includegraphics[width=\textwidth]{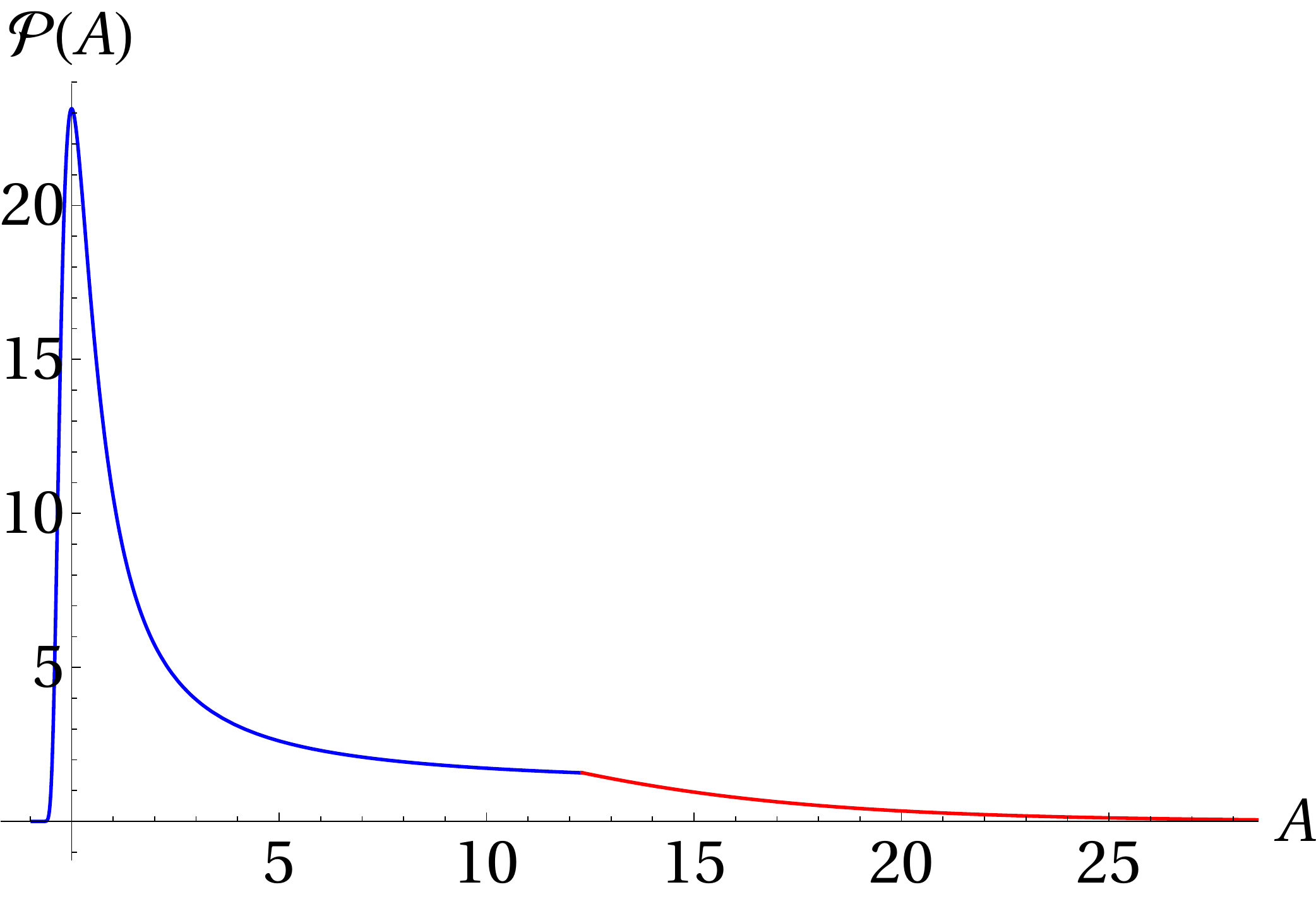}
\caption{$\mathcal{P}(A)$ for $\alpha=0$}
\end{subfigure}
\begin{subfigure}[t]{0.32\textwidth}
 \includegraphics[width=\textwidth]{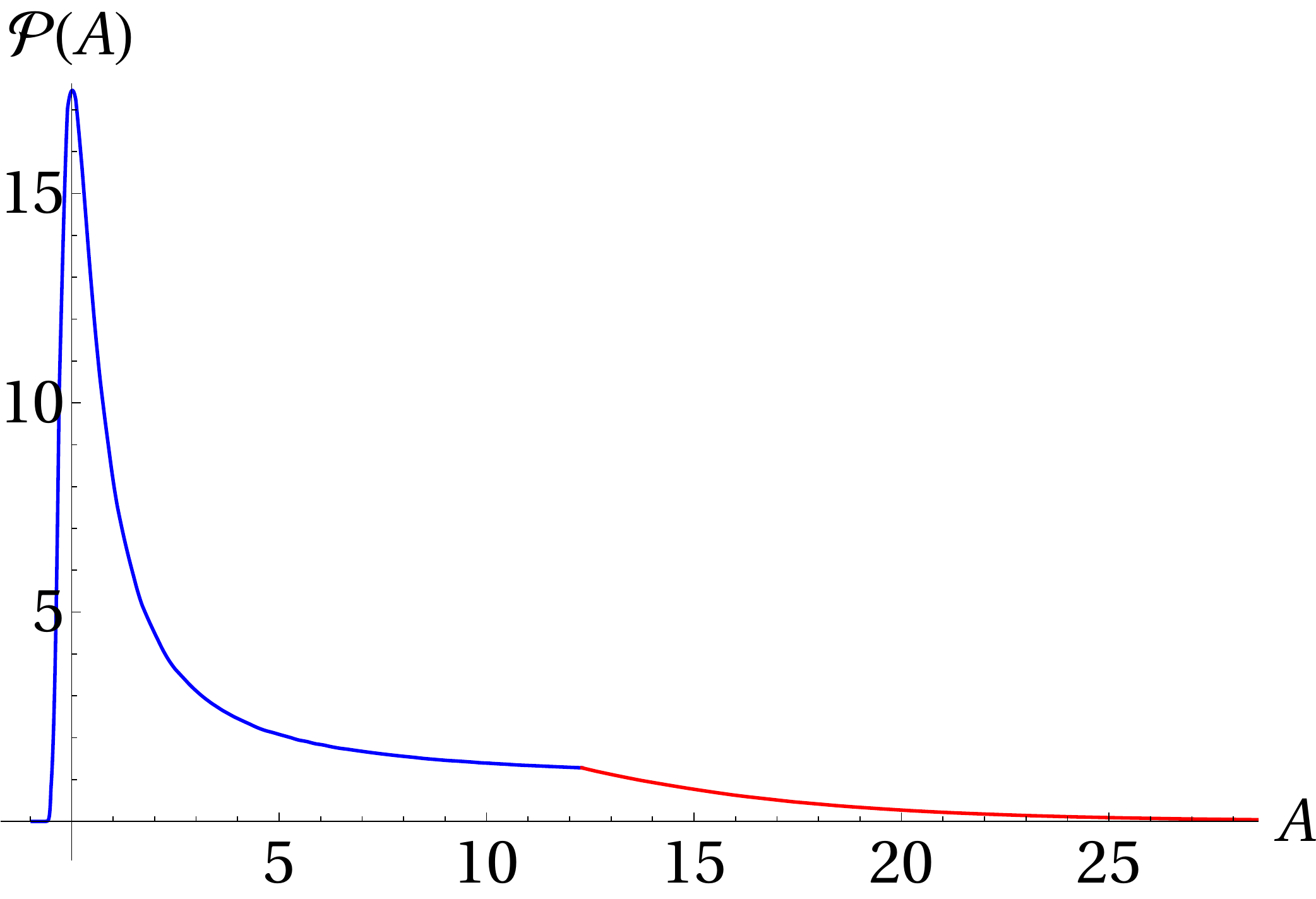}
\caption{$\mathcal{P}(A)$ for $\alpha=i/2$}
\end{subfigure}
\begin{subfigure}[t]{0.32\textwidth}
 \includegraphics[width=\textwidth]{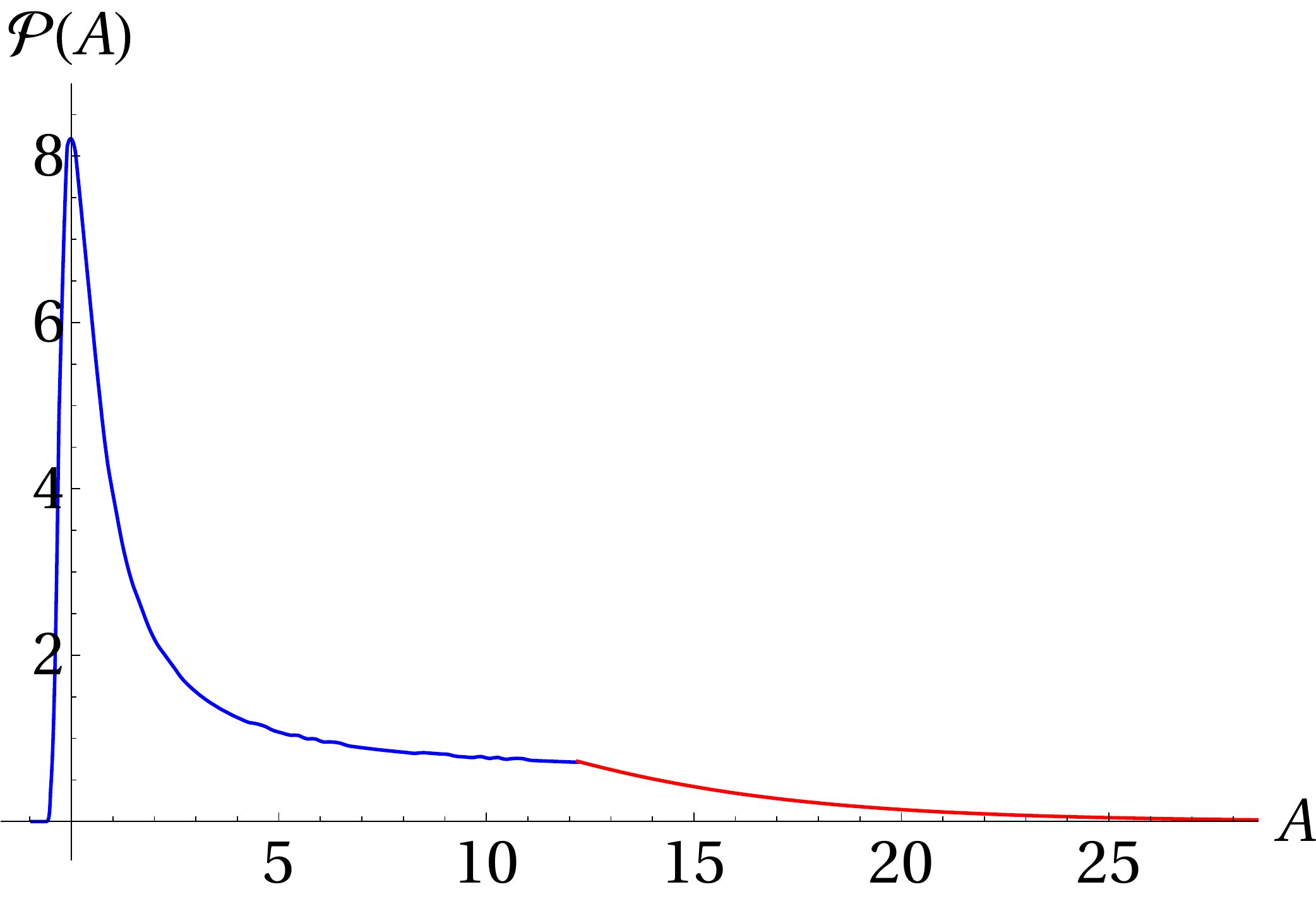}
\caption{$\mathcal{P}(A)$ for $\alpha=i$}
 \end{subfigure}
\caption{Slices of the probability distributions for $B=0$ for three different values of $\alpha$. The blue curves show the NUT contributions whereas the dominant Bolt contribution is given in red.}
\label{fig:PBdS2}
\end{figure}

We show a two-dimensional slice $P(A,\alpha)$ of the probability distribution for $B=0$  in Fig. \ref{fig:PBdS}. As expected the distribution is normalizable and peaks at the pure de Sitter history with a round sphere boundary and zero scalar field. 
In Fig. \ref{fig:PBdS2} we show slices of constant $\alpha$ of this distribution $P(A)$ for three different values of the coefficient $\alpha$ specifying the asymptotic scalar profile. The Bolt solutions provide the dominant contribution to the probabilities at large squashings $A$. This is the dS counterpart of the Hawking-Page like phase transition in the AdS domain of the wave function. The total probability of histories associated with Bolt saddle points is small however and decreases for increasing scalar field.

The probability distribution over the classical ensemble can also be used to compute the total probability in this model that an asymptotically classical universe emerges from a regular bounce in the past and therefore lies in the quasiclassical realm throughout its entire history. This is obtained by integrating the probability distribution over the domain in the $(A,B)$-plane shown in the left of Fig. \ref{fig:parspace}. This corresponds, for a given scalar value $\Phi_f$ on a constant $a$ surface, to bouncing histories when classically extrapolated backwards. We plot the probability $P_{\rm bounce}$ as a function of $\alpha$ in the right panel of Fig. \ref{fig:parspace}, where we restriced to a single squashing. This shows that the total probability of a non-singular origin is significant in this model when the scalar field is everywhere relatively small. However, it sharply decreases outside this regime and vanishes for histories in the large field regime near the edge of the inflationary regime of the potential.

\section{Holographic Wave Function}

In \cite{Hawking:2017wrd} the holographic form \eqref{dSCFT} of the wave function was studied in a vector toy model dual to the anisotropic minisuperspace of eternally inflating cosmologies we considered in the previous section. Here we compare this with our bulk results of Section \ref{dscft1}.

In its holographic form \eqref{dSCFT} the arguments of the wave function enter as sources in the dual. For the geometry this means one must evaluate the dual partition function on the double-squashed sphere \eqref{eqn:metric} as before. For the bulk scalar this means it now sources a scalar deformation by an operator ${\cal O}$ of dimension one, with coupling $\alpha$. This is a relevant operator which in our dual $O(N)$ vector toy model induces a flow from the free to the critical $O(N)$ model. Moreover the coefficient $\alpha$ is imaginary in the dS domain of the theory as discussed above. Hence one is led to evaluate the partition function of the critical $O(N)$ model as a function of the squashing parameters $A$ and $B$ and an imaginary mass deformation $\alpha \equiv \tilde{m}^2$.

The deformed, critical $O(N)$ model is obtained from a double trace deformation $f(\phi\cdot \phi)^2/(2N)$ of the free model \eqref{eqn:Zfree} with in addition a source $\rho f \tilde{m}^2$ turned on for the single trace operator $\mathcal{O} \equiv (\phi \cdot \phi)$. In the limit $f\rightarrow \infty$ the theory flows from its unstable UV fixed point where the source has dimension one to its critical fixed point with a source of dimension two \cite{Klebanov2002}. To see this, we introduce an auxiliary variable $\tilde{m}^2=\frac{m^2}{\rho f} +\frac{\mathcal{O}}{\rho}$ and write \eqref{eqn:Zfree} as
\begin{align}
	Z_{\rm free}[A,B,m^2]=\int \mathcal{D} \tilde{m}^2 e^{-\frac{N}{2f}\int d^3x\sqrt{g} (m^2-\rho f \tilde{m}^2)^2} Z_{\rm crit}[A,B,\tilde{m}^2]\ ,  \label{eqn:ZfreeifoZcrit}
\end{align}
with 
\begin{align}
	Z_{\rm crit}[A,B,\tilde{m}^2] =  \int \mathcal{D}\phi e^{-I_{\rm free}+N \int d^3x\sqrt{g}\left[\rho f \tilde{m}^2\mathcal{O} -\frac{f}{2} \mathcal{O}^2\right]} \ .
\end{align}
Inverting \eqref{eqn:ZfreeifoZcrit} yields $Z_{\rm crit}$ as a function of $Z_{\rm free}$:
\begin{align}
	Z_{\rm crit}[A,B,\tilde{m}^2] = e^{\frac{Nf\rho^2}{2}\int d^3x \sqrt{g} \tilde{m}^4}  \int \mathcal{D}m^2 e^{N\int d^3x\sqrt{g} \left(\frac{m^4}{2 f} -\rho \tilde{m}^2 m^2\right)}Z_{\rm free}[A,B,m^2] \ . \label{eqn:ZcritifoZfree}
\end{align}
The value of $\rho$ can be fixed by comparing two-point functions in the bulk with those in the boundary theory \cite{Anninos2012}. For the $O(N)$ model this implies $\rho=1$ \cite{Anninos2013}. 

To compute $Z_{\rm crit}[A,B,\tilde m^2]$ one can first calculate the partition function of the free mass deformed $O(N)$ vector model on a double squashed sphere and then evaluate \eqref{eqn:ZcritifoZfree} in a large $N$ saddle point approximation. The factor outside the path integral in \eqref{eqn:ZcritifoZfree} diverges in the large $f$ limit. This can be canceled by adding the appropriate counterterms. The saddle point equation then becomes, for homogeneous deformations, 
\begin{align}
\frac{2 \pi^2}{\sqrt{(1+A)(1+B)}} \left(\frac{m^2}{f}  - \tilde{m}^2\right) =- \frac{\partial \log Z_{\rm free}[A,B,m^2]}{\partial m^2} \ .\label{eqn:saddleSPN}
\end{align}
Solving this for $m^2$ in the large $f$ limit and inserting the result in \eqref{eqn:ZcritifoZfree} yields $Z_{crit}[A,B,\tilde m^2]$.  

\begin{figure}[ht!]
\centering
  \begin{subfigure}[t]{0.31\textwidth}
    \includegraphics[width=\textwidth]{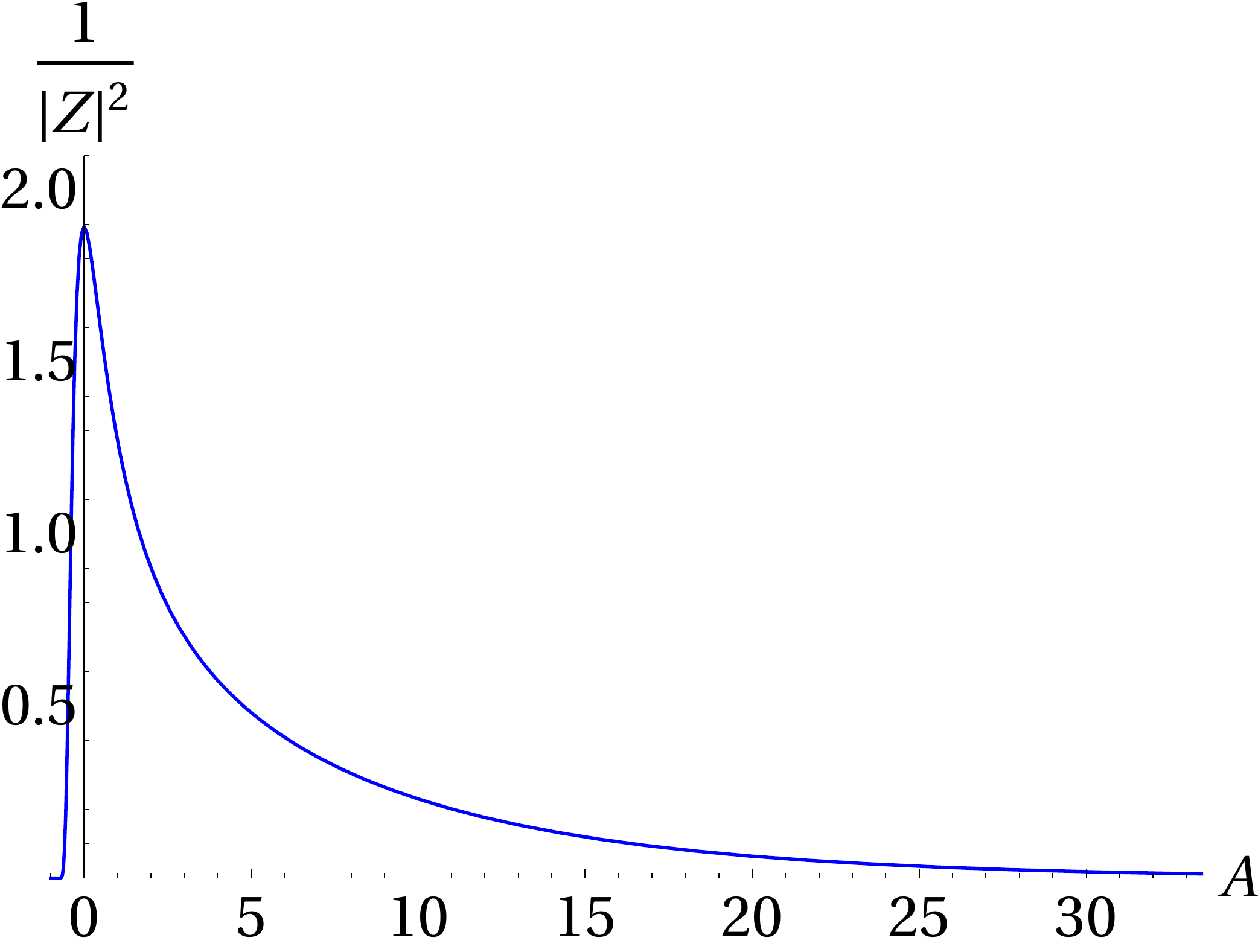}
    \caption{$\tilde{m}^{2}=0.0$}
  \end{subfigure}
    \begin{subfigure}[t]{0.31\textwidth}
    \includegraphics[width=\textwidth]{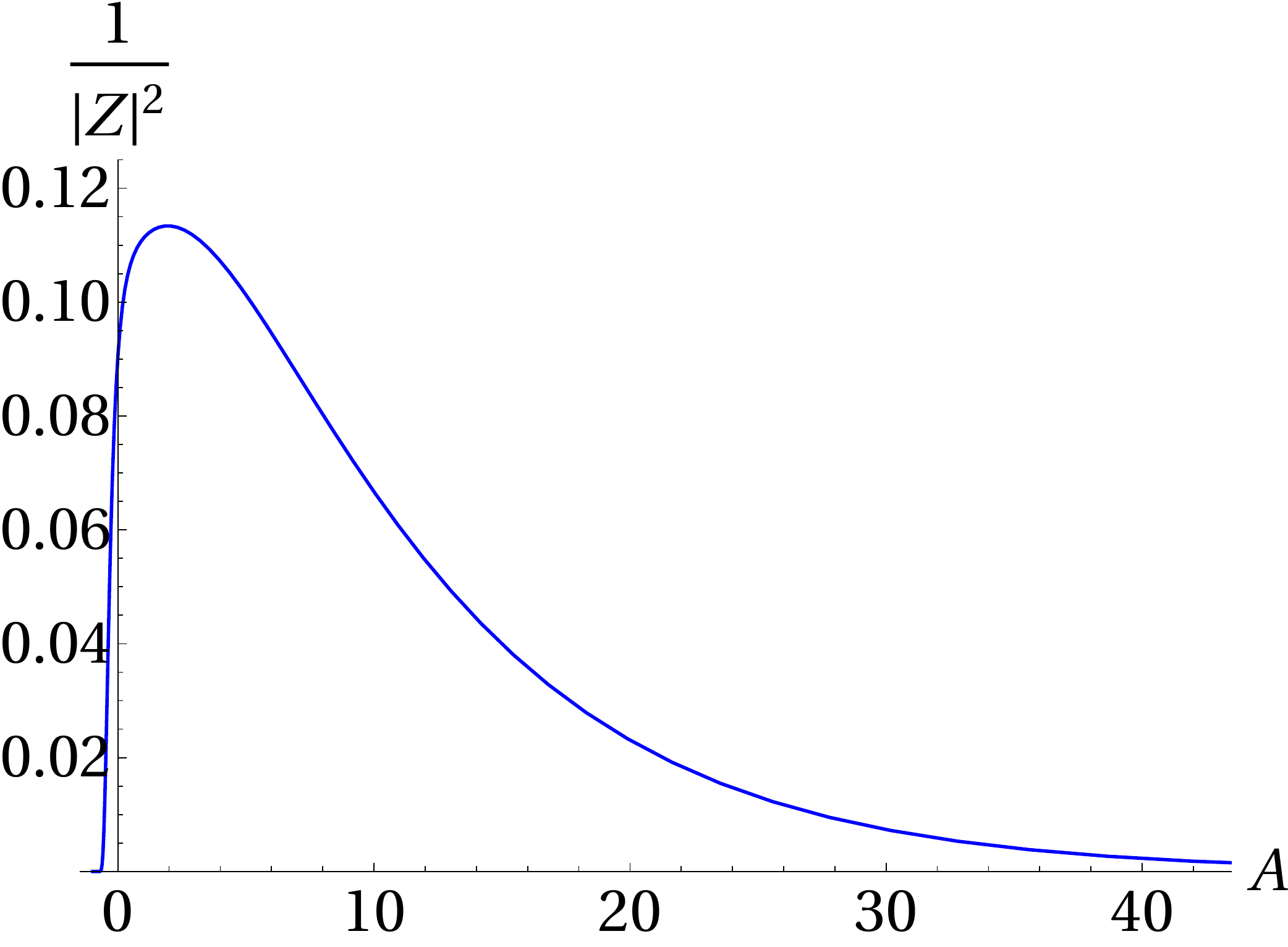}
    \caption{$\tilde{m}^{2}=0.05i$}
  \end{subfigure}
	    \begin{subfigure}[t]{0.31\textwidth}
    \includegraphics[width=\textwidth]{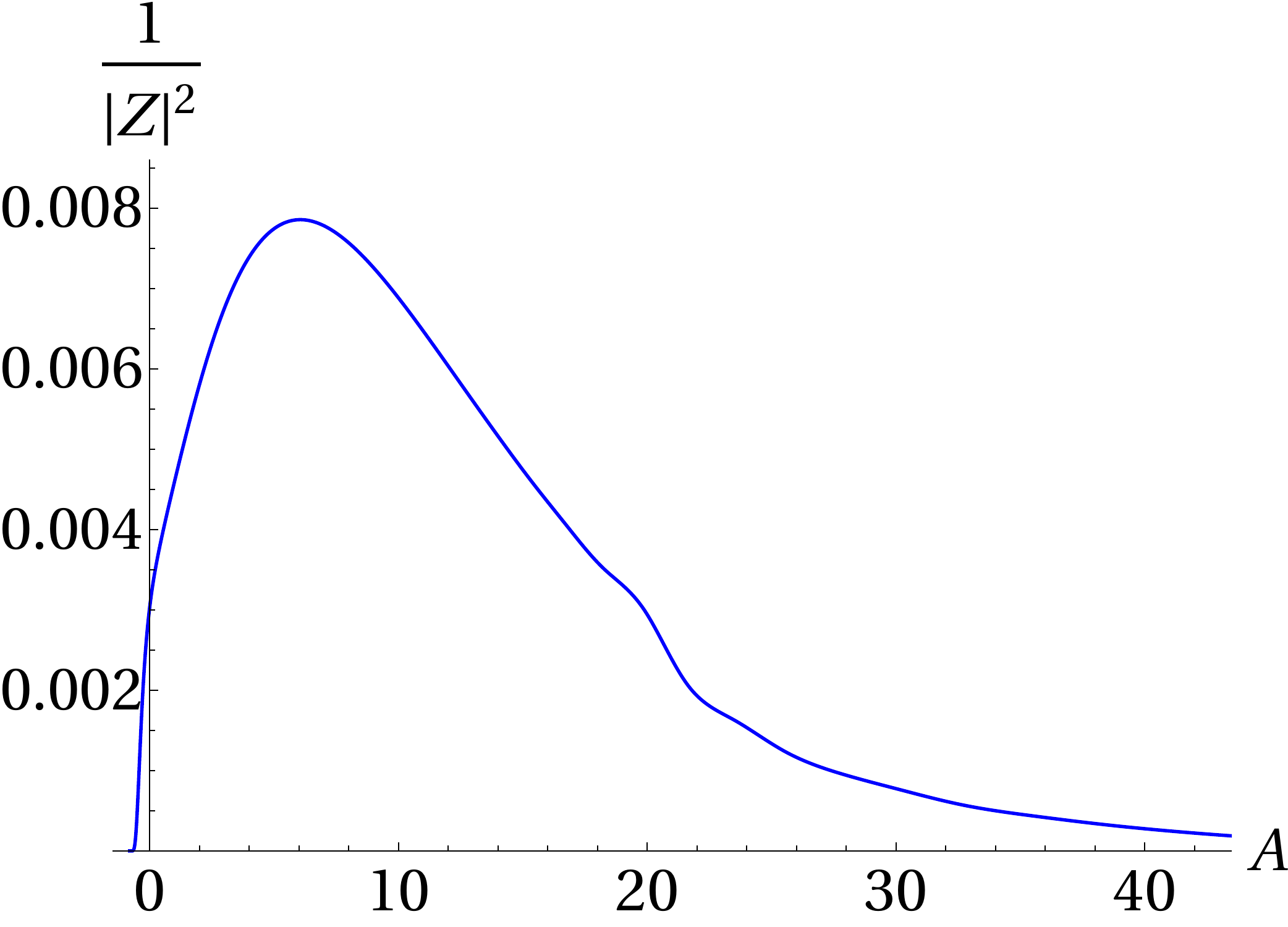}
    \caption{$\tilde{m}^{2}=0.1i$}
    \end{subfigure}
\caption{Three one-dimensional slices of the holographic probability distribution in a vector toy model dual of eternal inflation, for three different values of $\tilde m^2$. Remarkably, the distribution qualitatively agrees with that obtained using bulk methods and shown in Fig. \ref{fig:PBdS}.}\label{fig:PvsBmsqSpNB}
\end{figure}

In \cite{Hawking:2017wrd} this procedure was implemented to compute $Z_{crit}[A,\tilde m^2]$ for a single squashing $A$ using the numerical techniques described in Section \ref{holo} above and in Appendix \ref{app:CFTreg}, and by numerically inverting \eqref{eqn:saddleSPN} to find the behavior of the complex deformation $m^2$ as a function of imaginary $\tilde{m}^2$. Inserting this in \eqref{eqn:ZcritifoZfree} yields the partition function $Z_{\rm crit}[A,B,\alpha]$. The resulting holographic probability distribution over $i\tilde m^2$ and $A$ turns out to be well behaved and normalizable, with a global maximum at zero squashing and zero deformation, corresponding to the pure de Sitter history. We illustrate the behavior of this distribution in Fig. \ref{fig:PvsBmsqSpNB} where we plot three one-dimensional slices of the distribution for three different values of $\tilde m^2$. A comparison with the analogous distribution obtained through bulk methods shown in Fig. \ref{fig:PBdS2} and in Fig. \ref{fig:PBdS} shows they qualitatively agree. One sees that when the scalar is turned on, the local maximum in Fig. \ref{fig:PvsBmsqSpNB} shifts slightly towards positive values of $A$ -- a feature which is absent in the bulk result.

A key feature of the distribution is that it exponentially suppresses regions of the configuration space where the combination $R/8 + m^2$ is negative. This includes in particular boundary geometries with negative scalar curvature $R(A) <0$. The holographic measure predicts the amplitude is low for such conformal boundary surfaces far from the round conformal structure. This can be traced to the fact that the partition function of the free $O(N)$ model diverges for sources for which $R/8 + m^2 \leq 0$, as we discussed in Section \ref{holo} above. Through \eqref{eqn:ZcritifoZfree} this implies the holographic measure \eqref{dSCFT} strongly suppresses these configurations\footnote{We attribute the exponentially small tail to the saddle point approximation of \eqref{eqn:ZcritifoZfree}.}. This also means it is very much plausible this result generalizes to two squashings and more general metric deformations indeed \cite{Hawking:2017wrd}. Based on this it was conjectured in \cite{Hawking:2017wrd} that global constant density surfaces in eternal inflation are globally smoother than what the usual semiclassical gravity analysis indicates. 

\section{Discussion}

We have studied gauge-gravity duality with squashed boundary geometries and in the presence of scalar excitations in the bulk. 
In AdS these describes scalar condensates, whereas in dS the scalar drives inflation.

The AdS bulk solutions we construct are generalizations of the AdS Taub-NUT/Bolt solutions to two anisotropy parameters and with a scalar field turned on. We have compared the thermodynamic properties of these solutions to the behavior of the partition of the deformed free $O(N)$ vector model defined on the double squashed three sphere. Even though the latter is dual to higher-spin gravity, we find that both theories exhibit a number of remarkable similarities. This includes the behavior of the one-point function of the scalar operator illustrated in Fig. \ref{fig:vevvssource}, which shows that the expectation value diverges for a finite value of the source. In the bulk this property depends on the detailed shape of the scalar potential for large field values. Another correspondence concerns the behavior of the free energy as a function of the squashings for zero scalar field \cite{Bobev2016}. A notable difference between both theories arises when the scalar is turned on: Whereas the gravitational action attains a local maximum at zero squashings for all values of the scalar condensate, the maximum on the CFT side shifts slightly towards positive squashings. 

 
The dS bulk solutions we construct describe anisotropic deformations of de Sitter space with a scalar field driving (eternal) inflation. 
They close off in a regular manner in the past and therefore yield valid saddle points of the no-boundary wave function. We have compared the resulting semiclassical measure with the holographic measure specified by the partition function of the interacting $O(N)$ vector model defined on a two-parameter family of squashed three spheres and deformed by a mass term. Again we find a remarkable agreement between both theories. In both cases the amplitude is low for conformal boundary surfaces far from the round conformal structure. This is in line with general field theory expectations and lends support to the conjecture of \cite{Hawking:2017wrd} that the exit from eternal inflation is reasonably smooth, producing universes that are relatively regular on the largest scales with globally finite surfaces of constant density. 

\bigskip

\noindent \textbf{ Acknowledgements:}
\noindent We thank Nikolay Bobev, Adam Bzowski, James Hartle, Stephen Hawking, Ruben Monten and Edgar Shaghoulian for useful discussions. Our work is supported in part by the European Research Council grant no. ERC-2013-CoG 616732 HoloQosmos and the KU Leuven C1 grant ZKD1118 C16/16/005.

\appendix

\section{Complex Anisotropic Scalar Domain Wall Solutions}
\label{App:AdSexpansions}
In these appendices we provide the technical details to find the anisotropic bulk solutions and their action in the  gravitational theories discussed in the main text.

\subsection{Equations of motion}

The equations of motion can be obtained by substituting the metric \eqref{eqn:doublesqansatz} into the action \eqref{eqn:GRaction} and by  varying this respectively with respect to $l_0$, $l_1$, $l_2$, $l_3$ and $\Phi$,\footnote{One can explicitly check that these equations satisfy the Einstein equations. More precisely,  the first equation here is equal to the $(r,r)$-component of the Einstein equation, the second equation is a linear combination of the $(\theta,\theta)$-component and $(\theta,\phi)$-component of the Einstein equation and the fourth equation is related to the $(\psi,\psi)$-component which is proportional to the $(\psi,\phi)$-component. All the other components of the Einstein equations are linearly dependent to the equations of motion presented here, or they are 0.}

\begin{align} \label{eom}
   & \frac{ {l_0}^2 { l_3}^2 }{ { l_1}^2   { l_2}^2 }  +\frac{{l_0}^2   { l_2}^2 }{ { l_1}^2   { l_3}^2 }+\frac{ {l_0}^2  { l_1}^2 }{ { l_2}^2   {l_3}^2 }-\frac{2 {l_0}^2   }{ {l_1}^2 }-\frac{2  {l_0}^2  }{ {l_2}^2 }-\frac{2 {l_0}^2   }{ {l_3}^2 }+4 {l_0}^2  V(\Phi) 
    +\frac{ 4{l_1}'   {l_2}' }{ {l_1}   {l_2} } +\frac{4 {l_1}'  {l_3}' }{ {l_1}   {l_3} } 	+\frac{4 {l_2}'   {l_3}' }{ {l_2}   {l_3} } -2 {\Phi'}^2=0\ , \nonumber  \\
  &   -\frac{4   {l_0}'   {l_1}' }{ {l_0}     {l_1} }  -\frac{4  {l_0}'   {l_3}' }{   {l_0}   {l_3} }-\frac{  {l_0}^2   {l_3}^2 }{ {l_1}^2   {l_2}^2 }+\frac{3{l_0}^2  
    {l_2}^2 }{ {l_1}^2   {l_3}^2 }-\frac{ {l_0}^2    {l_1}^2 }{ {l_2}^2   {l_3}^2 }-\frac{2{l_0}^2   }{ {l_1}^2 }+\frac{2 {l_0}^2   }{ {l_2}^2 }-\frac{2 {l_0}^2  }{ {l_3}^2 }
    +4  {l_0}^2   V(\Phi)  \nonumber  \\ &+\frac{4  {l_1}'' }{ {l_1} }  
      +\frac{4 {l_1}'   {l_3}' }{ {l_1}   {l_3} }  +\frac{4  {l_3}'' }{ {l_3} }  +2{\Phi'}^2=0\ ,\nonumber \\
  & -\frac{4  {l_0}'    {l_2}' }{      {l_2} }  -\frac{4 {l_0}'   {l_3}' }{      {l_3} }-\frac{{l_0}^2       {l_3}^2 }{ {l_1}^2   {l_2}^2 }-\frac{ {l_0}^2    
    {l_2}^2 }{ {l_1}^2   {l_3}^2 }+\frac{3  {l_0}^2      {l_1}^2 }{ {l_2}^2   {l_3}^2 }+\frac{2  {l_0}^2    }{ {l_1}^2 }-\frac{2  {l_0}^2    }{ {l_2}^2 }-\frac{2{l_0}^2 
       }{ {l_3}^2 } 
     +4   {l_0}^2    V(\Phi) \nonumber  \\ & +\frac{4  {l_2}'' }{ {l_2} } 
      +\frac{ 4{l_2}'   {l_3}' }{ {l_2}   {l_3} }    +\frac{4    {l_3}'' }{ {l_3} }+2{\Phi'}^2=0\ ,\nonumber \\
  & -\frac{ 4 {l_0}'  {l_1}' }{      {l_1} }  -\frac{4   {l_0}' {l_2}' }{      {l_2} }+\frac{3 {l_0}^2        {l_3}^2 }{ {l_1}^2   {l_2}^2 }-\frac{  {l_0}^2    {l_2}^2 }{ {l_1}^2   {l_3}^2 }-\frac{  {l_0}^2     {l_1}^2 }{ {l_2}^2   {l_3}^2 }-\frac{2  {l_0}^2    }{ {l_1}^2 }-\frac{2  {l_0}^2    }{ {l_2}^2 }+\frac{2   {l_0}^2   }{ {l_3}^2 }
   +4  {l_0}^2    V(\Phi) \nonumber  \\ & +\frac{4  {l_1}'' }{ {l_1} } 
     +\frac{ 4 {l_1}'   {l_2}' }{ {l_1}   {l_2} } +\frac{4 {l_2}'' }{ {l_2} }  +2{\Phi'}^2=0\  ,\nonumber \\
 &  {l_0}^2  \frac{\partial   V(\Phi)}{\partial \Phi}   +  \frac{{l_0}' \Phi'}{{l_0}}   -\frac{ {l_1}' \Phi'}{l_1}-\frac{ {l_2}' \Phi'}{l_2} -\frac{ {l_3}' \Phi'}{l_3} -\Phi''=0 \ .
\end{align}
These equations of motion are valid for both the AdS and dS domain of the wave function, for this reason the variables are understood to be a function of $\tau$, defined in \eqref{eqn:tau} and $'$ means a derivative with respect to $\tau$. The AdS equations of motion get retrieved by setting $\tau=r$, and the Lorentzian dS solutions lie along the line $\tau = t + i \pi /2$. In the rest of this appendix we will use the same gauge as in the main text, namely $l_0=1$. 

\subsection{Solutions}

\subsubsection{IR NUT}

For the NUT solutions we know that around the NUT, denoted here by $\tau_*$, the metric should look like $\mathbb{R}^4$
\begin{align}
ds^2= d\tau^2 + \frac{(\tau-\tau^*)^2}{4}(\sigma_1^2+\sigma_2^2+\sigma_3^2) \ .
\end{align}
Therefore we can expand the fields around $\tau = \tau^*$  with the following Ansatz
\begin{align}
 \label{nutexp}
&\Phi(\tau) =\Phi_0+ \Phi_k (\tau-\tau^*)^k, \quad
 &l_1(\tau)=\frac{1}{2}( \tau-\tau^*)  + \beta_{k+1} (\tau-\tau^*)^{k+1}\;, \nonumber \\
 &l_2(\tau)=\frac{1}{2}( \tau-\tau^*)  + \gamma_{k+1} ( \tau-\tau^*)^{k+1} \;, 
 &l_3(\tau)=\frac{1}{2}( \tau-\tau^*)  + \delta_{k+1} ( \tau-\tau^*)^{k+1}\;, 
\end{align}
where $k$ runs from 1 to $\infty$.  
By plugging in this Ansatz into the equations of motion \eqref{eom} we get the following leading order terms
\begin{equation} \label{icnut}
\begin{split}
l_1(\tau)& =\frac12 (\tau-\tau^*) +\beta_3 (\tau-\tau^*)^3 +\frac{1}{1920}\bigg( -4V(\Phi_0)^2-576V(\Phi_0) \gamma_3 - 6912{\gamma_3}^2 \\
& +144 V(\Phi_0) \beta_3 -6912 \gamma_3\beta_3 +4608 {\beta_3}^2 -3   \left(\frac{\partial V(\Phi_0)}{\partial\Phi_0}\right)^2\bigg)(\tau-\tau^*)^5  +\mathcal{O}\left((\tau-\tau^*)^7\right) \ , \\
l_2(\tau) & = \frac12 (\tau-\tau^*) +\gamma_3 (\tau-\tau^*)^3 +\frac{1}{1920}\bigg( -4V(\Phi_0)^2+144V(\Phi_0) \gamma_3  + 4608{\gamma_3}^2 \\
&   -576 V(\Phi_0) \beta_3 -6912 \gamma_3\beta_3 +6912 {\beta_3}^2 -3   \left(\frac{\partial V(\Phi_0)}{\partial\Phi_0}\right)^2\bigg)(\tau-\tau^*)^5 +\mathcal{O}\left((\tau-\tau^*)^7\right) \ , \\
l_3(\tau) & = \frac12 (\tau-\tau^*) -\left(\frac{1}{12}V(\Phi_0)+\beta_3+\gamma_3 \right)  (\tau-\tau^*)^3 +\frac{1}{1920}\bigg( 16V(\Phi_0)^2 +624V(\Phi_0) \gamma_3 \\
&  + 4608{\gamma_3}^2 +624 V(\Phi_0) \beta_3 +16128 \gamma_3\beta_3 +4608 {\beta_3}^2 -3 \left(\frac{\partial V(\Phi_0)}{\partial\Phi_0}\right)^2\bigg)(\tau-\tau^*)^5 \\
&  +\mathcal{O}\left((\tau-\tau^*)^7\right) \ , \\
\Phi(\tau) & = \Phi_0 +\frac18 \frac{\partial V(\Phi_0)}{\partial\Phi_0} (\tau-\tau^*)^2 + \bigg(\frac{1}{288} V(\Phi_0)   \frac{\partial V(\Phi_0)}{\partial\Phi_0} +\frac1{192}  \frac{\partial V(\Phi_0)}{\partial\Phi_0}\frac{\partial^2 V(\Phi_0)}{\partial\Phi_0^2} \bigg) (\tau-\tau^*)^4 \\ &  +\mathcal{O}\left((\tau-\tau^*)^6\right) \ .
\end{split}
\end{equation}
This expansion is controlled by the three real parameters $\beta_3$, $\gamma_3$ and $\Phi_0$ which are ultimately related to the two squashing parameters $A$ and $B$ together with the coefficients $A_3$ and $B_3$ of the subleading terms and the two free parameters in the UV expansion of the scalar field $\alpha$ and $\beta$, at the asymptotic boundary. 

\subsubsection{IR Bolt}
%
We can do the same thing for the Bolt solutions. In this case we know that the metric should look like $\mathbb{R}^2\times S^2$ around the Bolt position $\tau_*$, that is
\begin{align}
ds^2= d\tau^2 + \frac{(\tau-\tau^*)^2}{4}\sigma_1^2+\beta_0^2\sigma_2^2+\gamma_0^2\sigma_3^2 \ .
\end{align}
 Therefore we take the following Ansatz for the expansion of the fields around $\tau=\tau^* $, 
 \begin{align}
&\Phi(\tau)=\Phi_0+ \Phi_k ( \tau-\tau^*)^k, \quad
& l_1(\tau)=\beta_0 + \beta_k ( \tau-\tau^*)^k  \;, \nonumber \\
 &l_2(\tau)=\gamma_0 + \gamma_k ( \tau-\tau^*)^k\;, 
 &l_3(\tau)=\frac{1}{2}( \tau-\tau^*)  + \delta_{k+1} ( \tau-\tau^*)^{k+1} \;,
\end{align}
with $k$ going from 1 to $\infty$. 
If we solve the equations of motion \eqref{eom} with this Ansatz, we get
\begin{align} \label{icbolt}
\begin{split}
l_1(\tau)=\gamma_0 +&\left(\frac1{4\gamma_0}-\frac{ V(\Phi_0)\gamma_0}{4}\right) (\tau-\tau^*)^2 -   \bigg( \frac{11}{192{\gamma_0}^3} -\frac{ V(\Phi_0)}{24\gamma_0}  +\gamma_4 +\frac{ \gamma_0}{32}  \left(\frac{\partial V(\Phi_0)}{\partial\Phi_0}\right)^2 \bigg)(\tau-\tau^*)^4 \\
& +\mathcal{O}\left((\tau-\tau^*)^6\right) \ ,\\
l_2(\tau) = \gamma_0 +&\left(\frac1{4\gamma_0}-\frac{ V(\Phi_0)\gamma_0}{4}\right) (\tau-\tau^*)^2 +\gamma_4 (\tau-\tau^*)^4  +\mathcal{O}\left((\tau-\tau^*)^6\right) \ , \\
l_3(\tau)= \frac12 (\tau&-\tau^*) -\frac{(\tau-\tau^*)^3}{12{\gamma_0}^2} + \bigg( \frac{V(\Phi_0)^2}{160} +\frac{53}{1920 {\gamma_0}^4} -\frac{V(\Phi_0)}{40{\gamma_0}^2}-\frac1{320}   \left(\frac{\partial V(\Phi_0)}{\partial\Phi_0}\right)^2\bigg) (\tau-\tau^*)^5 \\
& +\mathcal{O}\left((\tau-\tau^*)^7\right) \ , \\
\Phi(\tau)=  \Phi_0 + &\frac14\frac{\partial V(\Phi_0)}{\partial\Phi_0}(\tau-\tau^*)^2 + \frac1{192 \gamma_0^2}\frac{\partial V(\Phi_0)}{\partial\Phi_0} \left(-4+3{\gamma_0}^2(2 V(\Phi_0)+ \frac{\partial^2V(\Phi_0)}{\partial \Phi_0^2})\right)(\tau-\tau^*)^4  \\ & +\mathcal{O}\left((\tau-\tau^*)^6\right) \ .
\end{split}
\end{align}
We chose to parametrize this expansion by the three independent real parameters $\gamma_0$, $\gamma_4$ and $\Phi_0$ which are again mapped to the squashing parameters $A$, $B$ and $\alpha$, $\beta$ in the UV. Notice that to get the NUT or Bolt double squashing results without scalar field we have to put $\Phi_0=0$ and $V(\Phi)=\Lambda$ in the initial conditions above, effectively reducing the above expansions and equations of motion to the ones discussed in \cite{Bobev2016}. 

\subsubsection{UV}
The asymptotic solutions are the same for both the NUT and the bolt. To find them, we look at the asymptotic form of the metric 
\begin{align}
ds^2 =d\tau^2  +e^{2\tau}(A_0\sigma_1^2+ B_0 \sigma_2^2+C_0\sigma_3^2) \ .
\end{align}
If we use that the scalar field potential around $\Phi=0$ behaves as $V(\Phi) \sim \Lambda - \Phi^2$, we can make the Ansatz of a Fefferman-Graham expansion
\begin{align} \label{uvans}
& \Phi(\tau)= \alpha e^{-\tau} + \beta e^{-2 \tau} +D_k e^{-(2+k)\tau} \ , 
&l_1(\tau)=A_0 e^{\tau} + A_k e^{(1-k) \tau} \ , \nonumber \\ 
&l_2(\tau)=B_0 e^{\tau} + B_k e^{(1-k) \tau} \ ,   
&l_3(\tau)=C_0 e^{\tau} + C_k e^{(1-k) \tau} \ , 
\end{align}
where the sum over $k$ goes over all positive integers.
The constants are determined by solving the equations of motion \eqref{eom}, order by order, giving the following consistent series expansion
\begin{equation} \label{bval}
\begin{split}
l_1(\tau)& = A_0 e^\tau + \frac1{16 A_0{B_0}^2{C_0}^2}\bigg(-5{A_0}^4 +2 {A_0}^2{B_0}^2   +3{B_0}^4 +2{A_0}^2 {C_0}^2 -6 {B_0}^2{C_0}^2 \\
& \hspace{10mm} +3{C_0}^4 - 2({A_0} {B_0} {C_0})^{4/3} {\alpha}^2\bigg)e^{-\tau} + A_3 e^{-2\tau} +\mathcal{O}(e^{-3\tau}) \ , \\
l_2(\tau)& =  B_0 e^\tau + \frac1{16 A_0{B_0}^2{C_0}^2}\bigg(3{A_0}^4 +2 {A_0}^2{B_0}^2   -5{B_0}^4 -6{A_0}^2 {C_0}^2 +2 {B_0}^2{C_0}^2 \\
& \hspace{10mm}  +3{C_0}^4 -2({A_0} {B_0} {C_0})^{4/3} {\alpha}^2\bigg)e^{-\tau} + B_3 e^{-2\tau} +\mathcal{O}(e^{-3\tau})\ , \\
l_3(\tau)& =  C_0 e^\tau + \frac1{16 A_0{B_0}^2{C_0}^2}\bigg(3{A_0}^4 -6 {A_0}^2{B_0}^2   +3{B_0}^4 +2{A_0}^2 {C_0}^2+2 {B_0}^2{C_0}^2  \\
& \qquad -5{C_0}^4 -2({A_0} {B_0} {C_0})^{4/3} {\alpha}^2\bigg)e^{-\tau}  - \left(\frac{A_3 C_0}{A_0}-\frac{B_3 C_0}{B_0}-\frac{2}{3A_0 B_0}  {\alpha}{\beta}\right) e^{-2\tau} +\mathcal{O}(e^{-3\tau}) \ , \\
\Phi(\tau) & = \frac{\alpha}{(A_0 B_0 C_0)^{1/3}}e^{-\tau} + \frac{\beta}{(A_0 B_0 C_0)^{2/3}}e^{-2\tau}  +\mathcal{O}(e^{-3\tau}) \ .
\end{split}
\end{equation}

We have performed this expansion up to eight order and have verified that it is controlled by the seven parameters $\{A_0,B_0,C_0,A_3,B_3,\alpha,\beta\}$. The coefficients $\alpha$ and $\beta$ appearing in the expansion of $\Phi$ are undetermined by the equations of motion, here we rescaled them to the most convenient convention making sure it is conform with the literature. Notice that when we are deep into the dS domain $\tau=t+i \pi/2$, which makes the scale factors imaginary, giving the Lorentzian metric from \eqref{eqn:UVmetric2}.

 Since the equations of motion \eqref{eom} are invariant under constant shifts of the radial coordinate, one can set $A_0=\frac{1}{4}$ by an appropriate shift of $\tau$. 
One can now identify $B_0$ and $C_0$ with the squashing parameters in \eqref{eqn:metric} as follows
\begin{equation}\label{eqn:alphabetaBC}
A = \frac{1}{4C_0^2} - 1\;, \qquad\qquad B = \frac{1}{4B_0^2} - 1\;.
\end{equation}
The parameters $A_3$, $B_3$ and $\beta$ are independent from the point of view of the UV expansion but are ultimately fixed in terms of $A$, $B$ and $\alpha$ by the regularity conditions that we imposed for the numerical solutions of the full nonlinear equations of motion.

\subsubsection{From IR to UV}

\begin{figure}[ht!]
\centering
\includegraphics[width=0.45\textwidth]{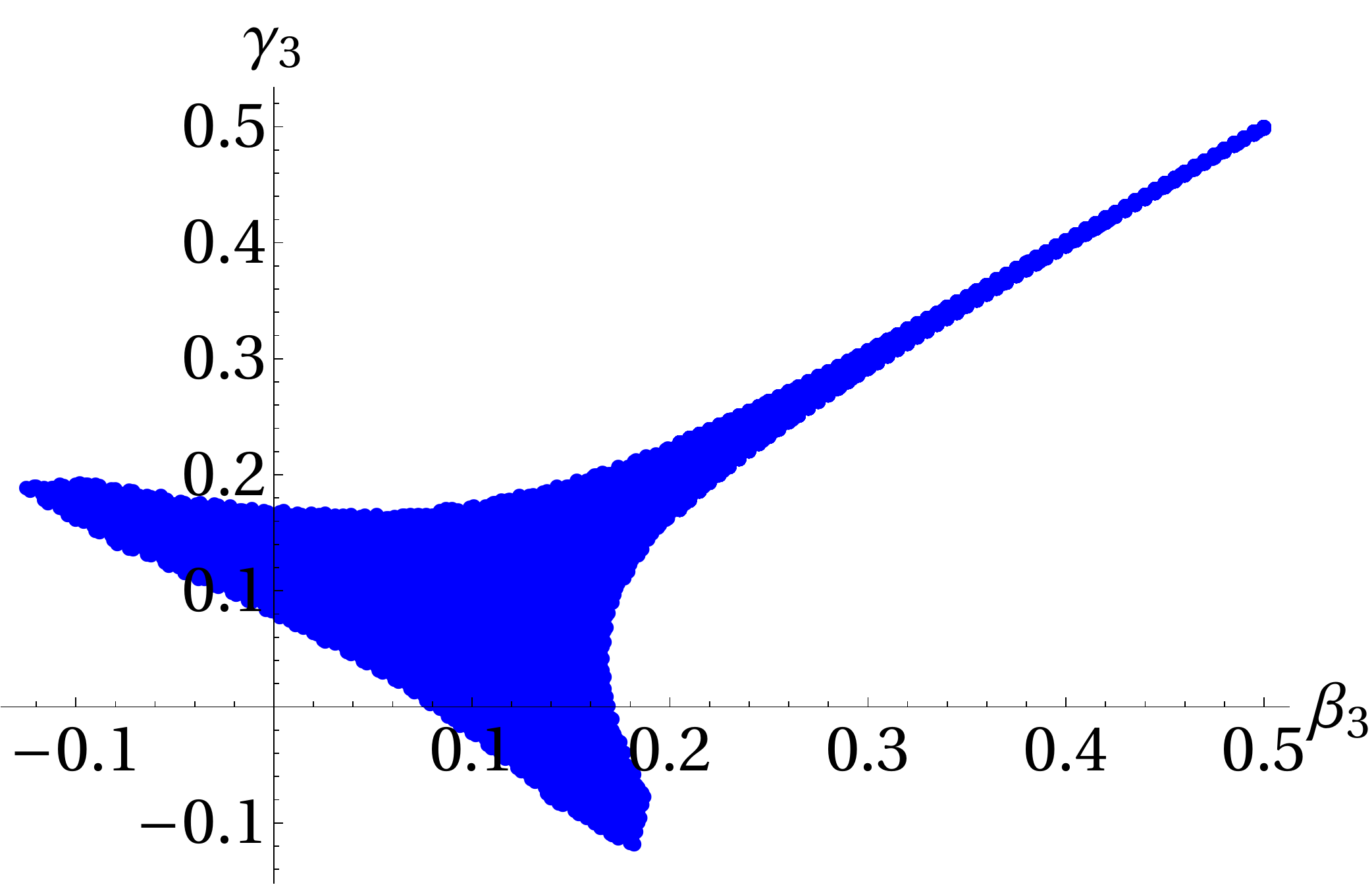}
\includegraphics[width=0.45\textwidth]{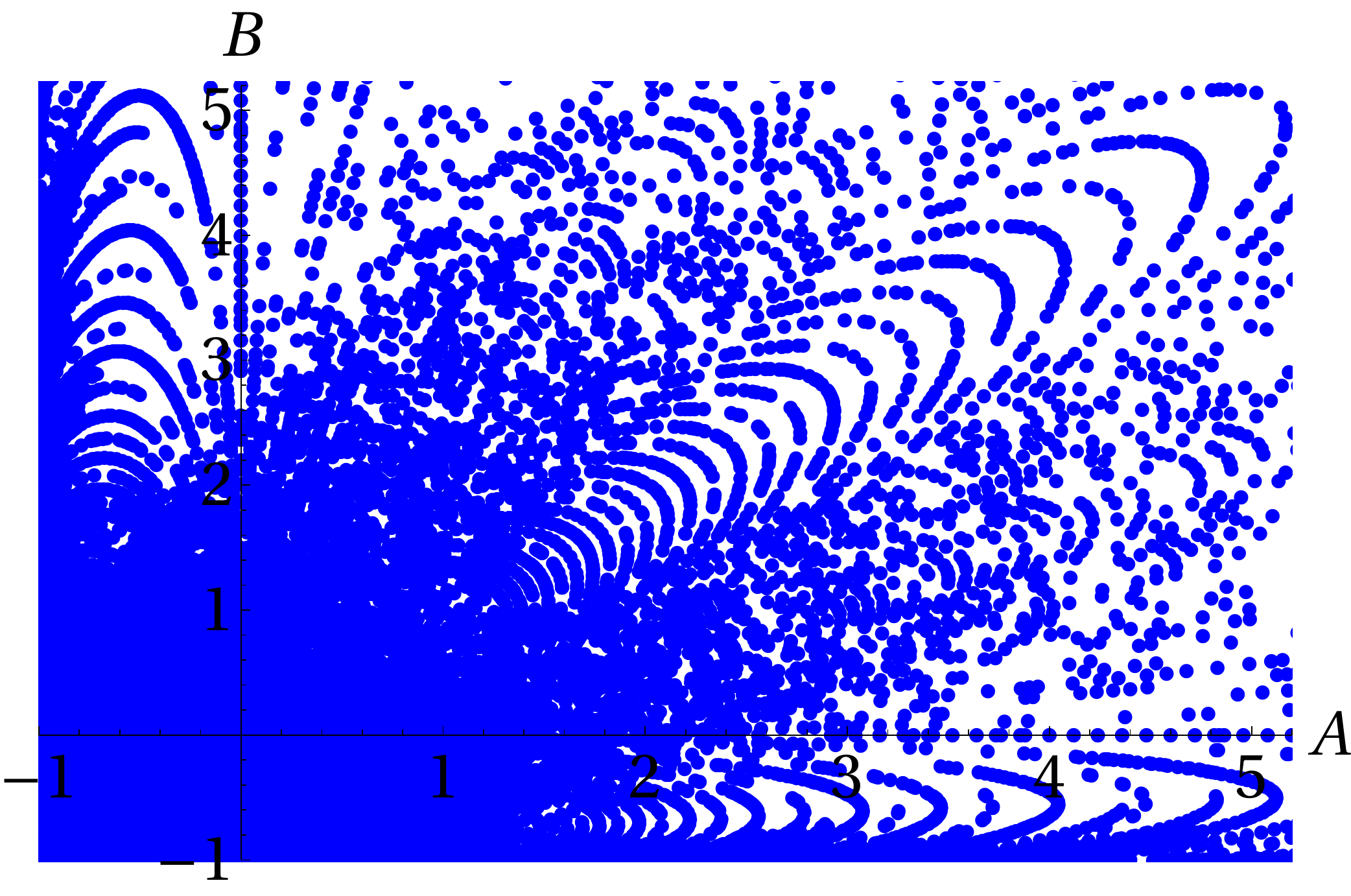}
\caption{The range of parameters for the AdS-Taub-NUT solutions with two squashings. \textit{Left panel}: the values of $\gamma_3$ and $\beta_3$ that lead to regular solutions. \textit{Right panel}: the resulting values of the squashing parameters $A$ and $B$.}\label{fig:rangesolsNUT}
\end{figure}

\begin{figure}[ht!] 
\centering
\begin{subfigure}[t]{0.45\textwidth}
\includegraphics[width=\textwidth]{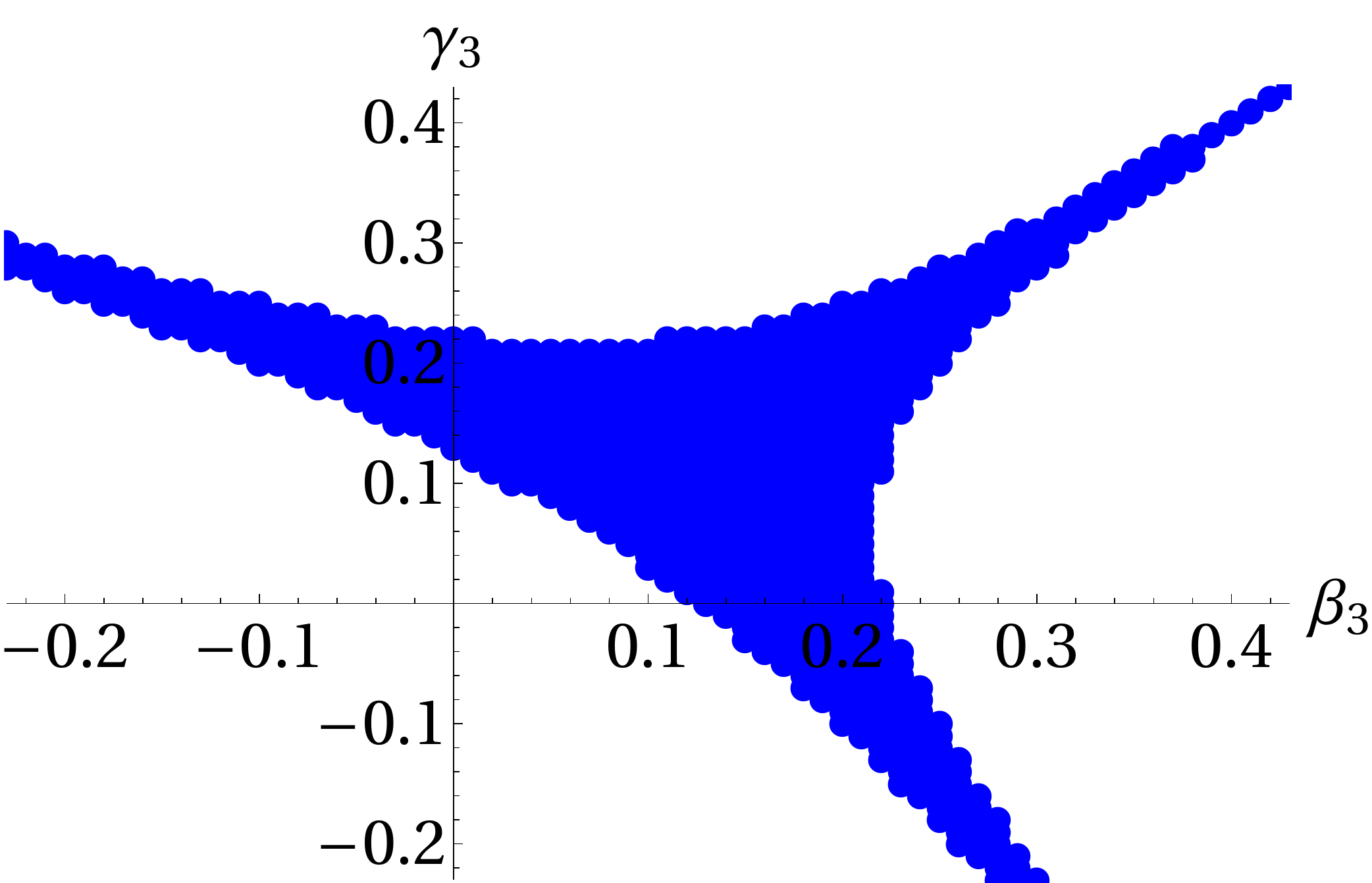}
\caption{$\Phi_0=1$}
\end{subfigure}
\begin{subfigure}[t]{0.45\textwidth}
 \includegraphics[width=\textwidth]{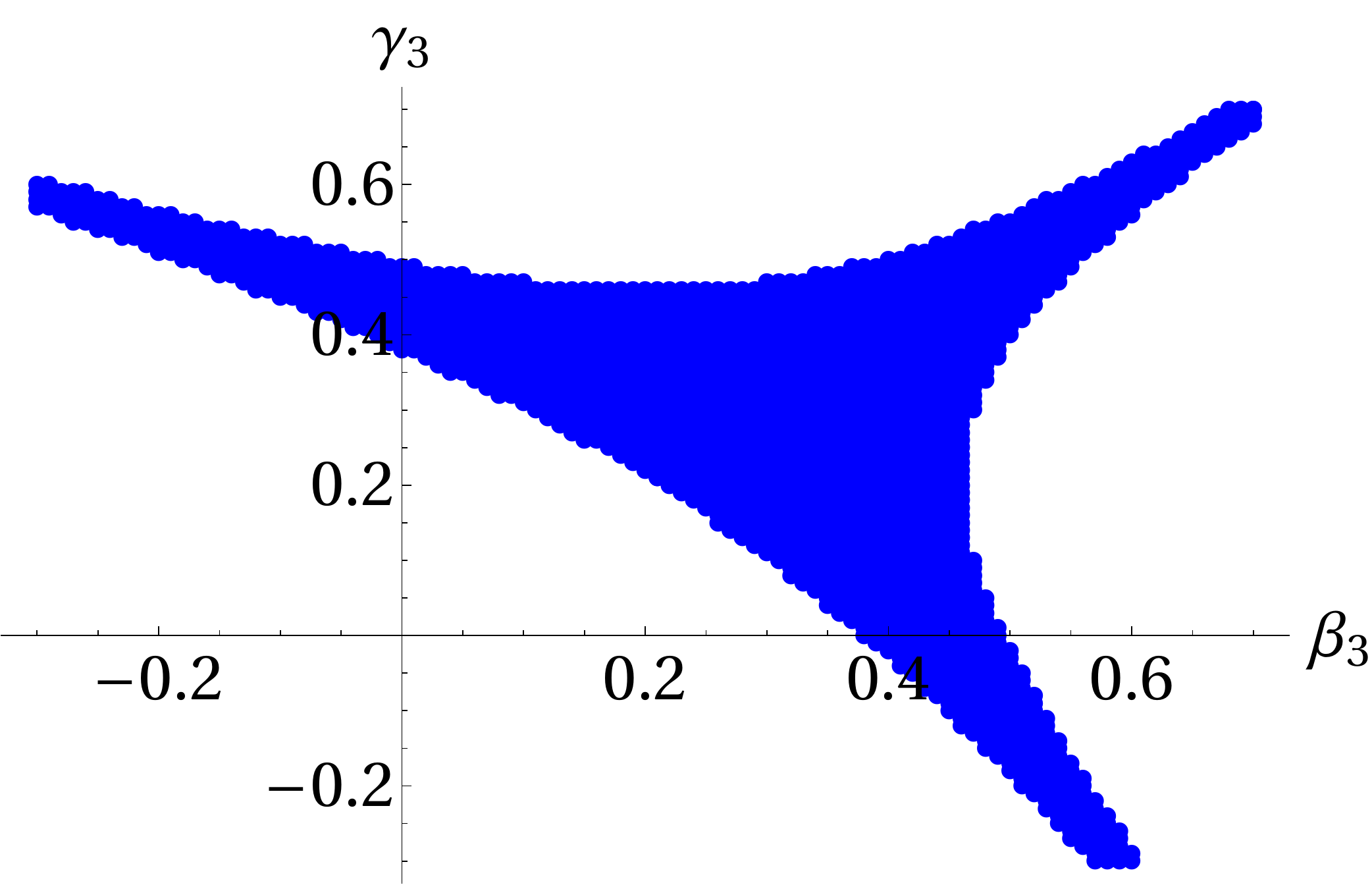}
 \caption{$\Phi_0=2$}
 \end{subfigure}
\caption{Initial conditions for $\gamma_3$ and $\beta_3$ that give rise to non-singular solutions for two different values of $\Phi_0$.}
\label{fig:icphi}
\end{figure}

It is worth discussing how we construct the numerical solutions of the full nonlinear equations of motion in \eqref{eom}. Let us start with the AdS-Taub-NUT solutions (taking $\tau=r$ in the above expansions). For these we picked real values for the parameters $\beta_3$, $\gamma_3$ and $\Phi_0$ in the IR expansion \eqref{icnut}. For each such value we then numerically integrated the equations of motion from $r=0$ to some large value of $r$. If the resulting numerical solution does not exhibit a singularity at an intermediate value of the radial coordinate $r$ we declared the solution to be asymptotically $AdS$ and read off the asymptotic parameters $B_0$, $C_0$ and $\alpha$ and $\beta$ in  \eqref{bval} which we then related to the squashing parameters $A$ and $B$ using \eqref{eqn:alphabetaBC}. 
As expected we find that there are no restrictions on the parameters $A$ and $B$, i.e. as we vary $\beta_3$ and $\gamma_3$ for a fixed $\Phi_0$ we can explore the whole $(A,B)$ plane. This is illustrated in Fig. \ref{fig:rangesolsNUT} in the case that there is no scalar field. If we take a non-zero value for $\Phi_0$ we will reach the same conclusion with the only difference  that the region in the ($\beta_3,\gamma_3$) plane that gives valid UV solutions shifts to higher values of $\beta_3$ and $\gamma_3$ when $\Phi_0$ increases as can be seen in Fig. \ref{fig:icphi}. 

The method to find the dS-Taub-NUT solutions is very similar to the AdS case, except that we now have to evaluate the equations of motion along a contour in the complex $\tau$-plane. Due to the special nature of the potential chosen here \cite{Hartle2008,Hertog2016}, the classical solutions all lie along a horizontal line at $\tau = i \pi / 2 +  t$. Evaluating the equations of motion along this line for large $\tau$ will learn us if the initial conditions give valid solutions that do not evolve into a singularity. The initial conditions  ($\beta_3$, $\gamma_3$) that give valid solutions are just minus the ones from the AdS solutions without a scalar field. However, when $\Phi_0$ gets increased, the initial conditions do not change significantly in this case.

%
%
The procedure we use to construct the AdS-Taub-Bolt solutions is again very similar. We start with the IR expansion in \eqref{icbolt}, vary the parameters $\gamma_0$ and $\gamma_4$ and integrate numerically the equations of motion. Finally, we read off the asymptotic parameters $B_0$ and $C_0$ from the behavior of the numerical solutions at large $r$ and deduce the corresponding values of $A$ and $B$ using the relation in \eqref{eqn:alphabetaBC}. However, there is an important difference between these solutions and the AdS-Taub-NUT solutions. For a fixed value of $B$ there are critical values of $A$ below/above which there are no AdS-Taub-Bolt solutions. This leads to curves in the $(A,B)$ plane and the AdS-Taub-Bolt solutions exist only for values of the squashing parameters that are below or above these critical curves.  Furthermore for every value of $(A,B)$ for which Bolt solutions exist there are two possible solutions of the equations of motion which we dub ``positive'' and ``negative''  branch. All of these features are extensions of the familiar behavior of the analytically known AdS-Taub-Bolt solutions with $B=0$ discussed in \cite{Chamblin1998, Emparan1999} and for which more details can be found in \cite{Bobev2016}.

The dS counterpart of these Bolt solutions can be found by making the AdS initial conditions imaginary. The other properties of the AdS-Bolt solutions carry over in a straightforward way, e.g. there is only a limited region in parameter space were these solutions exist, and there is always a positive and negative branch. 

\subsection{Holographic Renormalization}

To evaluate the action, it is easiest to use the on-shell version of \eqref{eqn:GRaction}:
\begin{align}
	I_{\rm E}=&-2\pi \int d \tau \; l_1(\tau) l_2(\tau) l_3(\tau) V(\Phi(\tau))\nonumber \\&   -2\pi \bigg( l_2(\tau) l_3(\tau)l_1'(\tau) +  l_1(\tau) l_3(\tau) l_2'(\tau) + l_1(\tau) l_2(\tau) l_3'(\tau)\bigg)_{\tau=\tau_{\rm c}} \ , \label{eqn:I_EOS}
\end{align}
where $\tau_{\rm c}$ is the cut-off radius at which we take the boundary $\partial\mathcal{M}$.

It is interesting to focus on the evaluation of the action for our AdS solutions (thus assuming that $\tau=r$, the radial AdS coordinate). Namely, these lead to the thermodynamical properties that were discussed in Section \ref{sec:AdSsols}. As usual for asymptotically locally AdS space, the value of the on-shell action diverges, and one needs to implement a regularization procedure. We apply the usual tools of holographic renormalization \cite{Henningson1998, Skenderis2002, Bianchi2001} which were used for the single squashed AdS-NUT/Bolt solutions without scalar deformations in \cite{ Emparan1999}. This procedure consists of regularizing the action by introducing a radial cut-off at $r=r_c$. In the next step one performs an asymptotic analysis of the terms in the action to see how the divergences behave. Then, one finds the counterterms which make the action in \eqref{eqn:I_EOS} finite by writing the diverging terms in the regularized action as fields on the boundary.

The asymptotic form of the original on-shell gravitational action in \eqref{eqn:I_EOS} reads\footnote{Notice that to expand the scalar field, we have to assume that it rolled down its potential such that the potential is approximated by $V(\Phi)\approx \Lambda - \Phi^2 + \ldots$.}
\begin{align}\label{eqn:Ser}
I_{\rm E}&= -\pi\bigg(4A_0 B_0 C_0 \e^{3r} -\frac1{4 A_0B_0C_0}\Big(A_0^4+B_0^4+C_0^4-2B_0^2 C_0^2 -2A_0^2 B_0^2 \nonumber\\
&-2A_0^2 C_0^2+ (A_0B_0C_0)^{4/3}\alpha^2/2\Big)\e^{r}+\mathcal{O}(1)\bigg)_{r=r_{\rm c}}\;,
\end{align}
where the constant part is determined by the IR behavior of the bulk. To cancel these divergences one can add the following covariant counterterms also found by \cite{Emparan1999} for pure gravity and \cite{deHaro2000} for gravity and scalar matter
\begin{align}
 S_{\rm ct}=\frac{1}{8 \pi } \int_{\partial \mathcal{M}} d^3x \sqrt{h}\left(2+\frac{\mathcal{R}}{2} +\frac{\Phi^2}2 \right) \   , \label{eqn:actionct}
\end{align}
where $\mathcal{R}$ is the scalar curvature of the boundary metric $h_{ij}$. 
Evaluating this counterterm action  yields
\begin{align}
S_{\rm ct}=\pi\frac{2({l_1}^2{l_2}^2+{l_2}^2{l_3}^2+{l_1}^2{l_3}^2)+2{l_1}^2{l_2}^2{l_3}^2(4+\Phi^2)-{l_1}^4-{l_2}^4-{l_3}^4}{2 l_1 l_2 l_3}\;.
\end{align}
Substituting our asymptotic expansions of the functions $l_i(r)$ and the scalar field \eqref{bval}  gives
\begin{align}\label{eqn:Scter}
S_{\rm ct}&= \pi\bigg(4A_0 B_0 C_0 \e^{3 r} -\frac1{4 A_0B_0C_0}\Big(A_0^4+B_0^4+C_0^4-2B_0^2 C_0^2 -2A_0^2 B_0^2 \nonumber\\
&-2A_0^2 C_0^2+ (A_0B_0C_0)^{4/3}\alpha^2/2\Big)\e^{r}-\frac{1}{12}\alpha\beta+\mathcal{O}(\e^{-r})\bigg)_{r=r_{\rm c}}\;.
\end{align}
As expected the sum 
\begin{equation}
I^{\rm ren}_{\rm E} = I_{\rm E}+S_{\rm ct}\;,
\end{equation}
remains finite in the $r=r_{\rm c}\to \infty $ limit and thus this sum can serve as a good regularized on-shell action.

Since our gravitational solutions are constructed numerically, evaluating the regularized on-shell action $S_{\rm ren}$ is tricky. The difficulty comes from the fact that one has to add a large positive and a large negative number and this could lead to numerical instabilities. To remedy this, we found it useful to employ the following strategy. From \eqref{eqn:Ser} we know how the on-shell action diverges at large values of $r$. We can thus evaluate numerically this on-shell action at large but finite values of $r$ and fit the resulting values to the function
\begin{align}\label{eqn:fitfun}
f= D e^{3r_c} + E e^{2r_c}+ F e^{r_c} + G + H e^{-r_c} + I e^{-2r_c} \ .
\end{align}
We can then read of the coefficients $D$, $E$, and $F$ and use the first three terms in \eqref{eqn:fitfun} as our numerical counterterm action that should be added to $I_{\rm E}$ to produce a finite result. If there is no scalar field, the value of $G$ is the final value for the renormalized action. 

In the case of the AdS theories we considered we have to do some more work if there is a non-zero scalar field. Because in the free $O(N)$ model we want to analyse a current of dimension $\Delta=1$, the scheme of alternate quantization comes into play and we have to evaluate the action in terms of $\beta$. To achieve this, we have to perform a Legendre transform by adding the following boundary term \cite{Klebanov1999}
\begin{align}
	S_-=-\frac{1}{8\pi}\int_{\partial \mathcal{M}} d^3x\sqrt{h} \Phi \pi_{\Phi} \ .
\end{align}
To have a well-defined boundary term that is invariant under shifts in $r$, the momentum $\pi_{\Phi}$ has to be understood as the renormalized momentum defined by \cite{Papadimitriou2007}
\begin{align}
	  \pi_{\Phi}=\frac1{\sqrt{h}}\frac{\delta (I_{\rm E}+ S_{\rm ct})}{\delta \Phi}=\partial_r \Phi +\Phi \ .
\end{align}
After plugging in the asymptotic expansions in this term, we get 
\begin{align}
S_-= \frac{\pi}{4} \alpha\beta + \mathcal{O}(e^{2r})\ . \label{eqn:Smin}
\end{align}

Thus to get the complete action, we have to add to $G$ the constant parts from \eqref{eqn:Scter} and \eqref{eqn:Smin}. As a consistency check of our numerical results we should find that the coefficient $E$ in \eqref{eqn:fitfun} is approximately $0$. We found that this value usual was of the order of $\mathcal{O}(10^{-10})$, but became bigger, up to order $\mathcal{O}(10^{-4})$, for squashings close to -1. 

By adding the extra boundary terms \eqref{eqn:Smin}, we obtain Neumann boundary conditions, which correspond to a multi-trace deformation of the dual QFT\cite{Papadimitriou2007}. The exact one-point function of the holographic dual with source $J=\beta$ can now be calculated by
 \cite{deHaro2000,Bianchi2001, Skenderis2002, Papadimitriou2007}
\begin{align}
\langle \mathcal{O}\rangle = \lim_{r_c\rightarrow \infty} \frac{\delta(I_{\rm E}+S_{\rm ct} +S_-)}{\delta \beta} = \alpha \ .
\end{align}

Notice that holographic renormalization corresponds to a minimal renormalization scheme, it is possible to add other finite counterterms that may give a contribution like $\Phi^3$, $R\Phi$. However, we require our renormalization scheme to be the same for the dS and AdS solutions. To evaluate the dS actions it is sufficient to evaluate the action along the path ${\cal C}'$ in the complex $\tau$ plane, see Fig. \ref{contour}. Due to the properties of the dS solutions the real part of the action will tend to a constant along the Lorentzian part of the contour, while all the divergent terms are encapsulated by the imaginary part of the action, which agree perfectly with the counterterms in \eqref{eqn:actionct} \cite{Hertog2011} eliminating the need for other finite counterterms.

\section{Numerical regularization of the CFT}
\label{app:CFTreg}

In this appendix, we give more details on the numerical techniques used to calculate the free energy of the $O(N)$ vector model. As already mentioned, our regularization scheme is based on the same technique as the one used in \cite{Anninos2012} and was succesfully applied to massless scalar fields on a double squashed sphere in \cite{Bobev2016} and to scalars and fermions in a number of odd dimensions in \cite{Bobev2017}. 

The goal will be to calculate the free energy in \eqref{eqn:LogZGeneral}
\begin{align}
F=\frac{N}{2}\log\left(\det \left[\frac{-\nabla^2+m^2+\frac{R}{8}}{\Lambda^2}\right]\right) \ ,
\end{align}
where $\Lambda$ is an energy cutoff. In general the free energy is a diverging quantity which we need to regularize. To do this, we use a heat-kernel type regulator \cite{Anninos2012,Vassilevich2003} 
\begin{align}
	\log \det\left[\frac{-\nabla^2+m^2+\frac{R}{8}}{\Lambda^2}\right] = \sum_i \int_{\Lambda^{-2}}^{\infty}\frac{dt}{t} e^{-t\lambda_i} \, ,\label{eqn:regulator}
\end{align}
where we have denoted the eigenvalues of the Laplacian operator by $\lambda_i$. This expression yields the determinant for modes whose energies are less than a ``soft'' cutoff $ \Lambda$, while cutting off the sum exponentially above this value. In particular, for $\lambda  \ll \Lambda^2$, one finds
\begin{align}
	- \int_{\Lambda^{-2}}^{\infty}\frac{dt}{t} e^{-t\lambda_i} =\log(\lambda_i/ \Lambda^{2})+\mathcal{O}(\lambda_i/ \Lambda^{2}) \ ,\label{eqn:upperincompleteGamma}
\end{align}
while for $\lambda  \gg  \Lambda^2$,
\begin{align}
	- \int_{\Lambda^{-2}}^{\infty}\frac{dt}{t} e^{-t\lambda_i} =-e^{-\lambda_i/ \Lambda^{2}}\left(\frac{\Lambda^{2}}{\lambda_i }+ \mathcal{O}\left(\frac{\Lambda^{4}}{\lambda_i^{2}}\right)\right) \ .
\end{align}
The integral can now be split into two pieces, one with low energy modes (IR) and another with high energy modes (UV)
\begin{align}\label{eqn:detIRplusUV}
	\log \det\left[\frac{-\nabla^2+m^2+\frac{R}{8}}{\Lambda^2} \right] =  \textrm{det}_{UV} + \textrm{det}_{IR} \, ,
\end{align}
where 
\begin{equation}
\begin{split}
	\textrm{det}_{UV} &\equiv  \sum_{i}\int_{\Lambda^{-2}}^{\delta} \frac{dt}{t}  m_i e^{-t \lambda_{i}} \ , \\
	\textrm{det}_{IR} &\equiv \sum_{i} \int_{\delta}^{\infty} \frac{dt}{t} m_i e^{-t \lambda_{i} }= \sum_{i}m_i \Gamma(0,\lambda_i \delta)\ . \label{eqn:detUV}
\end{split}
\end{equation}
Here, $\Gamma(a,z)$ is the incomplete Euler Gamma function, $\delta$ is an arbitrary positive real number that we can change to get a better convergence, and $m_i$ is the multiplicity of the eigenvalue $\lambda_i$.

The sum over the IR modes converges for large $\lambda_i$, and can therefore be done numerically if the maximum number of eigenvalues is chosen large enough. The divergences are all contained in det$_{UV}$. These have to be controlled and subtracted. If the eigenvalues $\lambda_i$ are known analytically, it is possible to apply the Euler-Maclaurin formula to estimate the behavior of the sum in det$_{UV}$ \cite{Anninos2012}.

If the eigenvalues are only known numerically, e.g. for the double squashed sphere, the divergences can be controlled by evaluating the sums in \eqref{eqn:detUV} for different values of $t$, which is possible because these sums converge for large enough eigenvalues. The maximum number of eigenvalues will be called $n_{\rm max}$.  We let $t$ go from a starting value $t_{\rm init}$ to $\delta$ with stepsize $\Delta t$. This we can fit and integrate, giving us det$_{UV}$ as a function of the cutoff.

In order to know which function we need to use for the fit, it is necessary to know the behavior of the integrand as a function of $t$. The divergences arise from covariant counterterms such as the metric and curvature scalar of the squashed sphere\footnote{Notice that in three dimensions there are no conformal anomalies.}
\begin{align}
{\rm divergences} = A \Lambda^3 \int d^3x\sqrt{g} +  \Lambda \int d^3x\sqrt{g}(B  R +C m^2) \ . \label{eqn:div}
\end{align}
Because $t \sim \Lambda^{-2}$ this means that the necessary fit function is of the form
\begin{align}
{\rm fit}= \frac{A_1 }{t^2} + \frac{A_2}{t^{3/2}} + \frac{A_3}{t} + \frac{A_4}{t^{1/2}} +\cdots \ .\label{fittdef}
\end{align}
From \eqref{eqn:div} it is clear that $A_1$ and $A_3$ should vanish. This can be used as a check of our numerical procedure. If the fitted values of $A_1$ and $A_3$ are small enough, we can trust our fit. To give a flavor of our results for the coefficients in \eqref{fittdef} , we present explicitly the values for $A_i$ for different values of $\alpha$ and $\beta$ in Table \ref{tbl:coeff}. To obtain the values in Table \ref{tbl:coeff}, we took for det$_{UV}$, $n_{\rm max}=1500$, and we start our fit from $t_{\rm init} = 10^{-4}$ with step size $\Delta t= 10^{-4}$. For the parameter $\delta$  we chose $\delta = 10^{-2}$. 

After the fit function in \eqref{fittdef} is obtained in this way, we have good control over the divergences in ${\rm det}_{UV}$ which we then subtract and, after that, evaluate the integral in \eqref{eqn:detUV}. The result is then used to obtain the finite regularized value of the free energy.
\begin{table*}
  \centering
  \begin{tabular}{r|r||c|c|c|c} 
   $\alpha$ &$\beta$&  $A_1$ & $A_2$  & $A_3$ & $A_4$   
   \\
    \hline \hline 
{$-0.5$} & $2.0$ & {$1.0006\cdot 10^{-11}$}&  {$0.361801$}&  {$2.1679\cdot 10^{-7}$} &  {$0.184913$} \\ \hline
$-0.8 $& $0.0$ & {$3.2026\cdot10^{-11}$}&  {$0.990832$}&  {$6.9398\cdot 10^{-7}$} &  {$0.31374$} \\ \hline
$1.2 $& $0.0$& {$1.21575 \cdot 10^{-14}$}&  {$0.298747$}&  {$2.61598\cdot 10^{-10}$} &  {$ 0.207765$}\\ \hline
$2.8 $&$-0.4$&  {$6.69228 \cdot 10^{-12}$}&  {$0.293459$}&  {$1.44942 \cdot 10^{-7} $}&  {$ 0.166992$}\\ \hline
$21.2 $&$10.1$&  {$-1.32555 \cdot 10^{-7}$}&  {$0.0282593$}&  {$-2.52508 \cdot 10^{-2}$} &  {$ -0.342511$} \\ \hline
$-0.2 $& $12.1$& {$4.57221 \cdot 10^{-11}$}&  {$0.136878$}&  {$9.95477 \cdot 10^{-7}$} &  {$ 0.0876136$} \end{tabular}
  \caption{The coefficients of the fit function in \eqref{fittdef} for different values of $\alpha$ and $\beta$ with fixed $m^2=-0.4$. Whenever the Ricci curvature becomes very negative the coefficient $A_1$ and $A_3$ differ more significantly from the expected zero value and thus our numerical results are less accurate.  }
\label{tbl:coeff}   
\end{table*}    

\renewcommand{\leftmark}{\MakeUppercase{Bibliography}}
\phantomsection
\bibliographystyle{JHEP}
\bibliography{biblio}

\end{document}